\documentclass[11pt]{article}
\pdfoutput=1
\usepackage{jcappub,}%dsfont}
\input{colordvi.tex}

\def\nn{\nonumber}
\def\({\left(}
\def\){\right)}
\def\[{\left[}
\def\]{\right]}

\def\exd{{\hbox{d}}}

\def\nn{\nonumber}
\def\({\left(}
\def\){\right)}
\def\[{\left[}
\def\]{\right]}

\def\bk{{\bf k}}
\def\bq{{\bf q}}
\def\bx{{\bf x}}
\def\bp{{\bf p}}

\definecolor{gold}{rgb}{1.0, 0.84, 0.0}
\definecolor{maroon}{rgb}{.25,0,0}
\definecolor{darkorange}{rgb}{1.0, 0.55, 0.0}
\definecolor{corn}{rgb}{0.98, 0.93, 0.36}
\definecolor{bronze}{rgb}{0.8, 0.5, 0.2}
\definecolor{darkgreen}{cmyk}{0.85,0.2,1.00,0.2}

\bibstyle{JHEP}

\title{Phenomenology of fermion production during axion inflation}

\author[a]{Peter Adshead,}
\affiliation[a]{Department of Physics, University of Illinois at Urbana-Champaign, Urbana, Illinois 61801, U.S.A.}
\author[a]{Lauren Pearce,}
\author[b]{Marco Peloso,}
\affiliation[b]{William I. Fine Theoretical Physics Institute, School of Physics and Astronomy, University of Minnesota, Minneapolis, MN 55455, U.S.A.}
\author[c]{Michael A. Roberts,}
\affiliation[c]{Amherst Center for Fundamental Interactions, Department of Physics, University of Massachusetts, Amherst, MA 01003, U.S.A.}
\author[c]{Lorenzo Sorbo}

\emailAdd{adshead@illinois.edu}
\emailAdd{lpearce@illinois.edu}
\emailAdd{peloso@umn.edu}
\emailAdd{mroberts@umass.edu}
\emailAdd{sorbo@physics.umass.edu}

\abstract{We study the production of fermions through a derivative coupling with a pseudoscalar inflaton and the effects of the produced fermions on the scalar primordial perturbations. We present analytic results for the modification of the scalar power spectrum due to the produced fermions, and we estimate the amplitude of the non-Gaussianities in the equilateral regime. Remarkably, we find a regime where the effect of the fermions gives the dominant contribution to the scalar spectrum while the amplitude of the bispectrum is small and in agreement with observation. We also note the existence of a regime in which the backreaction of the fermions on the evolution of the zero-mode of the inflaton can lead to inflation even if the  potential of the inflaton is steep and does not satisfy the slow-roll conditions. 
}

\begin{document}

\begin{flushright}  ACFI-T18-05, UMN--TH--3713/18  \end{flushright}

\maketitle
\flushbottom
%\tableofcontents

%\baselineskip=17.63pt 

%%%%%%%%%%%%%%%%%%%%%%%%
\section{Introduction}%%
\label{sec:Intro}%%%%%%%
%%%%%%%%%%%%%%%%%%%%%%%%

There is strong evidence for a phase of accelerated expansion --- inflation --- in the early Universe \cite{Guth:1980zm,Linde:2005ht}. The observed density fluctuations are red-tilted, adiabatic, and Gaussian to a high degree \cite{Ade:2015xua, Ade:2013ydc}, in accord with generic predictions of single-field, slow-roll models of inflation. Slow-roll inflation predicts an amplitude of tensor perturbations (gravitational waves) which is smaller than that of the scalar ones, perhaps quite a bit smaller.  The imprint of primordial gravitational waves, in the form of a primordial $B$-mode polarization of the Cosmic Microwave Background (CMB), has not yet been measured. Current measurements restrict the tensor-to-scalar ratio to $r\lesssim0.07$ \cite{Ade:2015tva, Array:2015xqh}, putting pressure on 
many large field models of inflation  \cite{Ade:2015lrj}. Future experiments will reach the sensitivity $\sigma_r \sim 0.001$ \cite{Abazajian:2016yjj}, which motivates the study of mechanisms for the inflationary expansion and for the production of the primordial perturbations that go beyond the most minimal slow-roll inflationary models. 

Axion, or natural, {inflation} is a class of models of inflation in which the flatness of the inflation potential is protected by an (approximate) shift symmetry~\cite{Freese:1990rb}. This class of models naturally gives rise to the relatively large-amplitude primordial gravitational waves targeted by next-generation experiments (see, for example, \cite{Silverstein:2008sg, McAllister:2008hb,Peloso:2015dsa,DAmico:2017cda}). The shift symmetry that protects the form of the potential is respected by the coupling of the axion to matter, and requires that the axion couples derivatively to other matter fields in order to facilitate reheating. The operators of lowest dimensionality that couple the inflaton to matter appear at mass-dimension 5 and are the coupling to a gauge field ${\Delta L} = \frac{\phi}{f} F {\tilde F}$ (where $\phi$ is the pseudoscalar inflaton, $f$ a mass scale often denoted as `axion decay constant', $F$ a gauge field strength, and ${\tilde F}$ its dual) and the coupling to a fermion field  ${\Delta L} = \frac{\partial_\mu \phi}{f} \, {\bar \psi} \gamma^\mu \gamma_5\psi$. It has long been known  that the coupling to gauge fields can lead to an exponentially large gauge field amplification during inflation  \cite{Garretson:1992vt}, which can in turn lead to rich phenomenological consequences  (see \cite{Pajer:2013fsa} for a review) such as steep inflation \cite{Anber:2009ua}, thermalized inflation \cite{Ferreira:2017lnd,Ferreira:2017wlx}, magnetic field production \cite{Prokopec:2001nc, Anber:2006xt,Caprini:2014mja, Fujita:2015iga, Adshead:2016iae,Caprini:2017vnn}, large non-Gaussianity \cite{Barnaby:2010vf, Barnaby:2011vw, Barnaby:2011qe}, chiral gravitational wave production \cite{Sorbo:2011rz,Cook:2013xea, Shiraishi:2013kxa}, and the generation of primordial black holes \cite{Linde:2012bt, Bugaev:2013fya, Erfani:2015rqv, Garcia-Bellido:2016dkw}. The coupling to fermions has been comparatively much less studied. Given the rich phenomenology of derivatively coupled gauge fields, and the relevance of axion-driven inflation to the next generation of CMB experiments, in this work we aim to take a step toward characterizing the phenomenology of derivatively coupled fermions during a phase of pseudoscalar-driven inflation.

It is typically assumed that fermions do not play an important role during inflation. Because of Pauli blocking, their energy in long wavelength modes is necessarily small --- the small phase space in the infrared cannot be compensated by a large occupation number. Furthermore, the gravitational creation of fermions by the expanding Universe depends on their being massive. On the one hand, massless fermions are conformal, and are thus not created gravitationally by the expansion \cite{Parker:1968mv}, while on the other hand heavy particles decouple and are not excited. As a consequence, the maximum gravitational production of a fermion field of mass $m$ in (quasi) de Sitter space with Hubble parameter $H$ is obtained when $m \approx H$. In this case, there is only one scale, $H$, controlling the system, and the energy density of the fermions $\sim H^4$  is too small to induce any observable effects (with the possible exception of super heavy-dark matter, see for example~\cite{Kuzmin:1998kk,Chung:2011ck}). 

The coupling of the fermions to the pseudoscalar inflaton introduces the additional scale  $\dot\phi/f$, which can be much larger than the Hubble parameter. In the regime on which we focus in this work, where $\dot\phi/f\gg H$, fermion states can be populated up to scales $\sim\dot\phi/f$, which implies that the total number density of fermions can be parametrically larger than $H^3$~\cite{Adshead:2015kza}. These fermions can therefore have large energy densities during inflation, and in this work we study the effects of their backreaction on the inflaton and on its fluctuations. 

The first studies of this system have been performed by one of us in ref.\ \cite{Adshead:2015kza}. In that study, solutions of the Dirac equation including the effects of the homogeneous rolling inflaton mode were obtained along with the occupation number of the fermions during inflation and the oscillations in the phase immediately following inflation \cite{Adshead:2015kza}. The effect of the axion-fermion coupling is to helically-bias the production of  fermions leading to a net helicity asymmetry. This helicity asymmetry then leads to the possibility that the (chiral, given the parity-violating nature of the system) fermions could lead to successful leptogenesis \cite{Adshead:2015jza}. Subsequently, ref.\ \cite{Anber:2016yqr} studied the possibility that these fermions source gravitational waves during inflation, and found that in this system a chiral component of the spectrum of primordial tensors is generated.  Finally, ref.\ \cite{Alexander:2017bxe} found that these chiral fermions can generate circularly polarized photons, or V-modes after reheating.

In the present work we study the axion-fermion system in a different field basis. Using an axial transformation of the fermions $\psi \to e^{-i\phi \gamma_5}\psi$, we show (see section~\ref{sec:Set_Up} below) that the Lagrangian with a pseudoscalar derivatively coupled to a fermion is equivalent to a Lagrangian in which the fermion has a time-dependent mass term $m\, \bar{\psi}\,e^{-2i\gamma_5\phi}\psi$ --- see eq.~(\ref{eq:lagr_psi}).  The main motivation for this  reformulation is that it makes the behavior in the limit $m\to 0$ clear. In this limit,  the coupling of $\psi$ to the inflaton manifestly vanishes.  In the basis used in ref.\ \cite{Adshead:2015kza}, the coupling also vanishes in this limit. However, this vanishing is only apparent \emph{after} integrating the interaction by parts, and then using the Dirac equation. While the physics in either basis must of course be the same, we demonstrate that in the limit of interest ($\dot\phi/f\gg H$) the Hamiltonian in the basis of ref.\ \cite{Adshead:2015kza} does not have a convenient perturbative expansion. Consequently, the conclusions of this work regarding the occupation numbers are different to the conclusions reached in the work ref.\ \cite{Adshead:2015kza}.

The main focus of this paper is on the contribution that the non-vacuum fermion modes give to the spectrum and bispectrum of the scalar metric perturbations. We find that, in the regime $\mu\equiv m/H\lesssim 1$, $\xi\equiv \dot\phi/(2fH)\gg 1$, the occupation number of the fermions scales as $\mu^2/\xi$ for momenta up to a cutoff $\approx H\,\xi$, so that the total number density scales as $\mu^2\,\xi^2$. The modification to the power spectrum scales as $\frac{H^4}{f^4}\mu^2\log\xi$, and the bispectrum  as $\frac{H^6}{f^6}\frac{\mu^2}{\xi}$. This implies that in the regime of large $\mu\,\xi$ the system can be in a regime where the two-point function of the scalar fluctuations dominates over the vacuum contribution while the parameter $f_{NL}$, which measures the departures from Gaussianity, is small and in agreement with constraints from observations.  This result is surprising if one considers that the origin of the sourced spectrum is quadratic in the (Gaussian) fermion field, which might lead one to expect strong non-Gaussianiaty.  However, this result can be explained in terms of the central-limit theorem: the numerous fermion modes that contribute to the sourced spectrum sum incoherently, leaving a (quasi) Gaussian signal. 

While most of our analysis is performed in the regime where the backreaction of the produced fermions on the background dynamics is negligible, we consider also the situation where this is not the case.  and we find that strong backreaction effects can allow slow-roll inflation even if the potential for $\phi$ does not obey the usual slow-roll conditions $|V'|\ll V/M_P$, $|V''|\ll V/M_P^2$. The argument we have just presented suggests that the perturbations should also be highly Gaussian in this regime. However, we leave the (challenging) analysis of perturbations in this regime to future work.

This paper is organized as follows. In section~\ref{sec:Set_Up} we discuss the quantization of $\psi$ on the time-dependent background provided by the (quasi) de Sitter geometry along with the rolling inflaton, and we evaluate the resulting occupation number. As expected, the parity-violating nature of the system implies different occupation numbers for the two helicities of $\psi$. We demonstrate that in the limit $m\to 0$ the occupation number of fermions of both helicities vanishes, as a consequence of the conformal and chiral symmetry of the system.  In section~\ref{sec:backr}, we study the backreaction of the fermions on the zero-mode, or homogeneous, inflaton background. In section~\ref{sec:ps} we analyze the modifications to the inflationary power spectrum induced by the presence of a nonvanishing occupation number for the fermions, while in section \ref{sec:bs} we study the bispectrum. In section \ref{sec:strong}, we explore the possibility of slow-roll inflation on steep potentials in the limit of very strong backreaction. The work is concluded by section \ref{sec:disc}, and by several appendices where we present the details of our computations.

%%%%%%%%%%%%%%%%%%%%%%%%%%%%%%%%%%%%%%%%%%%%%%%%
\section{Fermion production during inflation}%%%
\label{sec:Set_Up}%%%%%%%%%%%%%%%%%%%%%%%%%%%%%%
%%%%%%%%%%%%%%%%%%%%%%%%%%%%%%%%%%%%%%%%%%%%%%%%

In this section we study the production of fermions during axion inflation and obtain solutions to the Dirac equation for a fermion coupled to the slowly-rolling ($\dot\phi=$constant) pseudoscalar. In particular, we  compute the resulting occupation number for the right- and the left-handed components of the fermion.

We consider the theory of a pseudoscalar inflaton $\phi$ interacting with a Dirac fermion $X$ through a derivative interaction with coupling constant $1/f$
\begin{align}
{\cal L}=a^4 \left\{ \bar X\left[i\,\left(\tilde{\gamma}^\mu\,\partial_\mu+\frac{3}{2}\,\frac{a'}{a}\tilde{\gamma}^0\right)-m-\frac{1}{f}\tilde\gamma^\mu\,\gamma^5\,\partial_\mu\phi\right]X + \frac{1}{2} \left( \partial \phi \right)^2 - V \left( \phi \right) \right\}\, .
\end{align}
Here the $\tilde\gamma$-matrices in  flat Friedmann-Lema{\^i}tre-Robertson-Walker spacetime with scale factor $a$ are related to those in Minkowski spacetime by $\tilde\gamma^\mu=\gamma^\mu/a$, while $\gamma^5=i\,a^4\,\tilde\gamma^0\tilde\gamma^1\tilde\gamma^2\tilde\gamma^3=i\,\gamma^0\gamma^1\gamma^2\gamma^3$. We neglect metric fluctuations\footnote{More precisely, we study scalar metric perturbations in the spatially flat gauge, neglecting the presence of the shift and lapse scalar factors which provide slow-roll suppressed contributions to the spectra.} and treat the background as fixed de Sitter spacetime.

Throughout this work we use conformal time and ``mostly minus'' signature for our metric, and  we use the Dirac representation for the $\gamma$ matrices. Specifically, 
\begin{eqnarray}
&&\gamma^0=\left(
\begin{array}{cc}
{1} & 0 \\
0 & -{1}
\end{array}
\right),\,\,\,
\gamma^i=\left(
\begin{array}{cc}
0 & \sigma_i \\
-\sigma_i & 0
\end{array}
\right),\,\,\,
\gamma^5=\left(
\begin{array}{cc}
0 & {1} \\
{1} & 0
\end{array}
\right).
\end{eqnarray}
The fermions are canonically normalized by redefining $Y=X\,a^{3/2}$, so that
\begin{align}\label{eq:lagr_Y}
{\cal L}=\bar Y\left[i\,\gamma^\mu\,\partial_\mu-m\,a-\frac{1}{f}\gamma^\mu\,\gamma^5\,\partial_\mu\phi\right]Y 
+ \frac{1}{2} a^2 \eta^{\mu \nu} \partial_\mu \phi \partial_\nu \phi - a^4 V \left( \phi \right) \,. 
\end{align}
Next, we perform one more redefinition of the fermion field,
\begin{equation} 
Y = {\rm e}^{-i \gamma^5 \phi / f} \, \psi\,,
\end{equation} 
which yields the Lagrangian
\begin{eqnarray}\label{eq:lagr_psi}
{\cal L} =  \bar \psi \left\{ i\,\gamma^\mu\,\partial_\mu-m\,a \left[ \cos \left( \frac{2 \phi}{f} \right) - i \gamma^5 \sin \left( \frac{ 2 \phi }{ f } \right) \right] \right \} \psi 
+ \frac{1}{2} a^2 \eta^{\mu \nu} \partial_\mu \phi \partial_\nu \phi - a^4 V \left( \phi \right) \,.
\end{eqnarray} 
The latter field redefinition is motivated by two considerations. First, as discussed in the introduction, by writing the Lagrangian in terms of $\psi$ it is apparent that the inflaton decouples from the fermion in the limit $m\to 0$. This decoupling is not as evident when the Lagrangian is in the form of eq.\ (\ref{eq:lagr_Y}). Second, in order to determine the occupation number for the fermions we resort to the usual technique of the Bogolyubov coefficients, which relies on the diagonalization of the portion of Hamiltonian that is quadratic in the fields. In the formulation of eq.\ (\ref{eq:lagr_Y}) the momentum conjugate to $\phi$, which is needed to compute the Hamiltonian, is given by $\Pi_\phi=a^2\,\dot\phi-\frac{1}{f}\bar{Y}\gamma^0\gamma^5 Y$, which contains a term that is quadratic in the fermion field (this should be compared with the simpler expression $\Pi_\phi=a^2\,\dot\phi$ obtained in the formulation in eq.\ (\ref{eq:lagr_psi})). This leads to a different definition of the quadratic part of the Hamiltonian which, in turn, leads to the unphysical result that certain modes of the fermion are excited by the rolling of the inflaton even in the limit $m\to 0$, where we would expect these degrees of freedom to decouple. These two issues are related, in the sense that the perturbation theory based on the quadratic fermion Hamiltonian obtained from eq.\ (\ref{eq:lagr_Y}) blows up at a finite time in the massless limit. We discuss these issues in greater detail in appendix~\ref{ap:old_basis}. In what follows, we work with the Lagrangian in eq.\ (\ref{eq:lagr_psi}).

In order to determine the Bogolyubov coefficients for the fermions $\psi$, and therefore their particle number, we must second quantize $\psi$ in the presence of the time dependent background induced by the rolling of $\phi$. To do so we first focus only on the $\psi$-dependent part of the Lagrangian, approximating $\phi$ as a homogeneous, time-dependent background: $\phi({\bf x},\,\tau)\simeq \phi_0(\tau)$. From the slow-roll condition $\dot\phi_0\simeq $constant,\footnote{We denote $\dot{}\equiv d/dt$, ${}'\equiv d/d\tau$, with $t$ cosmic time and $\tau$ conformal time. We note that $\dot{\phi}_0$ varies at second order in slow roll, but we disregard this small effect here.} we have
\begin{align}
\phi_0(\tau)=-\frac{\dot\phi_0}{H}\,\log\left(\tau/\tau_{\rm in}\right)\,,
\end{align}
where $\tau_{\rm in}$ is related to the initial value of $\phi$. We have verified that, consistent with the fact that the fermion is derivatively coupled to the inflaton, our results do not depend on the value of $\tau_{\rm in}$.

Relegating the details of our derivation to appendix \ref{ap:new_basis}, we find that, by decomposing
\begin{align}\label{eq:deco_psi}
\psi= \int\frac{\exd^3 k}{(2\pi)^{\frac{3}{2}}}e^{i\bk\cdot\bx}\,\sum_{r=\pm}\left[ {U}_r(\bk,\,\tau)\,a_r(\bk)+ {V}_r(-\bk,\,\tau)\,b_r^\dagger(-\bk)\right] \,, 
\end{align}
with
\begin{align}
U_r(\bk,\,\tau) & =   \frac{1}{\sqrt{2}}  \left( \begin{array}{c}
\chi_r(\bk)\, {u}_r( x )\\
r\chi_r(\bk)\, {v}_r( x )
\end{array} \right)  \,,\;\; {V}_r \left( \bk \right) = C \, {\bar {U}}_r \left( \bk \right)^T \,,\,\qquad C = i \gamma^0 \gamma^2 = \left( \begin{array}{cc} 0 & i \sigma_2 \\ i \sigma_2 & 0 \end{array} \right) \,, \nonumber\\ 
\chi_r (\bk) \equiv & \frac{\left(k+r\,\sigma\cdot\bk\right)}{\sqrt{2\,k\,(k+k_3)}}\,\bar{\chi}_r,\,\qquad\qquad
\bar\chi_+=\left(\begin{array}{c}
1 \\
0
\end{array}\right),
\,\quad
\bar\chi_-=\left(\begin{array}{c}
0 \\
1
\end{array}\right)\,, 
\end{align}
(where $k_3$ is the $z$-component of the vector $\bk$), the mode functions are given by
\begin{align}\label{eq:def_uv}
{u}_r( x )= & \frac{1}{\sqrt{2 x}} \left[ {\rm e}^{i r {\hat \phi} \left( x \right)} \, s_r \left( x \right) +  {\rm e}^{-i r {\hat \phi} \left( x \right)} \, d_r \left( x \right) \right], \,\nn\\
{v}_r( x ) = & \frac{1}{\sqrt{2 x}} \left[ {\rm e}^{i r {\hat \phi} \left( x \right)} \, s_r \left( x \right) -  {\rm e}^{-i r {\hat \phi} \left( x \right)} \, d_r \left( x \right) \right] \,, 
\end{align} 
which satisfy the normalization condition $|u_r|^2+|v_r|^2=2$, with
\begin{align}\label{eq:def_sd}
s_r \left( x \right)=   {\rm e}^{-\pi r \xi} \, W_{\frac{1}{2}+2 i r \xi ,\, i \sqrt{\mu^2+4 \xi^2}}(- 2 i x)\,,\qquad
 d_r \left( x \right)= - i  \, \mu \, {\rm e}^{-\pi r \xi} \, {W}_{-\frac{1}{2}+2 i r \xi ,\, i \sqrt{\mu^2+4 \xi^2}}( - 2 i x ) \,, 
\end{align}
where $W_{\mu,\,\lambda}(z)$ denotes the Whittaker W-function. We have also defined
\begin{align}
{\hat \phi} \left( x \right)\equiv \frac{\phi_0}{f} = - 2 \xi \log \left( x / x_{\rm in} \right) \,, 
\end{align}
with
\begin{eqnarray} 
x \equiv - k \tau\,,\quad x_{\rm in}\equiv -k\tau_{\rm in}\,,\quad \mu\equiv \frac{m}{H}\,,\quad \xi\equiv \frac{\dot\phi_0}{2fH}\,. 
\end{eqnarray} 
We  work in de Sitter spacetime, $a(\tau)=-1/H\tau$,  and disregard subleading corrections in slow roll.

With the knowledge of the mode functions, we diagonalize the quadratic Hamiltonian for the fermions, which reads
\begin{align}
H^{(2)}_{\psi} = \int \exd^3 x \, \bar \psi \left[ - i\,\gamma^i\,\partial_i + m_R \, - i \gamma^5 \, m_I \right] \psi \,, 
\end{align}
where the subscript ${}_\psi$ indicates that we are considering only the fermionic part of the Hamiltonian  and  the superscript ${}^{(2)}$ indicates that we are considering only the part of the Hamiltonian that is quadratic in the field fluctuations (in section~\ref{sec:ps} below we consider the cubic and quartic part of the Hamiltonian). The quantities $m_R$ and $m_I$ are defined as
\begin{equation}
m_R \equiv m\,a \, \cos \left( \frac{2 \phi_0}{f} \right) \;\;,\;\; m_I \equiv  m\,a \, \sin \left( \frac{2 \phi_0}{f} \right)\,.
\end{equation} 
Long but straightforward computations show that $H^{(2)}_{\psi}$ takes the form
\begin{align} 
H_{\psi}^{(2)} & = \sum_{r=\pm} \int \exd^3 k \left( a_r^\dagger \left( \bk \right) ,\,  b_r \left( - \bk \right) \right) 
\left( \begin{array}{cc} 
A_r & B_r^* \\ 
B_r & - A_r 
\end{array} \right) \, 
\left( \begin{array}{c} 
a_r \left( \bk \right) \\ 
b_r^\dagger \left( - \bk \right) 
\end{array} \right)\,, \nonumber\\ 
A_r &\equiv \frac{1}{2} \left[ m_R \left( \vert {u}_r  \vert^2 -  \vert {v}_r  \vert^2 \right) 
 +  k \left(  {u}_r^*  {v}_r  +  {v}_r^*  {u}_r  \right)  
 - i \, r \, m_I \left(   {u}_r^*  {v}_r  -  {v}_r^*  {u}_r  \right)  \right] \,,  \nonumber\\ 
B_r & \equiv   \frac{r \, e^{ir\varphi_\bk}}{2}  \left[   2 \, m_R {u}_r  \, {v}_r  - k \left(  {u}_r^2  -  {v}_r^2  \right)   - i \, r \, m_I \left( {u}_r^2  +  {v}_r^2  \right) \right] \,, 
\label{Hfree}
\end{align} 
with $e^{i\varphi_\bk}\equiv (k_1+i\,k_2)/\sqrt{k_1^2+k_2^2}$. Note that eq.\ \eqref{Hfree} implies that whenever $B_r$ is nonvanishing, the operators $a_r^\dagger$ and $b_r^\dagger$ do not create energy eigenstates.  Therefore, they should not be interpreted as ladder operators associated with a single-particle state. The matrix appearing in the first line of eq.\ (\ref{Hfree}) can be diagonalized as
\begin{eqnarray}
\left(\begin{array}{cc}
A_r & B_r^* \\ 
B_R & -A_r 
\end{array}\right)=
\left(\begin{array}{cc}
\alpha_r^* & \beta_r^* \\
-\beta_r & \alpha_r
\end{array}\right)
\left(\begin{array}{cc}
\omega &  0 \\
0 & -\omega
\end{array}\right)
\left(\begin{array}{cc}
\alpha_r & -\beta_r^* \\
\beta_r & \alpha^*_r
\end{array}\right)\,,\qquad \omega \equiv \sqrt{k^2 + m_R^2 + m_I^2}\,,
\end{eqnarray}
where the Bogolyubov coefficients $\alpha_r$ and $\beta_r$ read
\begin{align}
\alpha_r &=  {\rm e}^{i r \varphi_{\bk}/2} 
\left[ \frac{1}{2} \sqrt{ 1 + \frac{m_R}{\omega} } \; {u}_r + 
\frac{1}{2} \sqrt{ 1 - \frac{m_R}{\omega} } \, {\rm e}^{-i r \theta} \, {v}_r \right] \,, \nonumber\\ 
\beta_r &=  r  \,   {\rm e}^{i r \varphi_{\bk}/2} \, 
\left[  \frac{1}{2} \sqrt{ 1 - \frac{m_R}{\omega} } \, {\rm e}^{i r \theta} \,  {u}_r -  
\frac{1}{2} \sqrt{ 1 + \frac{m_R}{\omega} } \,  {v}_r \right]\,,
\label{Nr} 
\end{align}
where ${\rm e}^{i \theta} \equiv (k + i m_I)/\sqrt{k^2+m_I^2}$. It is then straightforward to see that the operators 
\begin{align}
\left(
\begin{array}{c}
\hat{a}_r(\bk)\\
\hat{b}_r^\dagger(-\bk)
\end{array}\right)=
\left(\begin{array}{cc}
\alpha_r & -\beta_r^* \\
\beta_r & \alpha^*_r
\end{array}\right)\cdot
\left(
\begin{array}{c}
{a}_r(\bk)\\
{b}_r^\dagger(-\bk)
\end{array}\right)
\label{ab-hat}
\end{align}
diagonalize the Hamiltonian, with $\hat{a}_r^\dagger(\bk)\,\hat{a}_r(\bk)$ and $\hat{b}_r^\dagger(\bk)\,\hat{b}_r(\bk)$ describing number operators of, respectively, particles and antiparticles with energy $\sqrt{k^2 + m_R^2 + m_I^2}=\sqrt{k^2+m^2a^2}$.

The occupation number of helicity-$r$ particles (and antiparticles) is then 
\begin{align}\label{eq:nr}\nn
N_r  \equiv  \vert \beta_r \vert^2  & =  \langle 0|\hat{a}_r^\dagger(\bk)\,\hat{a}_r(\bk)|0\rangle =  \langle 0|\hat{b}_r^\dagger(\bk)\,\hat{b}_r(\bk)|0\rangle  
\\ & = \frac{1}{2} - \frac{m_R}{4 \omega} \left( \vert {u}_r \vert^2 - \vert {v}_r \vert^2 \right) 
- \frac{k}{2 \omega} \, {\rm Re } \left( {u}_r^* {v}_r \right) 
- \frac{r \, m_I}{2 \omega} \, {\rm Im } \left( { u}_r^* {v}_r \right)\,.
\end{align} 

%%%%%%%
%%%%%%%
\begin{figure}
\centering
\includegraphics[width=\textwidth]{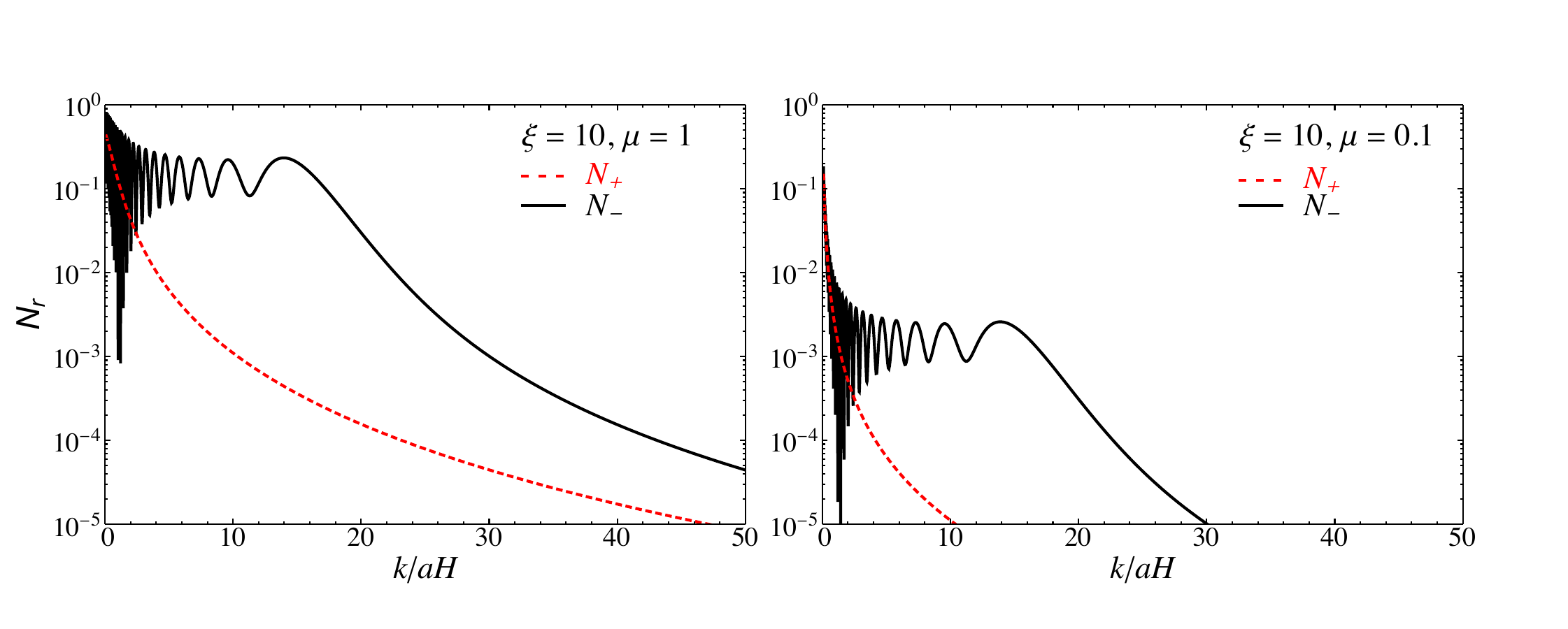}
\caption{The occupation number of $r=-1$ (top, solid curves) and $r=+1$ (bottom, dashed curves) fermions, $N_r$ as a function of the momentum $k$, for $\xi=10$ and $\mu=1$ (left panel) and $\mu=0.1$ (right panel).}
\label{fig:nr1}
\end{figure}
%%%%%%%
%%%%%%%

We now discuss the main properties of the functions $N_r(k)$. In figure~\ref{fig:nr1}, we show the occupation number of the $r=+1$ and $r=-1$ fermions at the end of inflation ($\tau=-1/H$) for $\xi=10$ with $\mu=1$ (left) and $\mu=0.1$ (right). First, let us focus on fermions with $r=+1$. For those particles the occupation number  drops rapidly to zero as $k$ gets larger than $m$. The reason for this behavior is that the presence of a nonvanishing mass leads to the breaking of conformality and the generation of fermions on the de Sitter background.  For momenta larger than $m$ the fermions are approximately conformal and the occupation number becomes smaller. This phenomenon is purely gravitational and affects both the left- and the right-handed modes, but for the modes with $r=-1$ it is overwhelmed by the effects of nonvanishing $\xi$. In fact, modes with $r=-1$ have nonvanishing occupation number for $k$ as large as $2\xi H$. We interpret this as the consequence of the fact that the excitation of those fermion modes is induced by the coupling to the pseudoscalar inflaton.\footnote{These results rely on the assumption that $\xi>0$. Changing the sign of $\xi$ has the effect of exchanging the occupation numbers of the $r=+$ and $r=-$ modes.} We also note that the occupation number of the interesting $r=-1$ mode displays high frequency oscillations as a function of the momentum $k$. The two panels of figure~\ref{fig:nr1} show that those oscillations happen around a value of the occupation number that is approximately given by $\mu^2/\xi$.

By evaluating analytically eq.~(\ref{eq:nr}) in various limits we observe that both $N_+$ and $N_-$ vanish as $\mu^2$ in the limit $\mu\to 0$. This is consistent with the decoupling of $\psi$ from the inflaton for $m=0$. More specifically, one obtains
\begin{align}
N_r\simeq \frac{\mu^2}{4\,x^2}\,,\qquad \mu\ll x\ll 1\,.
\end{align}

In the regime of moderate $\mu\lesssim 1$ and large $\xi\gg 1$ we find that the occupation number of the $r=-1$ modes is oscillating about a constant that is well approximated by $\mu^2/\xi$ for modes with $x\lesssim \xi$ before dropping as $\xi^2\,\mu^2/x^4$ for $x\gtrsim \xi$. As a consequence, for this range of parameters the total number density of the modes with $r=-1$ scales, for $\mu\lesssim 1$, as $\frac{\mu^2}{\xi}\times \xi^3\sim \mu^2\,\xi^2$ that can be parametrically larger than unity per Hubble volume.

\begin{figure}
\centering
\includegraphics[width=\textwidth]{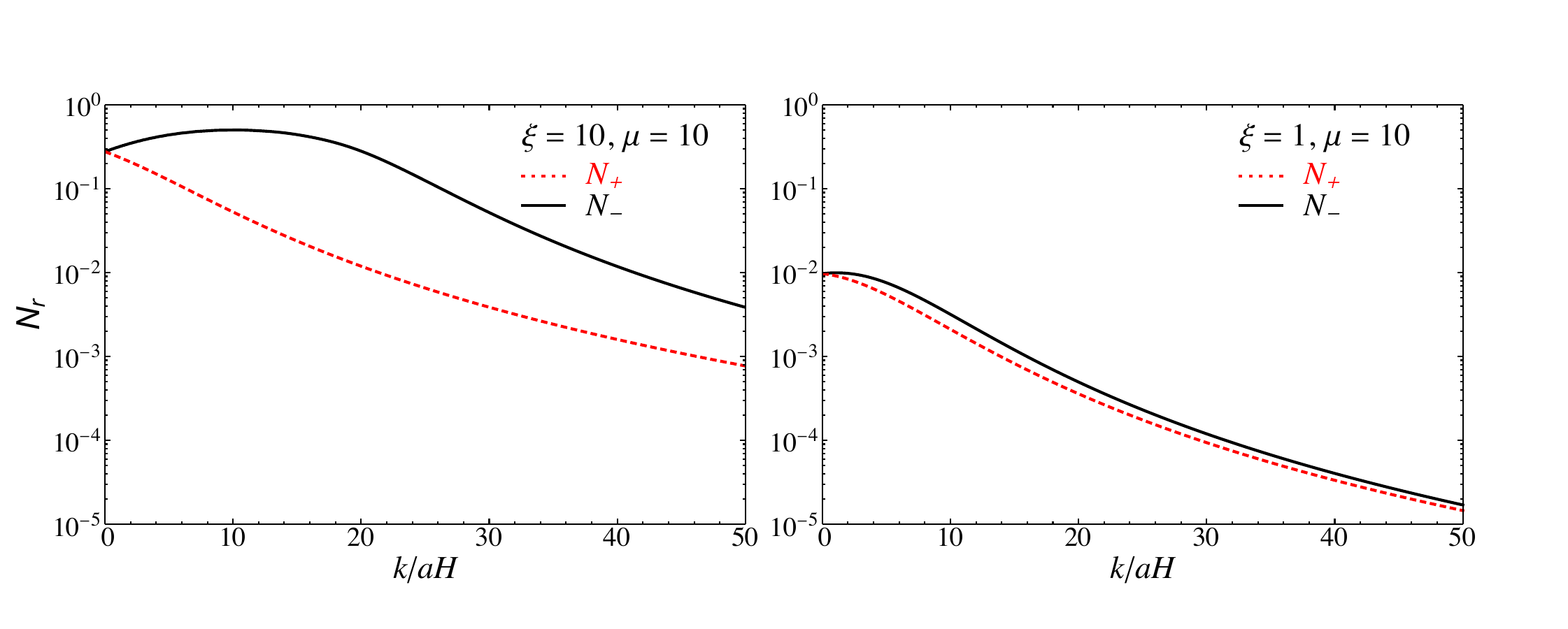}
\caption{Same as figure~\ref{fig:nr1}, but for $\xi=10$, $\mu=10$ (left) and for $\xi=1$, $\mu=10$ (right).}
\label{fig:nr2}
\end{figure}

Moving to the regime of large $\mu$, in figure~\ref{fig:nr2} we  show the occupation number for the $r=+1$ and $r=-1$ modes for $\mu=10$ and $\xi=10$ (left) and for $\mu=10$, $\xi=1$ (right). Remarkably, even if the occupation number for the $r=+1$ modes is  smaller than that of the $r=-1$ ones, both occupation numbers are of order unity despite the fact that the mass of the fermions is much larger than the Hubble scale. This means that the coupling to the inflaton prevents the decoupling of  fermions with $m\gg H$ (for comparison, the occupation number of fermions with $\mu=10$ and $\xi=0$, not plotted, is at most of the order of $10^{-5}$). Of course, the occupation number of the fermions decreases (as $\sim \xi^2/\mu^2$) when $\mu$ becomes much larger than $\xi$.

A numerical evaluation of the total number density of $r=-1$ fermions yields 
\begin{align}
\int \exd^3k\, N_-(k)\simeq 52\,H^3\mu^2\,\xi^2\,,\qquad \xi\gg \mu\,,\qquad \xi\gg 1. 
\end{align}

The main conclusion of this section is that a nonvanishing value of $\xi$ leads to nontrivial behavior of the fermions. Chiral fermions are copiously produced even if $m\ll H$ (as long as $\mu^2\,\xi^2$ is large enough), and even very heavy fermions with $m\gg H$ can be produced with large occupation numbers as long as $\mu\lesssim \xi$. We now move on to compute the  effect of these fermions on the inflaton.

%%%%%%%%%%%%%%%%%%%%%%%%%
\section{Backreaction}%%%
\label{sec:backr}%%%%%%%%
%%%%%%%%%%%%%%%%%%%%%%%%%

In this section we examine the backreaction of the produced fermions on the homogeneous, or background, inflaton. The equation of motion for the inflaton, derived from the Lagrangian~(\ref{eq:lagr_psi}), reads
\begin{align}\label{eq:full_el}
\phi''+2\frac{a'}{a}\phi'-\Delta\phi+a^2\,V'(\phi)=\frac{2\,m}{f\,a}\bar \psi \left[ \sin \left( \frac{2 \phi}{f} \right) +i \gamma^5 \cos \left( \frac{ 2 \phi }{ f } \right) \right] \psi  \,.
\end{align} 
We seek to establish the conditions under which the backreaction of the produced fermions on the background dynamics is negligible. We do this by imposing that the right hand side of the equation above, evaluated in the Hartree approximation, is much smaller than $a^2\,V'(\phi)$. Using manipulations analogous to those of section~\ref{sec:Set_Up} above, the quantity
\begin{align}\nn
{\cal B}&\equiv\frac{2\,m}{f\,a}\,\langle\bar\psi\left[\sin\left(\frac{2\phi}{f}\right)+i\,\gamma_5\,\cos\left(\frac{2\phi}{f}\right)\right]\psi\rangle \\ & =\frac{2}{f\,a^2}\,\int \frac{\exd^3 q\,\exd^3p}{(2\pi)^3}e^{i(\bq-\bp)\bx}\langle\bar\psi(\bp)\left[m_I+i\,\gamma_5\,m_R\right]\psi(\bq)\rangle\,,
\label{eq:B_def}
\end{align}
can be written as
\begin{align}\label{eq:calb}
{\cal B}&=4\frac{m\,H^3}{f}\,a^2\,\sum_rr\int \frac{y\,\exd y}{2\pi^2}\,{\rm {Im}}\left\{s_r\,d_r^*\right\}\,.
\end{align}
Quite remarkably, the integral can be computed  analytically (the details are in appendix~\ref{ap:backreaction}) after regulating it with a hard cut-off  at a finite and large $\Lambda$.  The integral turns out to have a logarithmic divergence for large $\Lambda$. As we discuss in greater detail in subsection~\ref{subsec:quartic} below, we can deal with the divergence either by simply subtracting the divergent part, or by adiabatic regularization. The result does not change in the limit of large $\xi$, and in the regime $\mu\lesssim 1\ll \xi$ we obtain
\begin{align}\label{eq:calb_final}
{\cal B}\simeq -\frac{8}{\pi}\,\frac{H^4}{f}\,a^2\,\mu^2\,\xi^2\,.
\end{align}

By imposing that the backreaction of the fermions on the zero mode of the inflaton is negligible, ${\cal B}\ll a^2\,V'(\phi)\simeq 3\,H\,\dot\phi\, a^2$, we derive a first condition on the parameter space of the model:
\begin{align}\label{eq:no_back}
\mu^2\,\xi\ll \frac{f^2}{H^2}\,,
\end{align}
where we emphasize that $f$ should be much larger than $H$ in order for the effective field theory to be valid at  energy scales of the order of $H$.

As a second condition for negligible backreaction we impose that the energy density of the produced fermions gives a negligible contribution to the expansion rate of the Universe. The energy density in fermions is computed in appendix \ref{ap:rho_psi} and reads
\begin{align}
\rho_\psi=16\,\pi^2 H^4  \mu^2 \xi^3\,,
\end{align}
so that by requiring it to be subdominant with respect to the energy in the inflaton one obtains the parametric constraint
\begin{align}
\mu^2\,\xi^3\ll \frac{M_P^2}{H^2}\,.
\end{align}
It is easy to check that during slow roll, $\dot\phi\ll H\,M_P$, this condition is satisfied as long as eq.~(\ref{eq:no_back}) holds.

%%%%%%%%%%%%%%%%%%%%%%%%%%%
\section{Power spectrum}%%%
\label{sec:ps}%%%%%%%%%%%%%
%%%%%%%%%%%%%%%%%%%%%%%%%%%

Fermions with a nonvanishing occupation number backreact on the fluctuations of the inflaton, therefore modifying the primordial scalar perturbations.  In this section we compute this effect to leading order. As we will show, at the level of approximation that we are using this modification is scale invariant, and therefore it is unobservable in the spectrum because it is degenerate with the vacuum contribution generated  by the inflationary expansion of the Universe. However, an observable effect is potentially generated in the bispectrum, which in single-field inflation is slow-roll suppressed to a currently unobservable level. A precise calculation of the bispectrum is very challenging, but in section~\ref{sec:bs} below we use the results of this section~\ref{sec:ps} to estimate its magnitude.

In order to focus on the physics, we only present the main steps of our calculation of the leading order correction to power spectrum in this section.  The details can be found in appendices~\ref{ap:quartic} and~\ref{ap:cubic}.  Discussions on our renormalization scheme are presented in appendix \ref{ap:renormalization}.

We compute the leading order modifications to the power spectrum of the fluctuations of the inflaton using the in-in formalism (see, e.g.\ \cite{Weinberg:2005vy}). To do so, we define the perturbation $\delta\phi(\bx,\,\tau)=\phi(\bx,\,\tau)-\phi_0(\tau)$ and we expand the interaction Hamiltonian to second order in $\delta\phi$
\begin{align}\label{eq:hint34} 
H_{\rm int} \supset& - \frac{2 a m}{f} \,  \int \exd^3 x \, {\bar \psi} \left[ \sin \left( 2 \, \frac {\phi_0}{f} \right) + i \, \gamma^5  \cos \left( 2 \, \frac {\phi_0}{f} \right) \right] \psi \, \delta \phi \nonumber\\
& - \frac{2 a m}{f^2} \,  \int \exd^3 x \, {\bar \psi} \left[ \cos \left( 2 \, \frac {\phi_0}{f} \right) - i \, \gamma^5  \sin \left( 2 \, \frac {\phi_0}{f} \right) \right] \psi \, \delta \phi^2\equiv H_{\psi}^{(3)}+H_{\psi}^{(4)}\,,
\end{align} 
where we have neglected the contribution from the inflaton self-interactions, whose effects are slow-roll suppressed. We then use $H_{\rm int}$ to compute the modification to the power spectrum
\begin{align}\label{P-int} 
\delta P_\zeta \left( \tau ,\, k \right) \Big\vert_{-k\tau \ll 1} &= \frac{k^3}{2 \pi^2} \, \frac{H^2 }{\dot\phi_0^2} \,  \sum_{N=1}^{\infty} \left( - i \right)^N \, \int^\tau \exd \tau_1 \dots \, \int^{\tau_{N-1}} \exd \tau_N \nonumber\\ 
&  \times \left\langle 
\left[ 
\left[ 
\cdots  
\left[ 
\delta \phi^{(0)} \left( \tau ,\, {\bk} \right)  \delta \phi^{(0)} \left( \tau ,\, {\bk}' \right)  ,\, H_{\rm int} \left( \tau_1 \right) 
\right] , \cdots 
\right] ,\,  H_{\rm int} \left( \tau_N \right) 
\right] 
\right\rangle' \;, 
\end{align}
where we have used the relation $\zeta=-H\,\delta\phi/\dot\phi_0$ between the fluctuations of the inflaton and the scalar perturbation of the metric, and the prime denotes the correlator stripped of the $\delta^{(3)}(\bk+\bk')$ associated with momentum conservation.

In evaluating the expression eq.\ (\ref{P-int}) we use the mode functions for $\psi$ found in section~\ref{sec:Set_Up} above, eqs.\ (\ref{eq:deco_psi}) through (\ref{eq:def_sd}). Regarding the mode functions of $\delta\phi$, we use those of a massless field in de Sitter space:
\begin{align}
\delta \phi^{(0)} \left(\bx,\, \tau \right) =&  \int \frac{\exd^3 k}{\left( 2 \pi \right)^{3/2}} \, {\rm e}^{i \bk \cdot \bx}\left[ \delta \phi_k^{(0)} \left( \tau \right) a_{\bk} +  \delta \phi_k^{(0)*} \left( \tau \right) a_{-\bk}^\dagger\right] \;, 
\label{phi-deco}
\end{align} 
with
\begin{equation}
\delta \phi^{(0)}_k \left( \tau \right) =  \frac{H}{\sqrt{ 2 k}} \left( i \, \tau + \frac{1}{k} \right) {\rm e}^{- i k \tau} \;. 
\label{df0}
\end{equation} 

\begin{figure}
\centering
\includegraphics[width=0.60\textwidth]{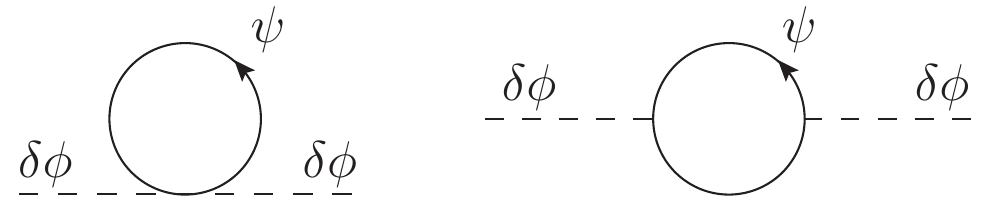}
\caption{The two diagrams that contribute at leading order to the two-point function of $\delta\phi$.}
\label{fig:diagrams}
\end{figure}

The two parts of the interaction Hamiltonian $H_\psi^{(3)}$ and $H_\psi^{(4)}$ describe a cubic $\bar\psi\,\psi\,\delta\phi$ vertex and a quartic $\bar\psi\,\psi\,\delta\phi^2$ vertex. Those two vertices can be used to draw the two diagrams shown in figure~\ref{fig:diagrams}, which contribute to eq.~(\ref{P-int}) at leading order in the $1/f$ expansion. We discuss these diagrams in the next two subsections.

%%%
\subsection{Quartic loop}\label{subsec:quartic}
%%%

The first diagram in figure~\ref{fig:diagrams} gives 
\begin{align} 
\delta P_\zeta^{(4)} & \left( \tau ,\, k \right)=  
i\,\frac{k^3}{2 \pi^2} \, \frac{{H}^2}{\dot\phi_0^2}\,\frac{2m}{f^2} \int^\tau \exd\tau_1 \,a(\tau_1)  \,  \int \frac{\exd^3 p \, \exd^3q \, \exd^3  w}{\left( 2 \pi \right)^3} \,
 \Bigg\langle \Bigg[ 
\delta \phi^{(0)} \left( \tau ,\, \bk \right)  \delta \phi^{(0)} \left( \tau ,\, \bk' \right)  ,\,  \nonumber\\ 
& {\bar \psi} \left( \tau_1 ,\, \bp \right) \left[ \cos \left( 2 \, \frac{\phi_0(\tau_1)}{f} \right) - i \, \gamma^5  \sin \left( 2 \, \frac{\phi_0(\tau_1)}{f} \right) \right]\, \psi \left( \tau_1 ,\, \bq \right) \nonumber \\
& \,  \qquad 
\delta \phi^{(0)} \left( \tau_1 ,\,{ \bf w} \right)  \, \delta \phi^{(0)} \left( \tau_1 ,\, \bp-\bq-{\bf w} \right) \Bigg] \Bigg\rangle',
\label{eq:quartic1}
\end{align} 
which, with some algebra and in the large scale limit $-k\tau\to 0$, can be simplified to
\begin{align} 
\delta P_\zeta^{(4)} \left( \tau ,\, k \right) 
 =  & \frac{2 H^5 m}{f^2 k^3 \pi^2 \dot{\phi}_0^2} \, \int^\tau \frac{ \exd \tau_1 }{ \tau_1 }   \,  
\left[ \cos \left( k \tau_1 \right) + k \tau_1 \sin \left( k \tau_1 \right) \right] 
\left[ \sin \left( k \tau_1 \right) - k \tau_1 \cos \left( k \tau_1 \right) \right] \nonumber\\ 
& \quad\quad \times\int \frac{\exd^3 p  }{\left( 2 \pi \right)^3}  \left\langle  {\bar \psi} \left(  \bp \right) \left[ \cos \left( 2 \, {\hat \phi} \right) - i \, \gamma^5  \sin \left( 2 \, {\hat \phi} \right) \right] \psi \left(  \bp \right)  \right\rangle_{\tau_1}'\,.
\label{eq:quartic2}
\end{align} 
By inserting the expressions for the mode functions of the fermions into this equation we finally obtain
\begin{align} \nn
\frac{\delta P_\zeta^{(4)} \left( \tau ,\, k \right) }{ P_\zeta^{(0)}  }  = &  \frac{4H^2 \mu}{f^2\pi^2} \, \int_x \frac{ \exd  x_1  }{ x_1^4 }   \,  
\left[ \cos \left( x_1 \right) +  x_1 \sin \left(   x_1 \right) \right] 
\left[   x_1 \cos \left(  x_1 \right)-\sin \left(  x_1 \right) \right]\\ & \qquad \times  \int \exd x_p \, x_p \, 
\sum_r \Re \left[ d_r^* \left( x_p \right) s_r \left( x_p \right) \right], 
\label{amp1}
\end{align} 
where we have normalized this contribution to $P_\zeta$ by the vacuum term $P_\zeta^{(0)}=H^4/(4\,\pi^2\,\dot\phi_0^2)$,  we have introduced the dimensionless integration variables $x_1\equiv -k\tau_1$ and $x_p\equiv -p\tau_1$, and where the functions $d_r(x)$ and $s_r(x)$ are given in eq.~(\ref{eq:def_sd}).

We  proceed to evaluate the two integrals that appear in eq.\ (\ref{amp1}). The integral in $\exd x_1$ diverges when the lower limit of integration $x$ is sent to $0$ (remember that $x=-k\tau=k/H$ as we want to evaluate the power spectrum at the end of inflation, $\tau=-1/H$). In fact one finds
\begin{align}
& \int_x^\infty \frac{ \exd  x_1  }{ x_1^4 }   \,  
\left[ \cos \left( x_1 \right) +  x_1 \sin \left(   x_1 \right) \right] 
\left[ x_1 \cos \left(  x_1 \right)-\sin \left(  x_1 \right) \right]\Big|_{x\to 0}\simeq \frac{1}{3}\log(x)+\frac{3\,\log 2+3\,\gamma_E-7}{9} \nonumber \\
& \qquad +{\cal O}(x^2)\,,
\end{align}
where $\gamma_E\simeq .577$ is the Euler-Mascheroni constant. This infrared divergence is a consequence of the fact that the fermions have a nonvanishing average density that keeps sourcing the fluctuations of the inflaton even when they are outside of the horizon. This divergence is regulated by the finite amount of $e$-foldings between the time when the inflaton mode leaves the horizon and the end of inflation. 

The integral in $\exd x_p$ is much more challenging, and is quadratically divergent in the ultraviolet. As it was the case for the integral in section~\ref{sec:backr}, it is possible to compute it analytically after introducing a UV regulator that sets the upper limit of integration to some finite and large $\Lambda$. We report the details of our calculation as well as the exact expression of the integral in appendix~\ref{ap:quartic}. The divergent part of the integral reads
\begin{align}\label{eq:intquart_div}
\int_0^\Lambda \exd x_p \, x_p \, 
\sum_r \Re \left[ d_r^* \left( x_p \right) s_r \left( x_p \right) \right]=\mu\left(\Lambda^2-\left(\mu^2-8\xi^2+1\right)\log\Lambda\right)+{\cal O}(\Lambda^0)\,.
\end{align}

Now, we have at least two different ways of dealing with this divergence. We can subtract from the exact integral its adiabatic part, or we can simply subtract by hand the part that diverges when $\Lambda\to \infty$. The adiabatic subtraction might be problematic, as the adiabatic contribution turns out to dominate the physical one at momenta  of order $H$~\cite{Durrer:2009ii}. Since these momenta contribute to the finite part of the integral, adiabatic subtraction can induce spurious components into our integral. 

As discussed in appendix~\ref{ap:renormalization}, adiabatic subtraction does indeed introduce spurious contribution which scales as $\mu\, \xi^2$ and is therefore large at large $\xi$.  However, this contribution is subdominant at sufficiently large $\xi$, as there is a physical contribution which scales as $\mu\, \xi^2 \log(\xi)$.  In this limit, and setting $\tau=-1/H$ to have quantities computed at the end of inflation, we get the simple result
\begin{align}\label{eq:p_quartic}
\frac{\delta P_\zeta^{(4)} \left( k \right) }{ P_\zeta^{(0)}  }\Big|_{\rm {end\ of\ inflation}} \simeq\frac{32\,m^2\xi^2\,\log\xi}{3\pi^2f^2}\,\log(H/k)\,.
\end{align}

%%%
\subsection{Cubic loop}
\label{sub:cubic}
%%%

The second diagram in figure~\ref{fig:diagrams} gives 
\begin{align}
& \frac{\delta P_\zeta^{(3)}(\tau,\,k)}{P_\zeta^{(0)}}=-\frac{2\,k^3}{H^2}\,\frac{m^2}{f^2}\int^\tau \exd \tau_1\,a(\tau_1)\int^{\tau_1}\exd \tau_2\,a(\tau_2)\,\int \frac{\exd^3 p\,\exd^3 q}{(2\pi)^{3}}\left(\delta(\bk+\bp-\bq)+\delta(-\bk+\bp-\bq)\right)\nonumber\\
&\qquad \times\left[\sin\left(\frac{2\,\phi_0(\tau_1)}{f}\right)+i\,\gamma^5\,\cos\left(\frac{2\,\phi_0(\tau_1)}{f}\right)\,\right]_{ij}\,\left[\sin\left(\frac{2\,\phi_0(\tau_2)}{f}\right)+i\,\gamma^5\,\cos\left(\frac{2\,\phi_0(\tau_2)}{f}\right)\,\right]_{ab}\nonumber\\
&\qquad \times\left(\delta\phi^{(0)}(\bk,\,\tau)\,\delta\phi^{(0)}(\bk,\,\tau_1)^*-\delta\phi^{(0)}(\bk,\,\tau_1)^*\,\delta\phi^{(0)}(\bk,\,\tau_1)\right)\nonumber\\
&\qquad  \times\left\{\delta\phi^{(0)}(\bk,\,\tau)\,\delta\phi^{(0)}(\bk,\,\tau_2)^*\,\langle\bar\psi(\bp,\,\tau_1)_i\,\psi(\bp,\,\tau_2)_b\rangle'\,\langle\psi(\bq,\,\tau_1)_j\,\bar\psi(\bq,\,\tau_2)_a\rangle'\right.\nonumber\\
&\qquad  -\left.\delta\phi^{(0)}(\bk,\,\tau_2)\,\delta\phi^{(0)}(\bk,\,\tau)^*\,\langle \bar\psi(\bq,\,\tau_2)_a\,\psi(\bq,\,\tau_1)_j\rangle'\,\langle\psi(\bp,\,\tau_2)_b\,\bar\psi(\bp,\,\tau_1)_i\rangle'\right\} \, ,
\end{align}
where we have already normalized to the vacuum power spectrum. With some work it is possible to evaluate the fermionic part (details can be found in appendix~\ref{ap:cubic}) and write
\begin{align}\label{eq:integral_p3}
&\frac{\delta P_\zeta^{(3)}(\tau,\,k)}{P_\zeta^{(0)}}=\frac{m^2}{2\,f^2\,k^3}\int^\tau \frac{\exd \tau_1}{\tau_1^2}\,\int^{\tau_1}\frac{\exd \tau_2}{\tau_2^2}\,\int \frac{\exd^3 p\,\exd^3 q}{(2\pi)^{3}p\,q}\,\left(\delta(\bk+\bp-\bq)+\delta(-\bk+\bp-\bq)\right)\nonumber\\
&\times\sum_{rs}\left(1+r\,s\,\frac{\bp\cdot\bq}{p\,q}\right)\,\left(\sin k\tau_1-k\tau_1\,\cos k\tau_1\right)\,\left\{\left(-i-k\tau_2\right)\,e^{ik\tau_2}\,\left(r\,s\, v_r(-p\tau_1)\, v_s(-q\tau_1)\, \right. \right.\nonumber\\
&\qquad \left. \left. +u_r(-p\tau_1)\,u_s(-q\tau_1)\right)\left(r\,s\,v_r^*(-p\tau_2)\,v_s^*(-q\tau_2)+u_r^*(-p\tau_2)\, u_s^*(-q\tau_2)\right)+h.c.\right\} \, ,
\end{align}
where we recall that $u_r$ and $v_r$ are given in eqs.~(\ref{eq:def_uv}) and~(\ref{eq:def_sd}).

The computation of the above integral is extremely challenging (even an estimate is challenging, as each term appearing in it is rapidly oscillating), so we must resort to a number of approximations. Again, details are presented in appendix~\ref{ap:cubic}, and here we simply outline our strategy.  First, we approximate the integrand assuming $p\gg k$, which implies $p\simeq q$. We expect this to generate at most an ${\cal O}(1)$ error in our final result.  At this point the integral still contains products of four Whittaker functions. To simplify the integral we Wick-rotate the time variables and we use simple approximations (that can be obtained in the limit $\xi\gg 1$ by dealing carefully with the branch cuts in the definition of the Whittaker functions) that bring the mode functions to the form
\begin{align}
s_r(-iy)\simeq A_{1,r}\,y^{-i\sqrt{\mu^2+4\xi^2}}\,e^y+B_{1,r}\,y^{i\sqrt{\mu^2+4\xi^2}}\,e^{-y}\,,\nonumber\\
d_r(-iy)\simeq A_{2,r}\,y^{-i\sqrt{\mu^2+4\xi^2}}\,e^y+B_{2,r}\,y^{i\sqrt{\mu^2+4\xi^2}}\,e^{-y}\,,
\label{eq:approx_modes}
\end{align}
with $A_i$ and $B_i$ constants that depend on $\mu$ and $\xi$.  We show in appendix~\ref{ap:cubic}  that in the $\mu\lesssim 1$, $\xi\gg 1$ regime of interest, the $r = s = +$ contribution is exponentially suppressed with respect to the $r = s = -$ contribution.

We finally recognize that the terms proportional to $A_1$ and $A_2$ correspond to the ``vacuum'' part of the modes we are considering, i.e.\ the part of modes that do not vanish and behave as positive frequency only as $p\to\infty$, and we subtract this part by hand from the mode functions, effectively keeping only the part proportional to $B_1$ and $B_2$ only. 

After these manipulations we obtain the scaling
\begin{align}\label{eq:p_cubic}
\frac{\delta P_\zeta^{(3)}(k)}{P_\zeta^{(0)}}\Big|_{\rm {end\ of\ inflation}}\propto \frac{m^2}{f^2}\,\mu^2\,\sqrt{\xi}\,|\log(k/H)|
\end{align}
which in the regime $\mu\lesssim 1,\,\xi\gg 1$  is subdominant with respect to contribution $\delta P_\zeta^{(4)}$ found in the previous subsection.

\subsection{Summary for the power spectrum}
\label{sub:sum-2}

We conclude this section by summarizing our main result: the first diagram in figure~\ref{fig:diagrams} dominates the modification to the power spectrum of scalar perturbations in this model, with  
\begin{align}\label{eq:deltapzeta} 
\delta P_\zeta \left( k \right)\big|_{\rm {end\ of\ inflation}}\simeq P_\zeta^{(0)} \,\frac{32\,m^2\,\xi^2\,\log\xi}{3\,\pi^2\,f^2}\,\log(H/k)\,.
\end{align} 

The scaling we find is consistent with the fact that the leading contribution to $\delta P_\zeta$ is approximately proportional to $1/f^2$ and to the total number of fermions $\sim\mu^2\,\xi^2$.  

\subsection{The spectral index}
\label{sub:sp_index}

From eq.~(\ref{eq:deltapzeta}) we can compute the scalar spectral index by assuming that each individual mode is evolving with constant values of $H$ and $\dot\phi$, while treating $H$ and $\dot\phi$ as time dependent when comparing different modes. 
This is justified by slow roll:  $H$ and $\dot\phi$ evolve adiabatically, on a times scale $\left( \epsilon \, H \right)^{-1}$,  while each mode evolves on the much faster timescale $H^{-1}$. Assuming for simplicity that the scalar spectrum is dominated by the sourced part $\delta P_\zeta \left( k \right)\big|_{\rm {end\ of\ inflation}}\gg P_\zeta^{(0)} $ , and setting $k=e^{Ht}H$, $t<0$, we have 
\begin{align}
P_\zeta&=\frac{H^4}{4\pi^2\,\dot\phi^2}\left[1+\frac{32\,m^2\,\xi^2\,\log\xi}{3\,\pi^2\,f^2}\log\left(H/k\right)\right]\nonumber\\
&\simeq \frac{H^4}{4\pi^2\,\dot\phi^2}\,\frac{32\,m^2\,\xi^2\,\log\xi}{3\,\pi^2\,f^2}\log\left(H/k\right)=\frac{8}{3\pi^4}\frac{m^2\,H^2\,\log\xi}{f^4}(-Ht)\,.
\end{align}
Then, the spectral index is obtained as
\begin{align}
n_s-1=\frac{d\log P_\zeta}{d\log k}=\frac{1}{H}\frac{d\log P_\zeta}{d t}=3\frac{\dot{H}}{H^2}+\frac{1}{Ht}+\frac{1}{H\,\log\xi}\frac{\dot\xi}{\xi}\,,
\end{align}
or, using the slow-roll relations $\dot{H}=-\epsilon H^2$, $|\dot\phi|=\sqrt{2\epsilon}\,H\,M_P$ an $\ddot\phi=\left(\eta-\epsilon\right)\,H\,\dot\phi$, 
\begin{align}
n_s-1=-3\epsilon-\frac{1}{N}+\frac{2\epsilon-\eta}{\log\xi}
\end{align}
where $N$ denotes the number of efoldings. The spectral index of the scalar perturbations can thus agree with the measured value ~\cite{Ade:2015xua} $n_s\simeq .97$ for reasonable values of $N$ and of the slow-roll parameters.

%%%%%%%%%%%%%%%%%%%%%%%%%%%
\section{Non-Gaussianity}%%
\label{sec:bs}%%%%%%%%%%%%%
%%%%%%%%%%%%%%%%%%%%%%%%%%%

\begin{figure}
\centering
\includegraphics[width=0.80\textwidth]{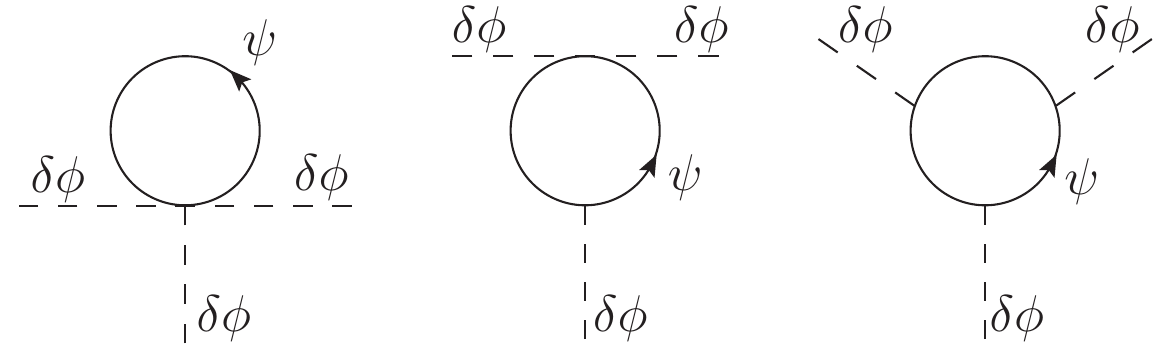}
\caption{The three diagrams that contribute at leading order to the three-point function of $\delta\phi$.}
\label{fig:3ptdiagrams}
\end{figure}

As we saw above, the calculation of the fermionic contribution to the two-point function of the inflaton is challenging, and for the cubic diagram we could only obtain what we consider to be a reasonable estimate. As one can expect, the calculation of the three-point function is even more challenging. There is a new operator, besides the cubic and the quartic interaction Hamiltonians $ H_\psi^{(3)}$ and $H_\psi^{(4)}$ given in eq.~(\ref{eq:hint34}) above, that contributes to the three-point function. It is a quintic interaction Hamiltonian
\begin{align}
H_\psi^{(5)}= - \frac{4ma}{3 f^3} \bar{\psi}  \left[ \sin \left( \dfrac{2 \phi_0}{f} \right) + i \gamma^5 \cos \left( \dfrac{2 \phi_0}{f} \right) \right] \psi\, \delta \phi^3\,,
\end{align}
which leads to a new $\bar\psi\,\psi\,\delta\phi^3$ vertex. Using the vertices generated from $ H_\psi^{(3)}$, $H_\psi^{(4)}$, and $ H_\psi^{(5)}$ we obtain, at leading order in $1/f$, the three diagrams of figure~\ref{fig:3ptdiagrams}.

%%%%%%%%%
%%%%%%%%%
%%%%%%%%%
%%%%%%%%%

%%
\subsection{The quintic diagram}

As we  argue below, the first of these diagrams gives the leading contribution to the bispectrum. Fortunately, this contribution can be calculated analytically; after some long calculations that we report in appendix~\ref{ap:nongaussianities}, we find the following expression
\begin{align}\nn
\langle \delta \phi(\bk_1,\,\tau)
\delta \phi(\bk_2,\,\tau)
\delta \phi(\bk_3,\,\tau) \rangle^\prime
= &  \dfrac{2 H^6  m}{3 f^3} \int^\tau \exd \tau_1 \, a(\tau_1) f(k_1,k_2,k_3,\tau_1) \\ &\times \int \dfrac{\exd^3 p}{(2\pi)^{9 \slash 2}} \sum_r \frac{r}{-p\tau_1} \Im\left\{(d^*_r(-p\tau_1) s_r(-p\tau_1)\right\} ,
\end{align}
where
\begin{align}
 f(k_1,k_2,k_3,\tau_1) &= \dfrac{1}{k_1^3 k_2^3 k_3^3} \cdot \left[ \tau_1 
\left(k_1 k_2 k_3 \tau_1^2- k_1 -k_2-k_3\right) \cos (\tau_1 (k_1+k_2+k_3)) \right. \nonumber \\
& \qquad \left. -\left(\tau_1^2 (k_1 k_2+ k_1 k_3+k_2 k_3)-1\right) \sin (\tau_1 (k_1+k_2+k_3)) \right]\,.
\end{align}

Since most of the dynamics occurs at momenta $-k\tau\sim\xi\gg 1$, which is well within the horizon, we expect the non-Gaussianities to be of equilateral shape. Therefore we estimate the magnitude of the bispectrum by setting $k_1=k_2=k_3\equiv k$. As in the two-point functions, the integral in $\exd \tau_1$ is logarithmically divergent as $-k\tau\to 0$, giving
\begin{align}\label{eq:quintic_before_long_integral}
\langle \delta \phi(\bk_1,\,\tau)
\delta \phi(\bk_2,\,\tau)
\delta \phi(\bk_3,\,\tau) \rangle^\prime\Big|_{\rm {equilateral}}& \simeq -\dfrac{4 H^5  m}{ f^3 k^6} 
\log (-k\tau) \cdot \dfrac{4 \pi}{(2\pi)^{9 \slash 2}} \nonumber \\
& \qquad \times \sum_rr\,  \int \exd y \, y \, \Im\left\{s_r^*(y)d_r(y)\right\}, 
\end{align}
where the integral in $\exd y$ is the same one appearing in eq.~(\ref{eq:calb}). This is as expected because the operator appearing here is the third derivative with respect to $\phi$ of the fermionic Lagrangian, which is identical in form to the one that contributes to eq.~(\ref{eq:calb}); that is to say, it is identical to the first derivative of that same part of the Lagrangian (times minus one). As we have discussed above, the integral in $\exd y$ in the equation above is divergent as $y\to\infty$; using the results of appendix~\ref{ap:backreaction} we obtain in the limit $\xi\gg 1$, $\mu\lesssim 1$
\begin{align}\label{eq:fnl_quintic_divergent}
\langle \delta \phi(\bk_1,\,\tau)
\delta \phi(\bk_2,\,\tau)
\delta \phi(\bk_3,\,\tau) \rangle^\prime\Big|_{eq}
&= -\dfrac{4 H^6  \mu^2}{(2\pi)^{7 \slash 2} f^3 k^6} 
\log (-k_*\tau) \,\left[8\pi\xi^2+12\xi\log\left(\xi/\Lambda\right)
\right. \nonumber \\
& \qquad \left. +{\cal O}(\xi)+{\cal O}(\Lambda^0)\right], 
\end{align}
where eventually we  drop the logarithmically divergent term.

\subsection{The remaining two diagrams}
\label{subsec:neglect}

Next, we would need to evaluate the remaining two diagrams in figure~\ref{fig:3ptdiagrams}. The calculation of the second of them  turns out to be of a complexity comparable to that of the cubic contribution to the power spectrum. The third one should be even more complicated, so that a complete evaluation of the bispectrum would be prohibitively difficult. We can, however, infer some scaling properties which allow us to claim that, compared to the first diagram, these two diagrams should give  negligible contributions  to $f_{NL}^{eq}$.

Our first observation is that each cubic vertex gives a contribution $\sim \frac{m}{f}\bar\psi\,\psi$, each quartic vertex gives a contribution $\sim \frac{m}{f^2}\bar\psi\,\psi$, and each quintic vertex gives a contribution $\sim \frac{m}{f^3}\bar\psi\,\psi$. Our next, and most crucial, observation is that $\bar\psi\,\psi$ oscillates with amplitude $\propto\frac{m}{\xi}$ for momenta up to $-k\tau\simeq \xi$. This implies that each fermionic loop integral, which goes as $\exd^3k$, gives a contribution $\sim\xi^3$.

Once we apply these scalings to the backreaction term of section~\ref{sec:backr} (which would be represented diagrammatically by a tadpole) we obtain a scaling $\frac{m}{f}\times \frac{m}{\xi}\times\xi^3\simeq \frac{m^2\,\xi^2}{f}$ which indeed is the result found in eq.~(\ref{eq:calb_final}). Next, we can check whether our scalings work in the calculation of the two diagrams of figure~\ref{fig:diagrams}. For the first (quartic) diagram we obtain the scaling $\frac{m}{f^2}\times\frac{m}{\xi}\times \xi^3\sim \frac{m^2\,\xi^2}{f^2}$ which agrees with result presented in eq.~(\ref{eq:p_quartic}). For the cubic diagram, on the other hand, we would expect the scaling $(\frac{m}{f})^2\times (\frac{m}{\xi})^2\times \xi^3\sim \frac{m^4}{f^2}\xi$ which is in disagreement; the result~(\ref{eq:p_cubic}) scales $\sim \frac{m^4}{f^2}\sqrt{\xi}$. A possible explanation of this disagreement is that this diagram contains a term $\sim (\bar\psi\,\psi)^2$ where each factor $\bar\psi\,\psi$ is {\em oscillating} with amplitude $\sim 1/\xi$, so that interference effects might reduce the overall amplitude of the integral.  Finally, as to the bispectra, we observe that the amplitude $\sim \frac{m^2}{f^3}\,\xi^2$ of the quintic diagram~(\ref{eq:quintic_before_long_integral}) agrees with the scalings outlined above, as it emerges as the product $\frac{m}{f^3}\times\frac{m}{\xi}\times\xi^3$. 

Based on the this discussion, the second diagram of figure~\ref{fig:3ptdiagrams} should scale as $\frac{m}{f}\times \frac{m}{f^2}\times(\frac{m}{\xi})^2\times \xi^3\sim \frac{m^4}{f^3}\,\xi$, and the third diagram in that figure should scale as $(\frac{m}{f})^3\times(\frac{m}{\xi})^3\times\xi^3\sim\frac{m^6}{f^3}$.  This may be further suppressed if the same phenomena that are reducing the amplitude of the cubic contribution to the spectrum are at work here.  Nonetheless, even without this suppression, both quantities are subdominant with respect to the contribution from the quintic diagram to the bispectrum. To sum up, the condition $\mu^2\ll \xi$ allows to neglect the second diagram of figure~\ref{fig:diagrams} in the computation of the power spectrum, and the second and third diagrams of figure~\ref{fig:3ptdiagrams} in the computation of the bispectrum.

\subsection{Summary for the bispectum}
\label{sub:sum-2}

To summarize, we argue that the parameter $f_{NL}^{eq}$ for this model, in the limit $\xi\gg 1$, $\mu\lesssim 1$ we are interested in, is obtained from eq.~(\ref{eq:fnl_quintic_divergent}) where the divergent part is dropped or renormalized away by adiabatic subtraction; we expect these to agree when $\xi \gg 1$, as happens for the contribution~(\ref{eq:p_quartic}) (see appendix~\ref{ap:renormalization}). 

Using the relation $\zeta=-H\,\delta\phi/\dot\phi_0$ and the relationship 
\begin{align}
- \dfrac{H^3}{|\dot{\phi}_0|^3} \langle \delta \phi \, \delta \phi \, \delta \phi \rangle' 
&= \dfrac{9}{10} (2\pi)^{5 \slash 2} f_{NL} \left[ \dfrac{H^2}{\dot{\phi}_0^2} \left( \dfrac{H}{2\pi} \right)^2\left(1+\frac{32\,m^2\,\xi^2\,\log\xi}{3\,\pi^2\,f^2}\,\log(H/k)\right) \right]^2 \dfrac{1}{k^6}
\end{align}
between the bispectrum and the parameter $f_{NL}$, where we have accounted for the fact that eq.~(\ref{eq:deltapzeta}) is also contributing to the power spectrum, we therefore obtain
\begin{align}\label{eq:fnl_final}
f_{NL}^{eq}\simeq \frac{\frac{160\, H^2 \mu^2  \xi^3 }{9 \pi f^2} \log(H/k)}{\left(1+\frac{32\,H^2\,\mu^2\,\xi^2\,\log\xi}{3\,\pi^2\,f^2}\,\log(H/k)\right)^2}\,.
\end{align}

\begin{figure}
\centering
\includegraphics[width=0.8\textwidth]{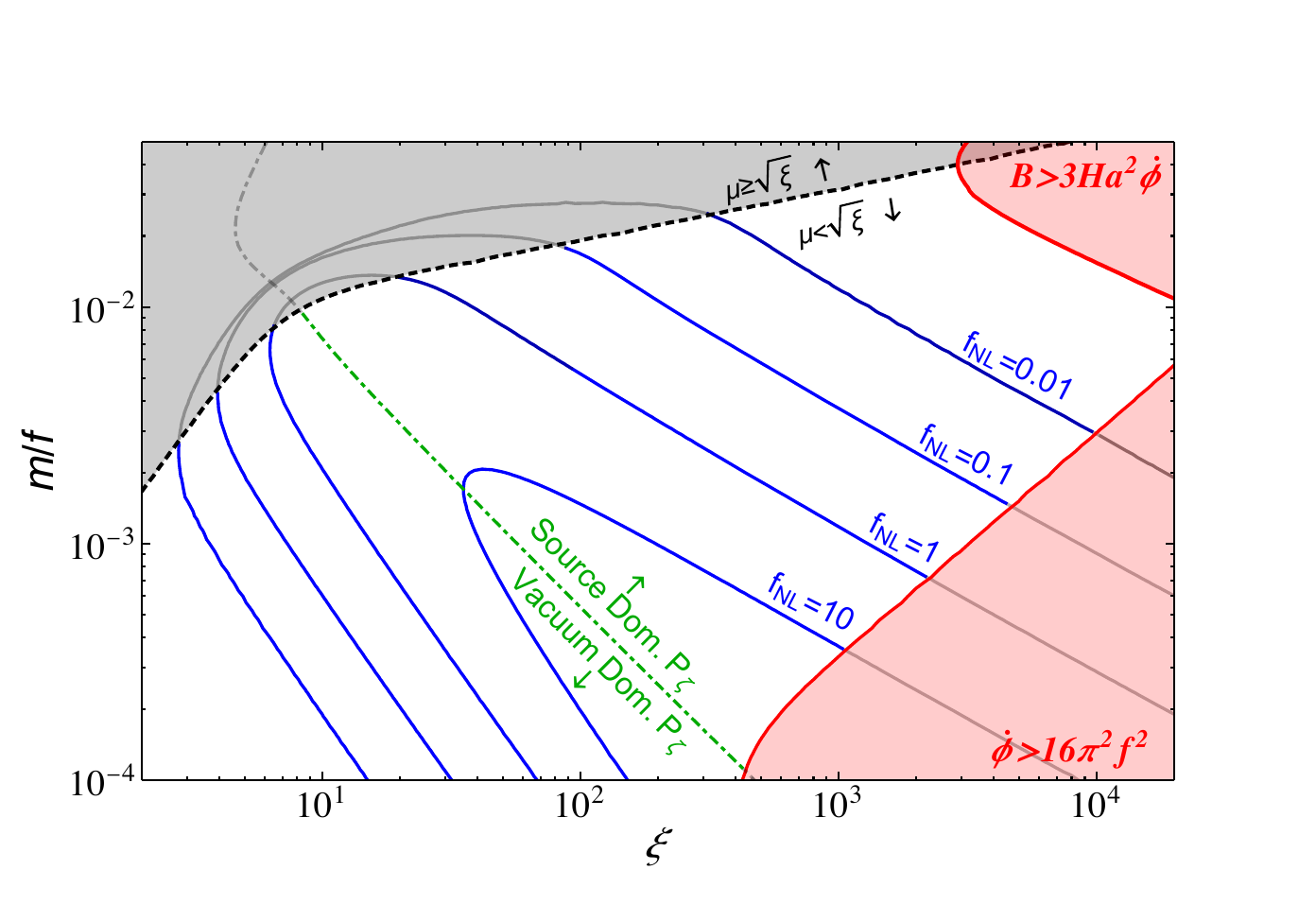}
\caption{Contour lines of non-linear parameter $f_{NL}$ evaluated on exactly equilateral configurations. We indicate the regions of parameter space in which the vacuum and the sourced perturbations dominate the scalar power spectrum. The 
figure also shows the region where $|\dot\phi| >(4\pi f)^2$ (where the effective quantum field theory description of the rolling axion with a fixed decay constant $f$ is inappropriate) and the region $\mu>\sqrt{\xi}$, where the diagrams that we have neglected in the calculation of the power spectrum and bispectrum are not negligible (see discussion in subsection~\ref{subsec:neglect}). The region where the motion of the inflaton is controlled by the backreaction of the produced fermions (where our analysis of the perturbations is invalid) lies in the top right corner of this plot.
}
\label{fig:fnl-plot}
\end{figure}

We show the value taken by the non-linear parameter as a function of parameter space in figure \ref{fig:fnl-plot}, where $H$ is determined as a function of $m/f$ and $\xi$ by imposing the measured normalization of the spectrum of scalar perturbations ($P_\zeta = 2.2 \cdot 10^{-9}$~\cite{Ade:2015lrj}) and we have taken $\log(H/k)=60$. In plotting figure~\ref{fig:fnl-plot} we have used the exact expressions of the quartic and quintic diagrams obtained from eqs.~(\ref{eq:full_quartic}) and (\ref{eq:quintic_exact}) respectively. We see that there is significant parameter space consistent with $f_{NL}^\mathrm{eq}=-4 \pm 43$~\cite{Ade:2015ava}. In the figure we also show the region where $\vert \dot{\phi}_0 \vert < (4\pi f)^2$, and the effective quantum field theory description of the rolling axion with a fixed decay constant $f$ is under control (we also need to impose $H < 4\pi f$; this condition is satisfied wherever $\vert \dot{\phi}_0 \vert < (4\pi f)^2$). We note that, for a fixed value of $m$, the non-Gaussianity first grows with growing $\xi$, and then it decreases. To understand this, we recall that fermion modes of chirality $r=-1$ are produced with momentum up to $\sim 2\, \xi H$, as we discussed after eq. (\ref{eq:nr}). Therefore, as $\xi$ increases we increase the number of fermion modes that are produced, and these then source the inflaton perturbations. The sourced perturbations are non-Gaussian, which explains the initial growth of the non-Gaussianity parameter with $\xi$. However, as $\xi$ keeps growing, the contributions from the various fermion modes add up in an uncorrelated way to each other, and their contribution becomes more and more Gaussian (due to the central limit theorem) as their number grows.\footnote{We note that this differs from the mechanism of non-Gaussian inflation perturbations sourced by a vector field~\cite{Barnaby:2010vf}. In that case the monotonic growth of non-Gaussianity with $\xi$ is due to the fact that the amplitude of the gauge modes grows exponentially with $\xi$.} This argument, and the trend in figure \ref{fig:fnl-plot}, leads us to argue that the perturbations should be Gaussian also in the regime of strong backreaction, where our computation of the perturbations is invalid. 

%%%%%%%%%
%%%%%%%%%
%%%%%%%%%
%%%%%%%%%

%%%%%%%%%%%%%%%%%%%%%%%%%%%%%%%%%%%%%%%%%%%%%%%%%%%%%%%%%%%%%%%%%%%%%%%%%
\section{Fermion production and inflation on a steep axionic potential}%%
\label{sec:strong}%%%%%%%%%%%%%%%%%%%%%%%%%%%%%%%%%%%%%%%%%%%%%%%%%%%%%%%
%%%%%%%%%%%%%%%%%%%%%%%%%%%%%%%%%%%%%%%%%%%%%%%%%%%%%%%%%%%%%%%%%%%%%%%%%

One might wonder whether the dissipation associated to the production of fermions in the regime of strong backreaction could allow for inflation on a steep potential, along the lines of~\cite{Anber:2009ua}. In particular, we would now like to address the question of whether fermion production would allow slow-roll inflation when the inflaton has an axionic potential $V_{\rm {ax}}(\phi)=\Lambda^4\left(\cos(\phi/f)+1\right)$ where the axion constant $f$ is much smaller than $M_P$. It is indeed well known that, for $f<M_P$, the axion potential $V_{\rm {ax}}$ is too steep to support successful slow-roll inflation. On the other hand, it is conjectured that potentials like $V_{\rm {ax}}$ with $f>M_P$ cannot be realized in UV-complete theories involving gravity, so that a mechanism able to produce enough inflation when $f<M_P$ would be of great interest.

We will therefore consider in this section a model  with a cosine-like potential $V(\phi)=V_{\rm {ax}}(\phi)$, so that $V'(\phi)\simeq V(\phi)/f$. Also, in this section only, we will denote the coupling between fermions and the inflaton as $\alpha/f$ with $\alpha$ a dimensionless coefficient. This implies that $\xi=\alpha\dot\phi/(2Hf)$. Given the level of approximation of the discussion in this section, we will set to one all the $O(1)$ and $O(2\pi)$ factors.

The equation for the zero mode of the inflaton, eq.~(\ref{eq:full_el}), including backreaction reads
\begin{align}
\ddot\phi_0+3H\dot\phi_0+V_{\rm {ax}}'(\phi_0)\simeq \frac{\alpha}{f}H^4\mu^2\xi^2=\frac{\alpha^3}{f^3}\mu^2H^2\dot\phi_0^2\,, 
\end{align}
where we recall that $\mu \equiv m/H$, with $m$ being the fermion mass, and that $\phi_0$ denotes the inflaton zero mode. 

We now assume that, in the regime of strong backreaction, the last term on the left hand side of the equation above is balanced by the term on the right hand side (rather than by the Hubble friction), so that, using $V_{\rm {ax}}'\simeq V_{\rm {ax}}/f\simeq H^2M_P^2/f$ we get the slow-roll equation
\begin{align}
\dot\phi_0=\frac{f\,M_P}{\alpha^{3/2}\mu}\,.
\end{align}

We now must impose a number of conditions for our theory to be valid.

\begin{enumerate}

\item First, in order for the effective field theory to make sense, we must work below the cutoff $f$  of the theory. This implies, in particular  $\dot\phi_0\ll f^2$, which gives the first constraint
\begin{align}\label{validity_eft}
\alpha^{3/2}\gg\frac{M_P}{f\,\mu}\gg 1\,,
\end{align}
where the second inequality emerges from requiring $f\ll M_P$ and the assumption $\mu\lesssim 1$.

\item Consistency requires $3H\dot\phi_0\ll V'(\phi_0)$, which gives
\begin{align}\label{neglect_3hdotphi}
\alpha^{3/2}\gg\frac{f^2}{H\,M_P\,\mu}\,.
\end{align}

\item Moreover, the energy density in fermions must be much smaller than that in the inflaton. The energy density of fermions can be computed directly (see appendix~\ref{ap:rho_psi}), and we obtain the scaling
\begin{align}
\rho_\psi\propto H^4\mu^2\xi^3\sim \frac{HM_P^3}{\alpha^{3/2}\mu}\,,
\label{eq:rho_psi_constraint}
\end{align}
which is subdominant with respect to the energy in the inflaton $\sim H^2M_P^2$, as long as 
\begin{align}\label{subdominant_rhopsi}
\alpha^{3/2}\gg\frac{M_P}{H\mu}\,.
\end{align}
Note that this condition is stronger than that of eq.~(\ref{validity_eft}), since $f\gg H$.

\item Of course, we must have inflation, which requires $\dot\phi_0^2\ll H^2M_P^2$; this gives the additional constraint
\begin{align}
\alpha^{3/2}\gg \frac{f}{H\mu}\,,
\end{align}
which is again automatically satisfied if eq.~(\ref{subdominant_rhopsi}) is valid, since $f\ll M_P$.

\item Finally, we must impose that we have a sufficient number of efoldings. To have $60$ efoldings the inflaton must span $\Delta\phi=60\,\dot\phi_0/H\ll f$ since the inflaton potential has period $\sim f$. This gives the condition 
\begin{align}\label{efoldings}
\alpha^{3/2}\gg 60\,\frac{M_P}{\mu H}\,,
\end{align}
which is stronger by a factor $60$ than eq.~(\ref{subdominant_rhopsi}).  

\end{enumerate}

To sum up, we have inflation with the inflaton motion controlled by  backreaction in our model if the two conditions~(\ref{neglect_3hdotphi}) and~(\ref{efoldings}) are satisfied. The arguments presented at the end of section \ref{sec:bs} lead to the conclusion that the perturbations in this regime are likely to be Gaussian to a high degree. However, the computations performed in this work are invalid in this regime. We expect  the study of perturbations in this regime to be rather challenging, and we postpone it to future work.

%%%%%%%%%%%%%%%%%%%%%%
\section{Discussion}%%
\label{sec:disc}%%%%%%
%%%%%%%%%%%%%%%%%%%%%%

In~\cite{Anber:2006xt} it was shown that the (derivative) coupling of a pseudoscalar inflaton to a gauge field would lead to the exponential amplification of the modes of one helicity of the gauge field. In the present paper we have discussed how an analogous coupling to fermions can lead to a fermion number density that is parametrically (although not exponentially) larger than unity in units of the Hubble radius. This can lead to a rich phenomenology, which is all the more interesting because fermions, because of Pauli blocking and because of conformality in the limit $m\to 0$, do not usually give any relevant effect during inflation.  Such a phenomenology is significantly different from that induced by the amplification of the modes of a gauge field  because the large number of produced particles lives, in the fermionic case, at large momenta, whereas in the case of the gauge field these modes live close to the horizon scale.

This work has focused, in particular, on the effect of the backreaction of the produced fermions on the zero mode and on the fluctuations of the inflaton in the regime $\mu\equiv m/H\lesssim 1$, $\xi\equiv \dot\phi/(2fH)\gg 1$. The backreaction on the zero mode turns out to be negligible for $\mu^2\,\xi\ll f^2/H^2$. In the regime in which the fermionic contribution to the power spectrum of scalar metric perturbations is subdominant with respect to the vacuum contribution, the cor frections to $P_\zeta$ and to the non-Gaussianity parameter $f_{NL}$ scale respectively as $(H^2/f^2)\mu^2\,\xi^2$ and $(H^3/f^3)\,\mu^2\,\xi^3$. Remarkably, however, in the regime $\xi\gg 1$ the fluctuations sourced by the fermion become approximately Gaussian, as a consequence of the fact that many fermion modes with large momenta contribute incoherently to the scalar perturbations. This leads to the interesting situation where the spectrum of perturbations is dominated by its sourced component, rather than by its vacuum one, and yet the non-Gaussianities are small and in agreement with observations. In this regime, the measured power spectrum does not yield the combination $H^4/\dot\phi^2$, but rather the combination $H^2\,m^2/f^4$. It will be interesting to evaluate the tensor-to-scalar ratio in this regime, a calculation that we leave to future work.

We have also shown that the backreaction of the fermions on the zero mode of the inflaton can be strong enough to allow for inflation even on a steep scalar potential. This is especially relevant in models of natural inflation with a cosine potential. Potentials with this shape are ubiquitous in string theory and enjoy properties of radiative stability that make them especially attractive. However, these potentials are conjectured to be always too steep to support successful slow-roll inflation. In the system considered in this paper, however, it would be possible to obtain 60 $e$-foldings of inflation on these steep axionic potentials thanks to the slow-down of the inflaton induced by the production of fermions. The  price to pay for this is a large value of the dimensionless parameter $\alpha$ that appears in section~\ref{sec:strong}. A similar mechanism was discussed in~\cite{Anber:2009ua}, with the fermions replaced by a gauge field. That work contained also an estimate of the amplitude of the primordial perturbations, which was found to be too large to agree with observations in the simple case of a single species of gauge field. We expect that an estimate of the spectrum of perturbations generated in the regime where fermions strongly backreact on the inflaton zero mode will be even more difficult, and we leave it to future work. However, given the peculiarities that we have encountered in this study, we can expect a different parametric dependence of the power spectrum that might lead to better agreement with observations.

%%%%%%%%%%%%%%%%%%%%%%%%%%%%%%%%%%%%%%%%%%%%%%%%%%%%%%%%%

\acknowledgments

The work of P.A. and L.P. was supported by the US Department of Energy through grant DE-SC0015655. P.A. and L.P.  thank the Aspen Center for Physics for its hospitality and support through National Science Foundation Grant No. PHY-1066293. L.P. gratefully acknowledges support from a Fortner Fellowship at the University of Illinois at Urbana-Champaign. The work of M.R. and L.S. is partially supported by the US-NSF grant PHY-1520292. The work of M.P. was supported in part by DOE grant DE-SC0011842 at the University of Minnesota.

\appendix

\section{The fermion mode functions and their occupation numbers}
\label{ap:new_basis}

In order to solve the equation of motion of the fermion 
\begin{align}
\left\{ i\,\gamma^\mu\,\partial_\mu-m\,a \left[ \cos \left( \frac{2 \phi}{f} \right) - i \gamma^5 \sin \left( \frac{ 2 \phi }{ f } \right) \right] \right \} \psi=0 \, ,
\end{align}
and to evaluate its occupation number in the $\psi$ basis, it is convenient to first solve the equation of motion of the fermion in the $Y$ basis 
\begin{equation} \label{eqn:YEOM}
\left[ i  \gamma^\mu \partial_\mu - a m - \frac{1}{f}  \gamma^0 \gamma^5 \partial_0 \phi_0  \right] Y  = 0 .\, 
\end{equation} 
Eq.\ \eqref{eqn:YEOM} follows from the Lagrangian in eq.~(\ref{eq:lagr_Y}). We decompose $Y$ as 
\begin{eqnarray}
Y (\bx,\,t)=\int\frac{\exd^3 k}{(2\pi)^{\frac{3}{2}}}e^{i\bk\bx}\,\sum_{r=\pm}\left[\tilde U_r(\bk,t)a_r(\bk)+ \tilde V_r(-\bk,t)b_r^\dagger(-\bk)\right],\nonumber
\label{psi-deco}
\end{eqnarray}
with 
\begin{eqnarray} 
\tilde U_r(\bk,t)=\frac{1}{\sqrt{2}}\left(
\begin{array}{c}
\chi_r(\bk)\,\tilde u_r(k,t)\\
r\chi_r(\bk)\,\tilde v_r(k,t)
\end{array}
\right),\quad  \tilde V_r(\bk,t)=\frac{1}{\sqrt{2}}\left(
\begin{array}{c}
\chi_r(\bk)\,\tilde w_r(k,t)\\
r\chi_r(\bk)\,\tilde y_r(k,t)
\end{array}
\right),
\end{eqnarray}
where $\chi_r(\bk)$ is a helicity-$r$ two-spinor, ${\bf\sigma}\cdot\bk\,\chi_r(\bk)=r\,k\,\chi_r(\bk)$, which we normalize as $\chi_r^\dagger \left( \bk \right) \, \chi_s \left( \bk \right) = \delta_{rs}$, and which can be written explicitly as 
\begin{eqnarray}
\chi_r(\bk)=\frac{\left(k+r\,\sigma\cdot\bk\right)}{\sqrt{2\,k\,(k+k_3)}}\,\bar{\chi}_r,\,\qquad\qquad 
\bar\chi_+=\left(\begin{array}{c}
1 \\
0
\end{array}\right),
\,\quad
\bar\chi_-=\left(\begin{array}{c}
0 \\
1
\end{array}\right)\,.
\end{eqnarray}
Courtesy of the invariance under charge conjugation, we can impose $\tilde V_r \left( \bk \right) = C \, {\bar{\tilde{U}}}_r \left( \bk \right)^T$, where $C = i \gamma^0 \gamma^2$, which implies 
\begin{equation}
\tilde w_r = \tilde v_{-r}^* \;\;,\;\; \tilde y_r = \tilde u_{-r}^* \,, 
\end{equation}
where we have used $i \sigma_2 \chi_r^* \left( \bk \right) = - r \chi_{-r} \left( \bk \right)$. 

The Dirac equation, after defining
\begin{align}
\mu\equiv \frac{m}{H}\,,\quad\xi\equiv \frac{\dot\phi_0}{2\,f\,H}\,,\quad x\equiv -k\tau \,, 
\end{align}
gives the following system 
\begin{eqnarray} 
 \partial_x \tilde u_r&=& i \, \frac{\mu}{x} \, \tilde u_r+i\,\left( 1 +  \frac{2\xi}{x}  \,r\right)  \,\tilde v_r\,,\nonumber\\
 \partial_x \tilde v_r&=&- i \, \frac{\mu}{x}   \, \tilde v_r+i\,\left( 1+\frac{2 \xi}{x}\,r\right)  \,\tilde u_r\,.
\end{eqnarray}
The system is solved by~\cite{Adshead:2015kza}
\begin{equation}
\tilde u_r = \frac{1}{ \sqrt{2 x}} \left( s_r + d_r \right) \;\;,\;\; 
\tilde v_r = \frac{1}{ \sqrt{2 x}} \left( s_r - d_r \right) \,,  
\label{eq:def_uvtilde} 
\end{equation}   
with
\begin{eqnarray}
 s_r =   {\rm e}^{-\pi r \xi} \, {W}_{\frac{1}{2}+2 i r \xi ,\, i \sqrt{\mu^2+4 \xi^2}}(- 2 i x)\,,\quad 
 d_r= - i  \, \mu \, {\rm e}^{-\pi r \xi} \,W_{-\frac{1}{2}+2 i r \xi ,\, i \sqrt{\mu^2+4 \xi^2}}(- 2 i x)\,, 
\end{eqnarray}
where ${W}_{\alpha,\beta}(z)$ denotes the  Whittaker function, and where the integration constants have been determined by imposing the normalization $\vert \tilde u_r \vert^2 + \vert \tilde v_r \vert^2 = 2$ and the positive frequency condition
\begin{align}
\lim_{x \to \infty} u_r \left( x \right) =  \lim_{x \to \infty} v_r \left( x \right) = 
{\rm e}^{i \left( x + 2 r \xi \ln \left( 2 x \right) - \frac{\pi}{4} \right) } .
\label{norma-uv} 
\end{align}
We can now use these results to compute the mode functions of the field $\psi$. Recalling that 
\begin{equation} 
Y = {\rm e}^{-i \gamma^5 \phi / f} \, \psi\,, 
\end{equation} 
and decomposing
\begin{eqnarray}
\psi=\int\frac{\exd^3 k}{(2\pi)^{\frac{3}{2}}}e^{i\bk\bx}\,\sum_{r=\pm}\left[U_r(\bk,t)a_r(\bk)+V_r(-\bk,t)b_r^\dagger(-\bk)\right]\,,
\end{eqnarray} 
we have 
\begin{eqnarray} 
{U}_r \left( \bk ,\tau \right) =  {\rm e}^{i \gamma^5 \phi / f } \,  \tilde U_r \left( \bk ,\tau \right)\,,\quad 
{V}_r \left( \bk , \tau \right) = {\rm e}^{i \gamma^5 \phi / f} \,  \tilde V_r \left( \bk , \tau \right)\,.
\end{eqnarray}  
Next, decomposing  ${U}_r \left( \bk ,\tau \right)$ and ${V}_r \left( \bk , \tau \right)$ as we did for $ \tilde U_r \left( \bk ,\tau \right)$ and $\tilde V_r \left( \bk , \tau \right)$ above, we obtain the relationship
\begin{eqnarray} 
{u}_r(k,\tau) & = & \cos \left( \frac{\phi}{f} \right) \tilde u_r \left( k , \tau \right) 
+ i \, r \, \sin \left( \frac{\phi}{f} \right)\tilde v_r \left( k , \tau \right) \;, \nonumber\\ 
{v}_r(k,\tau) & = & \cos \left( \frac{\phi}{f} \right) \tilde v_r \left( k , \tau \right) 
+ i \, r \, \sin \left( \frac{\phi}{f} \right) \tilde u_r \left( k , \tau \right)\,,  
\label{tildeuv-uv}
\end{eqnarray} 
which gives the expression of $u_r$ and $v_r$ in terms of $s_r$ and $d_r$ presented in eq.~(\ref{eq:def_uv}) in the main text.
It is  straightforward to see that the normalization condition $\vert \tilde u_r \vert^2 + \vert \tilde v_r \vert^2 = 2$ implies  that also $\vert u_r \vert^2 + \vert v_r \vert^2 = 2$. Moreover, one can see that the positive frequency condition~(\ref{norma-uv}) implies
\begin{equation}
\lim_{x \rightarrow \infty} {\tilde u}_r \left( k, \tau \right) = \lim_{x \rightarrow \infty} {\tilde v}_r \left( k, \tau \right) 
=  {\rm e}^{i \left( x + 2 r \xi \ln \left( 2 x \right) - \frac{\pi}{4} \right) } \, {\rm e}^{ - 2 \xi i r \ln \left( x / x_{\rm in} \right)} 
=   {\rm e}^{i \left( x + 2 r \xi \ln \left(  2 x_{\rm in} \right) - \frac{\pi}{4} \right)   } \,, 
\end{equation} 
where we have used $\phi=(\dot\phi_0/f)\log(x_{\rm {in}}/x)$, and this shows  that the subdominant $\log (x)$ term has disappeared from the exponent. 

We can now use these results to diagonalize the Hamiltonian for the fermions. We define 
\begin{equation}
m_R \equiv m\, a \, \cos \left( \frac{2 \phi_0}{f} \right) \;\;,\;\; m_I \equiv  m \,a \, \sin \left( \frac{2 \phi_0}{f} \right) \,, 
\end{equation} 
so that the fermionic part the Hamiltonian reads
\begin{align}
H_{\rm free} = \int \exd^3 x \, \bar \psi \left[ - i\,\gamma^i\,\partial_i + m_R \, - i \gamma^5 \, m_I \right] \psi \,. 
\end{align}

Using the decomposition~(\ref{eq:deco_psi}), performing long algebraic manipulations, and using properties such as
\begin{align} 
\chi_r(-\bk)=-r\,e^{ir\varphi_\bk}\,\chi_{-r}(\bk)\,, \quad e^{i\varphi_\bk}\equiv\frac{k_1+i\,k_2}{\sqrt{k_1^2+k_2^2}}\,;\qquad i \sigma_2 \chi_r^* \left( \bk \right) = - r \chi_{-r} \left( \bk \right)\,,
\end{align}
we eventually obtain
\begin{align} 
H_{\rm free} & =   \int \exd^3 k \left( a_r^\dagger \left( \bk \right) ,\,  b_r \left( - \bk \right) \right) 
\left( \begin{array}{cc} 
A_r & B_r^* \\ 
B_r & - A_r 
\end{array} \right) \, 
\left( \begin{array}{c} 
a_r \left( \bk \right) \\ 
b_r^\dagger \left( - \bk \right) 
\end{array} \right) \;, \nonumber\\ 
A_r &= \frac{1}{2} \left[ m_R \left( \vert {u}_r  \vert^2 -  \vert {v}_r  \vert^2 \right) 
 +  k \left(  {u}_r^*  {v}_r  +  {v}_r^*  {u}_r  \right)  
 - i \, r \, m_I \left(   {u}_r^*  {v}_r  -  {v}_r^*  {u}_r  \right)  \right]  \,, \nonumber\\ 
B_r & =    \frac{r \, e^{ir\varphi_\bk}}{2}  \left[   2 \, m_R {u}_r  \, {v}_r  - k \left(  {u}_r^2  -  {v}_r^2  \right)   - i \, r \, m_I \left( {u}_r^2  +  {v}_r^2  \right) \right]\,.
\label{eq:Hfree}
\end{align} 

We next diagonalize the Hamiltonian. We find that the matrix in eq.~(\ref{eq:Hfree}) above has eigenvalues $\pm\omega$, with
\begin{align}
\omega \equiv\sqrt{k^2 + m_R^2 + m_I^2} \;, 
\end{align}
so that the diagonalization will be realized by finding two numbers $\alpha_r$ and $\beta_r$  for which
\begin{eqnarray}
\left(\begin{array}{cc}
A_r & B_r^* \\ 
B_R & -A_r 
\end{array}\right)=
\left(\begin{array}{cc}
\alpha_r^* & \beta_r^* \\
-\beta_r & \alpha_r
\end{array}\right)
\left(\begin{array}{cc}
\omega &  0 \\
0 & -\omega
\end{array}\right)
\left(\begin{array}{cc}
\alpha_r & -\beta_r^* \\
\beta_r & \alpha^*_r
\end{array}\right)\,. 
\end{eqnarray}
This transformation can be interpreted as a definition of the operators that create and annihilate the quanta that diagonalize the Hamiltonian at the time $t$, see eq. (\ref{ab-hat}).  (The coefficients $\alpha$ and $\beta$ are conventionally denoted as  Bogolyubov coefficients.) The above equation is solved by
\begin{align} 
\alpha_r &= {\rm e}^{i r  \varphi_{\bk}/2+i \lambda_r} 
\left[ \frac{1}{2} \sqrt{ 1 + \frac{m_R}{\omega} } \, {u}_r + 
\frac{1}{2} \sqrt{ 1 - \frac{m_R}{\omega} } \, {\rm e}^{-i r \theta} \, {v}_r \right]\,, \nonumber\\ 
\beta_r &=  r  \,   {\rm e}^{i r \varphi_{\bk}/2- i \lambda_r} \, 
\left[  \frac{1}{2} \sqrt{ 1 - \frac{m_R}{\omega} } \, {\rm e}^{i r \theta} \,  { u}_r -  
\frac{1}{2} \sqrt{ 1 + \frac{m_R}{\omega} } \,  {v}_r \right]\,,\qquad
{\rm e}^{i \theta}\equiv \frac{k + i m_I}{\sqrt{k^2+m_I^2} } \;, 
\label{Nr} 
\end{align} 
where $\lambda_r$ is arbitrary and real. We get the occupation number 
\begin{align} 
N_r =  \vert \beta_r \vert^2 =& \frac{1}{2} - \frac{m_R}{4 \omega} \left( \vert {u}_r \vert^2 - \vert {v}_r \vert^2 \right) 
- \frac{k}{2 \omega} \, {\rm Re } \left( {u}_r^* { v}_r \right) 
- \frac{r \, m_I}{2 \omega} \, {\rm Im } \left( {u}_r^* {v}_r \right)\nonumber\\
= & \frac{1}{2}  - \frac{\mu}{2 x \sqrt{x^2+\mu^2}}  \, {\rm Re } \left[ s_r^* \, d_r \right]   - \frac{1}{4 \sqrt{x^2+\mu^2}} \,  \left[ \vert s_r \vert^2 - \vert d_r \vert^2 \right]\,,
\end{align} 
where the first line corresponds to the expression~(\ref{eq:nr}) in the main text.

\section{Motivating the change of basis}
\label{ap:old_basis}

In this appendix we further motivate our choice of using the $\psi$ variables in the basis of (\ref{eq:lagr_psi}) rather than the $Y$ variables in the basis of (\ref{eq:lagr_Y}) to compute the fermion production. We do so by computing and diagonalizing the fermion Hamiltonian in the $Y$ basis. Starting from  (\ref{eq:lagr_Y}), we have the conjugate fields 
\begin{equation}
\Pi_\phi \equiv \frac{\partial L}{\partial \dot{\phi}} = a^2 \, \dot{\phi} - \frac{1}{f} Y^\dagger \gamma^5 Y \;\;,\;\; 
\Pi_Y \equiv  \frac{\partial L}{\partial \dot{Y}} - i \, Y^\dagger \;, 
\end{equation} 
and the free Hamiltonian 
\begin{equation}
H_{\rm free} = i \int \exd^3 x \, Y^\dagger \, \partial_0 \, Y \;. 
\end{equation} 

Next, we perform the decomposition (\ref{psi-deco}), we insert it in this Hamiltonian, and we repeat the same steps that led to the expression (\ref{eq:Hfree}) in the $\psi$ basis, obtaining 
\begin{align} 
H_{\rm free} & =   \int \exd^3 k \left( a_r^\dagger \left( \bk \right) ,\,  b_r \left( - \bk \right) \right) 
\left( \begin{array}{cc} 
{\tilde A}_r & {\tilde B}_r^* \\ 
{\tilde B}_r & - {\tilde A}_r 
\end{array} \right) \, 
\left( \begin{array}{c} 
a_r \left( \bk \right) \\ 
b_r^\dagger \left( - \bk \right) 
\end{array} \right) \;, \nonumber\\ 
{\tilde A}_r &= \frac{1}{2} \left[ a \, m \left( \vert {\tilde u}_r  \vert^2 -  \vert {\tilde v}_r  \vert^2 \right) 
 +  \left( k + r \, \frac{\phi'_0}{f} \right) \left(  {\tilde u}_r^*  {\tilde v}_r  +  {\tilde v}_r^*  {\tilde u}_r  \right)  
 \right]  \,, \nonumber\\ 
{\tilde B}_r & =    \frac{r \, e^{ir\varphi_\bk}}{2}  \left[   2 a \, m \, {\tilde u}_r  \, {\tilde v}_r  - \left( k + r \, \frac{\phi_0'}{f} \right) \left(  {\tilde u}_r^2  -  {\tilde v}_r^2  \right)   \right]\,.
\label{Hfree-tilde}
\end{align}

We diagonalize it as we did for the Hamiltonian in the $\psi$ basis 
\begin{eqnarray}
\left(\begin{array}{cc}
{\tilde A}_r & {\tilde B}_r^* \\ 
{\tilde B}_r & -{\tilde A}_r 
\end{array}\right)=
\left(\begin{array}{cc}
{\tilde \alpha}_r^* & {\tilde \beta}_r^* \\
-{\tilde \beta}_r & {\tilde \alpha}_r
\end{array}\right)
\left(\begin{array}{cc}
{\tilde \omega}_r &  0 \\
0 & -{\tilde \omega}_r
\end{array}\right)
\left(\begin{array}{cc}
{\tilde \alpha}_r & -{\tilde \beta}_r^* \\
{\tilde \beta}_r & {\tilde \alpha}^*_r
\end{array}\right)\,,  
\end{eqnarray}
and find 
\begin{eqnarray}
{\tilde \alpha}_r &=& \frac{{\rm e}^{i r \phi_{\bk}/2}}{2} \left[ \left( 1 + \frac{a m}{{\tilde \omega}_r} \right)^{1/2} {\tilde u}_r + \sigma \left( 1 - \frac{a m}{{\tilde \omega}_r} \right)^{1/2} \, {\tilde v}_r \right] \;, \nonumber\\ 
{\tilde \beta}_r &=& \frac{r \, {\rm e}^{i r \phi_{\bk}/2}}{2} \left[ \sigma \left( 1 - \frac{a m}{{\tilde \omega}_r} \right)^{1/2} {\tilde u}_r -  \left( 1 + \frac{a m}{{\tilde \omega}_r} \right)^{1/2} \, {\tilde v}_r \right] \;, \nonumber\\ 
{\tilde \omega}_r &=& \sqrt{a^2 m^2 + \left( k + r \, \frac{\phi_0'}{f} \right)^2} \;\;,\;\; 
\sigma = {\rm sign } \left(  k + r \, \frac{\phi'}{f} \right) \;. 
\end{eqnarray} 

The free fermion Hamiltonian is obtained by disregarding the inflaton perturbations. We see that in this basis the free Hamiltonian for one helicity vanishes at a finite time in the massless fermion limit, when ${\tilde \omega}_r = 0$. This is not a singularity of the system, as the total Hamiltonian does not vanish at this instant, but only a sign that  the expansion into free and interacting Hamiltonian is not under perturbative control in the $Y$ basis for $m \rightarrow 0$. Therefore, unphysical effects might be expected if one uses this basis to compute the fermion production for $m \rightarrow 0$. Indeed, let us assume $\phi_0' > 0$ (in the opposite case, the results for the two chiralities are interchanged). 
Inserting the solutions (\ref{eq:def_uvtilde}) into $N_r = \vert {\tilde \beta}_r \vert^2$, and studying the small $\mu$ limit, we find that the occupation number for positive chirality fermions is of ${\mathcal O } ( \mu^2 )$. On the other hand, the one for negative chirality fermions has a sudden transition when $\sigma$ changes sign. At sufficiently early times, the physical momentum of a mode satisfies $p > {\dot{\phi}_0}/{f}$ and $\sigma > 0$. It is easy to verify that $N_{-1} = {\mathcal O } ( \mu^2 )$ in this regime. As $p$ drops below 
$ {\dot{\phi}_0}/{f}$ and $\sigma$ changes sign, one finds  $N_{-1} = 1 - {\mathcal O } ( \mu^2 )$. This sudden transition, and the final saturation of the occupation number $N_{-1}$ in the $\mu \rightarrow 1$ regime, are unphysical results, related to the breaking of the perturbative expansion based on the fact that the unperturbed Hamiltonian (\ref{Hfree-tilde}) vanishes when  $p = {\dot{\phi}_0}/{f}$ for $\mu  = 0$. More in general, the computation in the $Y$ basis goes out of perturbative control for small masses. For this reason, we compute the fermion production in the $\psi$ basis. 

\section{Backreaction calculations}
\label{ap:backreaction}

In this appendix we present the derivation of eq.~\eqref{eq:calb}, along with the analytic evaluation of its integral.  Beginning  with eq.~\eqref{eq:B_def} we evaluate the fermionic expectation value using eq.~\eqref{eq:deco_psi}, finding
\begin{align}
\mathcal{B} &= 
\dfrac{2 m}{f a(\tau) } \int \dfrac{\exd^3 p \, \exd^3q}{(2\pi)^3} e^{i(\bp - \bq) \cdot \bx} \delta(\bp - \bq)
\sum_r V^\dagger_{r,-\bp} (\tau) 
\left[ \sin \left( \dfrac{2 \phi_0}{f} \right) 
+ i \gamma_5 \cos \left( \dfrac{2 \phi_0}{f} \right) \right] 
V_{r,-\bp}(\tau) \nonumber \\
&= \dfrac{2 m}{f a(\tau)} \int \dfrac{\exd^3p}{(2\pi)^3} \sum_r
[-ir (\tilde u^* \tilde v - \tilde v^* \tilde u)]_{r,p,\tau} \,, 
\end{align}
in terms of the functions $\tilde u$ and $\tilde v$ which diagonalize the unrotated Hamiltonian (see eq.~\eqref{eq:def_uvtilde}).  This can be found either by using eq.~\eqref{tildeuv-uv} or by writing $\psi$ in terms of $Y$ prior to evaluating the expectation value.  We introduce the variable $y = - p \tau$ and substitute $a = -1 \slash H \tau$ to find
\begin{align}
\mathcal{B} &= -\dfrac{2 m H}{f  \tau^2} \cdot \dfrac{4 \pi}{(2\pi)^3} \cdot 
\int \exd y \, y \sum_r (ir) (d^* s - s^* d) = \dfrac{4 m H}{f  \tau^2} \cdot \dfrac{1}{2 \pi^2} \sum_r r \int \exd y\,y \,\Im(d^*s) \,, 
\label{eq:B_intermediate}
\end{align}
in agreement with eq.~\eqref{eq:calb}.  The remaining integral can be evaluated analytically; note that it will be used both for the backreaction and for the non-Gaussianity calculation in appendix \ref{ap:nongaussianities}, whose results are discussed in section~\ref{sec:bs}.

\begin{figure}
\centering
\includegraphics[scale=.8]{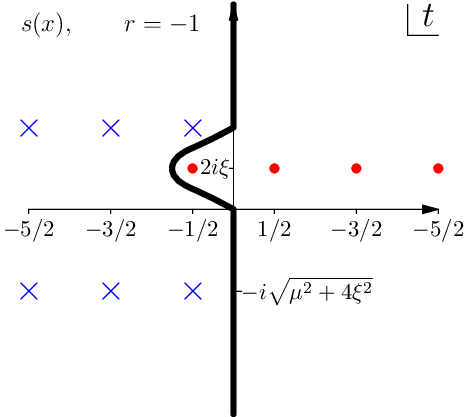} 
\hspace*{1cm}
\includegraphics[scale=.8]{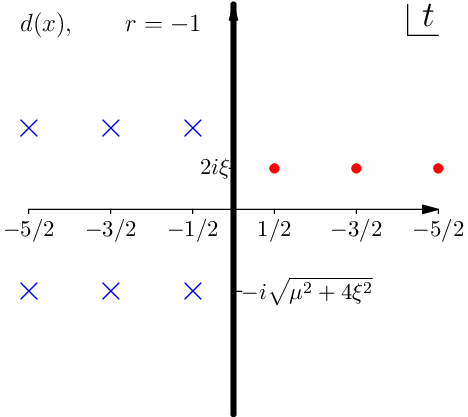} 
\caption{Location of poles for $s_{r=-1}(x)$ and $d_{r=-1}(x)$ and appropriate contour.  The blue crosses correspond to the poles of $\Gamma(1\slash 2 \pm b + t)$ while the red dots correspond to the poles of $\Gamma(-a - t)$, with $b = i \sqrt{\mu^2 + 4 \xi^2}$ and $a = 1 \slash 2 + 2 i r \xi$ (for $s$) and $a = - 1 \slash 2 + 2 i r \xi$ (for $d$).  For $r = +1$ the red poles reflect across the real axis.}
\label{fig:poles}
\end{figure}

The Mellin-Barnes representation of the Whittaker function is
\begin{align}
W_{a,b}(z) &= \dfrac{e^{-z \slash 2}}{2 \pi i} \int_{-i \infty}^{i \infty} \dfrac{\Gamma \left( \dfrac{1}{2} + b + t \right) \Gamma \left( \dfrac{1}{2} - b + t \right) \Gamma \left( - a - t \right) }{\Gamma \left( \dfrac{1}{2} + b - a \right) \Gamma \left( \dfrac{1}{2} - b - a \right)} z^{-t} \, \exd t \,, 
\end{align}
where the contour of integration is deformed to separate the poles of $\Gamma \left({1}/{2} + b + t \right) \Gamma \left( {1}/{2} - b + t \right) $ from the poles of $\Gamma \left( - a - t \right)$ when needed.  The location of the poles and integral path for our $s$ and $d$ functions is shown in figure~\ref{fig:poles}.  Applying this gives
% LONG CALC STARTS HERE
\begin{align}
&s_r(x) d_r^*(x) \nonumber \\
&= e^{- \pi r \xi} \cdot \dfrac{e^{i x}}{2\pi i}
\int_{-i\infty}^{i \infty} \exd t \,
\left( - 2 i x \right)^{-t} \nonumber \\
& \qquad \cdot 
\dfrac{
\Gamma \left( \dfrac{1}{2} + i \sqrt{ \mu^2 + 4 \xi^2} + t \right)
\Gamma \left( \dfrac{1}{2} - i \sqrt{ \mu^2 + 4 \xi^2} + t \right)
\Gamma \left( - \dfrac{1}{2} - 2 i r \xi - t \right)
}
{
\Gamma \left(- 2 i r \xi + i \sqrt{ \mu^2 + 4 \xi^2} \right)
\Gamma \left(- 2 i r \xi - i \sqrt{ \mu^2 + 4 \xi^2} \right)
}
 \nonumber \\
& \cdot i \mu e^{- \pi r \xi} 
\cdot \dfrac{e^{-ix}}{-2 \pi i}
\int_{i \infty}^{-i \infty} \exd s\, 
\left(  2 i x \right)^{-s} \nonumber \\
& \qquad \cdot \dfrac{
\Gamma \left( \dfrac{1}{2} - i \sqrt{ \mu^2 + 4 \xi^2} + s \right)
\Gamma \left( \dfrac{1}{2} + i \sqrt{ \mu^2 + 4 \xi^2} + s \right)
\Gamma \left( \dfrac{1}{2} + 2 i r \xi - s \right)
}
{
\Gamma \left(1 + 2 i r \xi - i \sqrt{ \mu^2 + 4 \xi^2} \right)
\Gamma \left(1 + 2 i r \xi + i \sqrt{ \mu^2 + 4 \xi^2} \right)
} \,, 
\label{eq:MellinBarnes}
\end{align}
using the fact that $\Gamma(z)^* = \Gamma(z^*)$.  The $\exd t$ and $\exd s$ integrals are along the imaginary axis, except where deformations are required, as described in figure~\ref{fig:poles}.

This in turn gives
\begin{align}
&\int y s_r(y) d_r^*(y) \, \exd y = - \dfrac{ i \mu e^{- 2\pi r \xi}}{(2\pi)^2} 
\int_m^\Lambda \exd y \int_{-i\infty}^{i \infty}  \exd t \, 
\int_{-i\infty}^{i \infty}  \exd s \, 2^{-(t+s)} y^{1 - (t+s)} i^{t - s} \nonumber \\
& \qquad \times
\dfrac{
\Gamma \left( \dfrac{1}{2} + i \mathfrak{a} + t \right)
\Gamma \left( \dfrac{1}{2} - i \mathfrak{a} + t \right)
\Gamma \left( - \dfrac{1}{2} - ir \mathfrak{b} - t \right)
}
{
\Gamma \left(- ir \mathfrak{b} + i \mathfrak{a} \right)
\Gamma \left(- ir \mathfrak{b} - i \mathfrak{a} \right)
}
 \nonumber \\
& \qquad \times
\dfrac{
\Gamma \left( \dfrac{1}{2} - i \mathfrak{a} + s \right)
\Gamma \left( \dfrac{1}{2} + i \mathfrak{a} + s \right)
\Gamma \left( \dfrac{1}{2} + ir \mathfrak{b} - s \right)
}
{
\Gamma \left(1 + ir \mathfrak{b} - i \mathfrak{a} \right)
\Gamma \left(1 + ir \mathfrak{b} + i \mathfrak{a} \right)
} \,, 
\end{align}
where we have defined 
\begin{align}
\mathfrak{a} \equiv \sqrt{ \mu^2 + 4 \xi^2}, \qquad
\mathfrak{b} \equiv 2 \xi,
\end{align}
and imposed cutoffs on the $\exd y$ (momentum) integral.  The $\exd y$ integral is now explicitly convergent; therefore we can interchange the order of the integrals and perform the $\exd y$ integral first.  We find
\begin{align}
\!\!\!\!\!\!\!\! \!\!\!\!  
\int y s_r(y) d_r^*(y) \, \exd y
&= - \dfrac{ i \mu e^{- 2\pi r \xi}}{(2\pi)^2} 
\int_{-i\infty}^{i \infty}  \exd t \, 
\int_{-i\infty}^{i \infty}  \exd s \, 2^{-(t+s)} i^{t - s}
\dfrac{\Lambda^{2 - (t+s)}}{2 - (t + s)}
\nonumber \\
& \qquad \times 
\dfrac{
\Gamma \left( \dfrac{1}{2} + i \mathfrak{a} + t \right)
\Gamma \left( \dfrac{1}{2} - i \mathfrak{a} + t \right)
\Gamma \left( - \dfrac{1}{2} - ir \mathfrak{b} - t \right)
}
{
\Gamma \left(- ir \mathfrak{b} + i \mathfrak{a} \right)
\Gamma \left(- ir \mathfrak{b} - i \mathfrak{a} \right)
}
 \nonumber \\
& \qquad \times
\dfrac{
\Gamma \left( \dfrac{1}{2} - i \mathfrak{a} + s \right)
\Gamma \left( \dfrac{1}{2} + i \mathfrak{a} + s \right)
\Gamma \left( \dfrac{1}{2} + ir \mathfrak{b} - s \right)
}
{
\Gamma \left(1 + ir \mathfrak{b} - i \mathfrak{a} \right)
\Gamma \left(1 + ir \mathfrak{b} + i \mathfrak{a} \right)
} \,, 
\end{align}
after taking the lower cutoff $m \rightarrow 0$.  The contours are closed with $\Re(t), \Re(s) \rightarrow  \infty$, which encloses poles at 
$0 = 2 - (t + s)$ along with those from the Gamma functions at $t = n_1 - 1/2 - i r \mathfrak{b}$ and $s = n_2 + 1/2 + i r \mathfrak{b}$ where $n_1$ and $n_2$ are positive integers.  These last two sets of poles correspond to those denoted with red dots for the $s$ and $d$ functions in figure~\ref{fig:poles} (recalling that the $d$ function is conjugated).  The residue of the Gamma functions at its poles, which are all simple poles, is given by
\begin{align}
\mathrm{Res}(\Gamma,-n) &= \dfrac{(-1)^n}{n!}.
\end{align}
Note that both contours are negatively oriented (i.e., clockwise).

We first consider the $\exd t$ integral, evaluating
\begin{align}
\dfrac{\mathcal{I}}{2\pi i} &\equiv
\int_{-i\infty}^{i \infty}  \exd t \, 
\dfrac{
\Gamma \left( \dfrac{1}{2} - i \mathfrak{a} + t \right)
\Gamma \left( \dfrac{1}{2} + i \mathfrak{a} + t \right)
\Gamma \left( -\dfrac{1}{2} + ir \mathfrak{b} - t \right)
}
{
\Gamma \left( -ir \mathfrak{b} - i \mathfrak{a} \right)
\Gamma \left( -ir \mathfrak{b} + i \mathfrak{a} \right)
}
2^{-(t+s)} i^{t - s}
\dfrac{\Lambda^{2 - (t+s)}}{2 - (t + s)} \, .
\end{align}
Below the contributions of the various poles to this integral are presented.  Although there are infinitely many poles, only a finite number contribute as $\Lambda \rightarrow \infty$.

\paragraph*{$n_1=0$ Pole:} First we evaluate the pole corresponding to $n_1=0$ which is at $t= -1 \slash 2 - i r \mathfrak{b}$, which contributes
\begin{align}
\mathcal{I}_{n_1=0} 
&= -2^{1 \slash 2 + i r \mathfrak{b} - s} 
i^{ - 1 \slash 2 - i r \mathfrak{b} - s}
\dfrac{\Lambda^{5 \slash 2 + i r \mathfrak{b} - s}}{5 \slash 2 + i r \mathfrak{b} - s} .
\end{align}
(This includes a negative sign from the orientation of the contour.)  From above, we note that because $\exd s$ integral has a poles at $n_2 + 1 \slash 2 + i r \mathfrak{b}$, the lowest $\exd s$ pole ($n_2 = 0$) will contribute a piece that diverges quadratically as $\Lambda \rightarrow \infty$.

\paragraph*{$n_1 = 1$ Pole:} The next pole is at $t = 1 \slash 2 - i r \mathfrak{b}$.  In order to simplify the residue of this pole, it is helpful to observe that $\Gamma(z+1) = z \Gamma(z)$, allowing us to write
\begin{align}
\dfrac{\Gamma \left( \dfrac{1}{2} - i \mathfrak{a} + t \right) \Gamma \left( \dfrac{1}{2} + i \mathfrak{a} +t \right)}{\Gamma(-i r \mathfrak{b} - i \mathfrak{a}) \Gamma( -i r \mathfrak{b} + i \mathfrak{a})} \bigg|_{t = 1 \slash 2 - i r \mathfrak{b}}
&= \mathfrak{a}^2 - \mathfrak{b}^2 = \mu^2.
\end{align}
Therefore the contribution from this pole is
\begin{align}
\mathcal{I}_{n_1=1} 
&=  \mu^2
2^{i r \mathfrak{b} - 1 \slash 2 -s}
i^{1 \slash 2 - i r \mathfrak{b}-s}
\dfrac{ 
\Lambda^{3 \slash 2 + i r \mathfrak{b} - s}
}{
3 \slash 2 + i r \mathfrak{b} - s
} \,, 
\end{align}
which has a linearly divergent piece as $\Lambda \rightarrow \infty$ at the first pole of the $\exd s$ integral.

\paragraph*{$n_1 = 2$ Pole:} The next pole is at $t = 3 \slash 2 - i r \mathfrak{b}$, and it contributes 
\begin{align}
\mathcal{I}_{n_1 = 2} 
&= - \dfrac{\mu^2}{2} \left[ 1 - 4 i r \xi + \mu^2 \right]
2^{i r \mathfrak{b} - 3 \slash 2 -s}
i^{3 \slash 2 - i r \mathfrak{b}-s}
\dfrac{ 
\Lambda^{1 \slash 2 + i r \mathfrak{b} - s}
}{
1 \slash 2 + i r \mathfrak{b} - s
} .
\end{align}
Since the $s$ poles begin at $1 \slash 2 + i r \mathfrak{b}$, this contribution is at worst log divergent in $\Lambda$.

\paragraph*{$n_1 \geq 3$ Poles:} The $n_1 = 3$ pole scales as $\Lambda^{-1 \slash 2 + i r \mathfrak{b} - s}$.  The poles of $\exd s$ integral are at $n_2 + 1 \slash 2 + i r \mathfrak{b}$.  Therefore, even for $n_2=0$, $\Lambda$ would have a negative exponent, and therefore, this contribution goes to zero as $\Lambda \rightarrow \infty$.  Similar reasoning applies to all poles with $n_1 \geq 3$, and therefore we may neglect them.

\paragraph*{$t = 2 -s$ Pole:} The final pole for the $\exd t$ integral arises from the simple pole resulting from the $\exd y$ integral.  Note that at this pole, $\Lambda$ has a zero power, so there is no effect from taking the regulator to infinity.   The contribution from this pole is
\begin{align}
\mathcal{I}_{t=2-s} 
&=
 - 
\dfrac{i^{2(1 -s)}}{4} \dfrac{\big| \Gamma \left( \dfrac{5}{2} - i \mathfrak{a} - s \right)\big|^2 \Gamma \left( - \dfrac{5}{2} + i r \mathfrak{b} + s \right)}{\Gamma(- ir \mathfrak{b} - i \mathfrak{a}) \Gamma( - i r \mathfrak{b} + i \mathfrak{a})}.
\end{align}

Therefore, after evaluating the $\exd t$ integral and dropping all terms which are negligible as $\Lambda \rightarrow \infty$, we have
\begin{align}
\!\!\!\!\!\!\!\! \!\!\!\!  
&\int y s_r(y) d_r^*(y) \, \exd y
=  -\dfrac{  \mu e^{- 2\pi r \xi}}{(2\pi)} 
\int_{-i\infty}^{i \infty}  \exd s \, 
\dfrac{
\Gamma \left( \dfrac{1}{2} - i \mathfrak{a} + s \right)
\Gamma \left( \dfrac{1}{2} + i \mathfrak{a} + s \right)
\Gamma \left( \dfrac{1}{2} + ir \mathfrak{b} - s \right)
}
{
\Gamma \left(1 + ir \mathfrak{b} - i \mathfrak{a} \right)
\Gamma \left(1 + ir \mathfrak{b} + i \mathfrak{a} \right)
} \nonumber \\
& \qquad \left(
2^{i r \mathfrak{b} + 1 \slash 2  - s} 
i^{ - 1 \slash 2 - i r \mathfrak{b} - s}
\dfrac{\Lambda^{5 \slash 2 + i r \mathfrak{b} - s}}{5 \slash 2 + i r \mathfrak{b} - s} 
- \mu^2
2^{i r \mathfrak{b} - 1 \slash 2 -s}
i^{1 \slash 2 - i r \mathfrak{b}-s}
\dfrac{ 
\Lambda^{3 \slash 2 + i r \mathfrak{b} - s}
}{
3 \slash 2 + i r \mathfrak{b} - s
} \right. \nonumber \\
& \qquad\qquad  \left.
+ \dfrac{\mu^2}{2} \left[ 1 - 4 i r \xi + \mu^2 \right]
2^{i r \mathfrak{b} - 3 \slash 2 -s}
i^{3 \slash 2 - i r \mathfrak{b}-s}
\dfrac{ 
\Lambda^{1 \slash 2 + i r \mathfrak{b} - s}
}{
1 \slash 2 + i r \mathfrak{b} - s
} \right. \nonumber \\
& \qquad\qquad \left. 
 + 
\dfrac{i^{2(1 -s)}}{4} \dfrac{\big| \Gamma \left( \dfrac{5}{2} - i \mathfrak{a} - s \right)\big|^2 \Gamma \left( - \dfrac{5}{2} + i r \mathfrak{b} + s \right)}{\Gamma(- ir \mathfrak{b} - i \mathfrak{a}) \Gamma( - i r \mathfrak{b} + i \mathfrak{a})}
\right).
\end{align}
For the terms involving $\Lambda$, the $\exd s$ integral can be performed in the same manner.  Although there are an infinite number of poles, again only a finite number contribute as $\Lambda \rightarrow \infty$.  Taking care with the double poles, we find
\begin{align}
& \sum_r \int r y s_r(y) d_r^*(y) \, \exd y
= \mu \left[ 
-4 \Lambda  \xi +\frac{1}{2} 
\left(
-12 i \xi  \log (2 \Lambda ) -12 i \gamma_E  \xi -5 i \xi  \left(\mu ^2+4 \xi ^2\right)+20 i \xi ^3+8 i \xi   \right. \right. \nonumber \\
& \left. \left. \quad +6 \pi  \xi +
\left(\mu ^2-8 \xi ^2+6 i \xi +1\right) H_{-i \left(\sqrt{\mu ^2+4 \xi ^2}-2 \xi \right)}
-\left(\mu ^2-8 \xi ^2-6 i \xi +1\right) H_{i \left(\sqrt{\mu ^2+4 \xi ^2}-2 \xi \right)}
\right. \right. \nonumber \\
& \quad \left. \left. 
-\left(\mu ^2-8 \xi ^2-6 i \xi +1\right) H_{-i \left(2 \xi +\sqrt{\mu ^2+4 \xi ^2}\right)}
+\left(\mu ^2-8 \xi ^2+6 i \xi +1\right) H_{i \left(2 \xi +\sqrt{\mu ^2+4 \xi ^2}\right)}  \right)
\right] \nonumber \\
&  + \sum_r r \mathcal{A}_r \,, 
\label{eq:intermediate1}
\end{align}
when $H_n$ denotes the $n$-th harmonic number, and where we have performed the sum over $r$.  The quadratic $\Lambda^2$ contribution was removed by performing this sum, as $\sum_{r = \pm 1} r \Lambda^2 = 0$.  As noted in eq.~\eqref{eq:B_intermediate} above, we are interested in the imaginary part of this quantity.  Note that taking the imaginary part will eliminate the linear divergence; therefore, the backreaction is logarithmically divergent in the UV.  The last line, which corresponds to the final $\exd t$ pole above, is independent of $\Lambda$ and is given by
\begin{align}
& \mathcal{A}_r = -\mu \int_{-i\infty}^{i \infty} \exd s  \frac{|i a-s+\frac{1}{2}|^2 
|i \mathfrak{a}-s+\frac{3}{2}|^2
e^{-\pi  (\mathfrak{b} r+i s)} }{8
\left(-i \mathfrak{b} r+s-\frac{5}{2}\right) \left(-i \mathfrak{b} r+s-\frac{3}{2}\right) \left(-i \mathfrak{b} r+s-\frac{1}{2}\right)}
\nonumber \\ 
& \qquad \cdot
\dfrac{\sinh (\pi  (\mathfrak{a}-\mathfrak{b} r)) \sinh (\pi  (\mathfrak{a}+b r)) }{\sin \left(\pi  \left(-i \mathfrak{b} r+s+\frac{1}{2} \right)\right)
\sin \left(\pi  \left(-i \mathfrak{a}+s+\frac{1}{2}\right)\right) 
\sin \left(\pi  \left(i \mathfrak{a}+s+\frac{1}{2}\right)\right) }
.
\label{eq:Ar}
\end{align}
Note that the infinite number of $\exd s$ poles contribute, as there is no regulator to eliminate them as $\Lambda \rightarrow \infty$.  Nonetheless, this integral can be evaluated with a trick introduced in~\cite{Kobayashi:2014zza,Frob:2014zka}.  We define the functions
\begin{align}
f_1(s) &\equiv -\dfrac{e^{-\pi  (\mathfrak{b} r+i s)} \sinh (\pi  (\mathfrak{a}-\mathfrak{b} r)) \sinh (\pi  (\mathfrak{a}+\mathfrak{b} r)) }{8 \sin \left(\pi  \left(-i \mathfrak{b} r+s+\frac{1}{2} \right)\right)
\sin \left(\pi  \left(-i \mathfrak{a}+s+\frac{1}{2}\right)\right) 
\sin \left(\pi  \left(i \mathfrak{a}+s+\frac{1}{2}\right)\right) } ,
 \nonumber \\
f_2(s) &\equiv \frac{|i \mathfrak{a}-s+\frac{1}{2}|^2 
|i a-s+\frac{3}{2}|^2
}{
\left(-i \mathfrak{b} r+s-\frac{5}{2}\right) \left(-i \mathfrak{b} r+s-\frac{3}{2}\right) \left(-i \mathfrak{b} r+s-\frac{1}{2}\right)}.
\end{align}
Note that $f_1(s-1) = f_1(s)$, and the integrand is equal to $f_1(s) f_2(s)$.  We write the second function as $f_2(s) = g(s) - g(s-1) + h(s)$, where
\begin{align}
g(s)
&= \frac{1}{2} \left(-i \mathfrak b r+s-\frac{1}{2}\right)^2+\left(\frac{3}{2}+4 i \mathfrak b r\right) \left(-i \mathfrak b r+s-\frac{1}{2}\right) \nonumber \\
& \quad + \frac{(\mathfrak a+\mathfrak b r) (\mathfrak a+\mathfrak b r+i) (\mathfrak a-\mathfrak b r) (\mathfrak a-\mathfrak b r-i)}{2 \left(-i \mathfrak b r+s-\frac{1}{2}\right)} \nonumber \\
& \quad -\frac{(\mathfrak a-\mathfrak b r) (\mathfrak a-\mathfrak b r+i) (\mathfrak a+\mathfrak b r) (\mathfrak a+\mathfrak b r-i)-\frac{1}{2} (\mathfrak a-\mathfrak b r) (\mathfrak a-\mathfrak b r-i) (\mathfrak a+\mathfrak b r) (\mathfrak a+\mathfrak b r+i)}{-i \mathfrak b r+s-\frac{3}{2}}, \nonumber \\
% ~~~~~~~~~~~~~
h(s) 
&= \dfrac{2 \left(\mathfrak{a}^2-3 \mathfrak{b}^2 r^2+3 i \mathfrak{b} r+1\right)}{- i \mathfrak{b} r + s - 5 \slash 2}.
\end{align}

\begin{figure}
\begin{center}
\includegraphics[scale=.6]{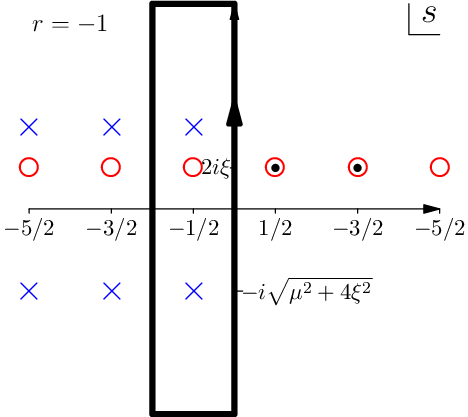}
\end{center}
\caption{The contour in \eqref{eq:fg_contour}, along with the poles in the integrand.  The two small black dots are the two poles of $g(s)$.  The poles of $f_1(s)$ result from the sine functions in the denominator and are represented with red circles ($s = n - 1 \slash 2 + i \mathfrak{b} r$) and blue crosses ($s = n - 1 \slash 2 \pm i \mathfrak{a}$).  This is for helicity $r = -1$; for $r = +1$, the black dots and red circles reflect over the real axis.}
\label{fig:weird_contour}
\end{figure}

Therefore, using $f_1(s) = f_1(s-1)$, the integrand can be written as $ f_1(s) f_2(s) = f_1(s) g(s) - f_1(s-1) g(s-1) + f_1(s) h(s)$ and the integral is
\begin{align}
\int_{-i \infty}^{i \infty} ds \, f_1(s) f_2(s)
&= \int_{-i \infty}^{i \infty} ds \, f_1(s) g(s) - \int_{-i \infty-1}^{i \infty-1} ds \, f_1(s) g(s) + \int_{-i \infty}^{i \infty} ds \, f_1(s) h(s) \,, 
\end{align}
where in the second line we have changed variables to $s^\prime = s-1$.  We further interchange the limits of integration in this term; then the first two terms can then be written as a single contour integral if one also includes integrations along a line from $i \infty$ to $i \infty - 1$ and along a line from $- i \infty -1$ to $- i \infty$.  This can be done because the integrand does indeed approach zero sufficiently fast as $s \rightarrow s_0 \pm i \infty$.  Therefore, the integral is equal to
\begin{align}
\int_{-i \infty}^{i \infty} ds \, f_1(s) f_2(s)
&= \int_{C} ds \, f_1(s) g(s) + \int_{-i \infty}^{i \infty} ds \, f_1(s) h(s),
\label{eq:fg_contour}
\end{align}
where the contour $C$ is shown in figure~\ref{fig:weird_contour}, along with the poles of the integrand. The poles of $g(s)$ are at $s = 1/2 + i \mathfrak{b} r$ and $3/2 + i \mathfrak{b} r$, which are not inside the loop.  $f_1$ has poles where the sine functions in its denominator are zero, which occurs at $s = n - 1/2 \pm i \mathfrak{a}$ and $s = n -1/2 + i \mathfrak{b} r$ where $n$ is any integer.  Therefore, three poles are enclosed in our contour, at $- 1/2\pm i \mathfrak{a}$ and $- 1/2 + i \mathfrak{b} r$.  The residue theorem gives an exact result which, when expressed in terms of $\mu$ and $\xi$, is

\begin{align}
&\sum_r \int_{C} ds \, r \Im(f_1(s) g(s)) =
\frac{1}{8}  
 \left[
2 \xi  \left(\mu ^2 \left(\frac{4 \left(\mu ^4-1\right)}{\left(\mu ^2+1\right)^2+16 \xi ^2}+\frac{\mu ^4+6 \mu ^2-40}{\left(\mu ^2+4\right)^2+64 \xi ^2}+5\right)+2\right)\right.
\nonumber \\
& \quad
\left.
+\frac{\left(-256 \left(5 \mu ^2+2\right) \xi ^4-16 \left(\mu ^2+1\right) \left(\mu ^2+2\right) \left(2 \mu ^2+5\right) \xi ^2+\left(\mu ^2-2\right) \left(\mu ^2+1\right)^2 \left(\mu ^2+4\right)\right) }{\left(\left(\mu ^2+1\right)^2+16 \xi ^2\right) \left(\left(\mu ^2+4\right)^2+64 \xi ^2\right)
\text{sinh}\left(2 \pi  \sqrt{\mu ^2+4 \xi ^2}\right)
} \right. \nonumber \\
& \qquad \left. \cdot 4 \sqrt{\mu ^2+4 \xi ^2} \sinh (4 \pi  \xi )
 \right] \,, 
\label{eq:quintic_last_integral_part1}
\end{align}
and we have also summed the imaginary part over $r$.

It remains to evaluate the last piece, $\int_{-i \infty}^{i \infty} ds \, f_1(s) h(s)$, with $f_1(s)$ and $h(s)$ given above.  We close this contour at $\Re(s) = - \infty$; note that the integrand goes to zero faster than $1 \slash R$ as $\Re(s) \rightarrow - \infty$.  In figure~\ref{fig:weird_contour} the red dots and blue crosses are the poles of $f_1$, and therefore are poles of this integrand as well.  This contour encloses all poles to the left of the imaginary axis.  ($h(s)$ has a pole at $s = 5 \slash 2 + i \mathfrak{b}$, which is not enclosed by our contour.)  Fortunately, although there are infinitely many poles that contribute to the integral, their contributions can be easily summed by separating the poles into the three groups, as described below.

\paragraph*{$n - \dfrac{1}{2} + i \mathfrak{b} r$ Poles:} First consider the poles $n -1/2 + i \mathfrak{b} r$ with $n \geq 0$.  The contribution of these poles is
\begin{align}
\sum_n 2 \pi i \mathrm{Res}_1(f_1(s) h(s))
&= \sum_{n\leq 0} \frac{\mathfrak{a}^2-3 \mathfrak{b}^2 +3 i \mathfrak{b} r+1}{2 (n-3)}.
\end{align}

\paragraph*{$n - \dfrac{1}{2} + i \mathfrak{a}$ Poles:} The contribution of these poles is
\begin{align}
\sum_n 2 \pi i \mathrm{Res}_2(f_1(s) h(s))
&= \sum_{n \leq 0} \frac{\text{csch}(2 \pi  \mathfrak{a}) e^{\pi  (\mathfrak{a}-\mathfrak{b} r)} \left(\mathfrak{a}^2-3 \mathfrak{b}^2 +3 i \mathfrak{b} r+1\right) \sinh (\pi  (\mathfrak{a}+\mathfrak{b} r))}{-2 i \mathfrak{a}+2 i \mathfrak{b} r-2 n+6} \,, 
\end{align}
where again the sum is over $n\leq 0$ because the contour is closed with $\mathrm{Re}(s) \rightarrow - \infty$.

\paragraph*{$n - \dfrac{1}{2} - i \mathfrak{a}$ Poles:} The contribution of these poles is
\begin{align}
\sum_n 2 \pi i \mathrm{Res}_3(f_1(s) h(s))
&= \sum_{n \leq 0} \frac{\text{csch}(2 \pi  \mathfrak{a}) e^{-\pi  (\mathfrak{a}+\mathfrak{b} r)} \left(\mathfrak{a}^2-3 \mathfrak{b}^2+3 i \mathfrak{b} r+1\right) \sinh (\pi  (\mathfrak{a}-\mathfrak{b} r))}{2 (i \mathfrak{a}+i \mathfrak{b} r-n+3)}.
\end{align}

Therefore the remaining integal reduces to the sum
\begin{align}
\int_{-i \infty}^{i \infty} ds \, f_1(s) h(s)
&= \sum_{n \leq 0} 
\frac{1}{2} 
\left(\mathfrak{a}^2-3 \mathfrak{b}^2+3 i \mathfrak{b} r+1\right) 
\left(\frac{1}{n-3}+\text{csch}(2 \pi  \mathfrak{a}) \right. \nonumber \\
& \times \left. \left(\frac{e^{\pi  (\mathfrak{a}-\mathfrak{b} r)} \sinh (\pi  (\mathfrak{a}+\mathfrak{b} r))}{-i \mathfrak{a}+i \mathfrak{b} r-n+3} 
 +\frac{e^{-\pi  (\mathfrak{a}+\mathfrak{b} r)} \sinh (\pi  (\mathfrak{a}-\mathfrak{b} r))}{i \mathfrak{a}+i \mathfrak{b} r-n+3}\right)\right).
\end{align}
Although the sum may naively appear divergent, it in fact converges.  This can be seen using the identity 
\begin{align}
1 &= \text{csch}(2 \pi  \mathfrak{a}) \left(e^{\pi  (\mathfrak{a}-\mathfrak{b} r)} \sinh (\pi  (\mathfrak{a}+\mathfrak{b} r))+e^{-\pi  (\mathfrak{a}+\mathfrak{b} r)} \sinh (\pi  (\mathfrak{a}-\mathfrak{b} r))\right) \,, 
\end{align}
to write
\begin{align}
\int_{-i \infty}^{i \infty} ds \, f_1(s) h(s)
&= 
\frac{1}{2} 
\left(\mathfrak{a}^2-3 \mathfrak{b}^2+3 i \mathfrak{b} r+1\right) \text{csch}(2 \pi  \mathfrak{a}) \nonumber \\
& \sum_{n \leq 0} 
\left( e^{\pi  (\mathfrak{a}-\mathfrak{b} r)} \sinh (\pi  (\mathfrak{a}+\mathfrak{b} r)) 
\left(\frac{1}{-i \mathfrak{a}+i \mathfrak{b} r-n+3} 
+ \dfrac{1}{n -3} \right) \right. \nonumber \\
& \qquad \left.   +e^{-\pi  (\mathfrak{a}+\mathfrak{b} r)} \sinh (\pi  (\mathfrak{a}-\mathfrak{b} r))\left( \frac{1}{i \mathfrak{a}+i \mathfrak{b} r-n+3} + \dfrac{1}{n - 3} \right) \right).
\end{align}
After taking the imaginary part and summing over $r$ one finds that the two sums can be easily done, giving
\begin{align}
&\sum_r \int_{-i \infty}^{i \infty} ds \, r \Im(f_1(s) h(s))
=
\frac{1}{4} \text{csch}(\pi  \mathfrak a) \text{sech}(\pi  \mathfrak a) 
\Big[\sinh (\pi  (\mathfrak a-\mathfrak b)) \cosh (\pi  (\mathfrak a+\mathfrak b))  \nonumber \\
& \qquad \left.\cdot \left(i  \left(\mathfrak a^2-3 \mathfrak b (\mathfrak b-i)+1\right) H_{i (\mathfrak a+\mathfrak b-2 i)} -i \left(\mathfrak a^2-3 \mathfrak b (\mathfrak b+i)+1\right) H_{-i (\mathfrak a+\mathfrak b+2 i)}+9 \mathfrak b\right) \right. \nonumber \\
& \left. 
+\sinh (\pi  (\mathfrak a+\mathfrak b)) \cosh (\pi  (\mathfrak a-\mathfrak b)) \right. \nonumber \\
& \qquad \cdot\left(i \left(\mathfrak a^2-3 \mathfrak b (\mathfrak b-i)+1\right) H_{2-i \mathfrak a+i \mathfrak b}-i \left(\mathfrak a^2-3 \mathfrak b (\mathfrak b+i)+1\right) H_{2+i \mathfrak a-i \mathfrak b}+9 \mathfrak b\right) \Big].
\end{align}

Because this integral will appear again in the calculation of the bispectrum, we present the final result for the integral,
\begin{small}
\begin{align}
&\sum_r \int y r \Im \left( s_r(y) d_r^*(y)\right) \, \exd y
=  \mu
\left[ -6 \xi  (\log (2 \Lambda )+\gamma_E)-\frac{3}{2} \sqrt{\mu ^2+4 \xi ^2} \sinh (4 \pi  \xi ) \text{csch}\left(2 \pi  \sqrt{\mu ^2+4 \xi ^2}\right)+7 \xi \right. \nonumber \\
&  \qquad \left. + \frac{1}{4} i \left(\mu ^2- 8 \xi^2 - 6 i \xi +1\right) 
\left[H_{-i \left(2 \xi +\sqrt{\mu ^2+4 \xi ^2}\right)} \left(\sinh (4 \pi  \xi ) \text{csch}\left(2 \pi  \sqrt{\mu ^2+4 \xi ^2}\right)+1\right)
\right. \right. \nonumber \\
& \qquad \qquad \left. \left.
+H_{i \left(\sqrt{\mu ^2+4 \xi ^2}-2 \xi \right)} 
\left(1-\sinh (4 \pi  \xi ) \text{csch}\left(2 \pi  \sqrt{\mu ^2+4 \xi ^2}\right)\right)\right] \right. \nonumber \\
& \qquad
\left. -\frac{1}{4} i \left(\mu ^2-8 \xi ^2+6 i \xi +1\right) 
\left[H_{i \left(2 \xi +\sqrt{\mu ^2+4 \xi ^2}\right)} \left(\sinh (4 \pi  \xi ) \text{csch}\left(2 \pi  \sqrt{\mu ^2+4 \xi ^2}\right)+1\right)
\right. \right. \nonumber \\
& \qquad \qquad \left. \left.
+H_{-i \left(\sqrt{\mu ^2+4 \xi ^2}-2 \xi \right)} \left(1-\sinh (4 \pi  \xi ) \text{csch}\left(2 \pi  \sqrt{\mu ^2+4 \xi ^2}\right)\right)\right]  
\right].
\label{eq:quintic_exact}
\end{align}
\end{small}
In the $\mu \ll 1 \ll \xi$ limit, this approaches 
\begin{align}
&\sum_r \int \exd y \,r \,\Im \left( s_r(y) d_r^*(y)\right) \, \exd y
\rightarrow - 4 \pi\mu\, \xi^2\,
\label{eq:quintic_limit}
\end{align}
after removing the $\log(\Lambda)$ contribution.  Substituting this into eq.~\eqref{eq:B_intermediate} gives eq.~\eqref{eq:calb_final}.

\section{Quartic loop integral}
\label{ap:quartic}

This appendix describes the calculation of the contribution to the power spectrum from the quartic loop, on the left of figure~\ref{fig:diagrams}.  The derivation of \eqref{eq:quartic2} from \eqref{eq:quartic1} is straightforward, using 
\begin{align}
\left< \delta \phi(\tau, \bk) \, \delta \phi(\tau^\prime, \bk^\prime) \right>
&= \dfrac{H^2}{2k^3} \left( 
k^2 \tau \tau^\prime +i k(\tau - \tau^\prime) + 1
 \right) 
\left( \cos[k(\tau^\prime-\tau)] + i \sin[k(\tau^\prime-\tau)] \right) \nonumber \\
& \qquad 
\cdot \delta^{(3)}(\bk + \bk^\prime) \,, 
\label{eq:phi_commutator}
\end{align}
and
\begin{align}
[ \delta \phi(\tau,\bk), \delta \phi(\tau^\prime, \bk^\prime)]
&=i \dfrac{H^2 }{k^3} \delta^{(3)}(\bk + \bk^\prime) 
\left[
\left( 1+ \tau \tau^\prime \right) \sin[k(\tau^\prime - \tau)]
+  k (\tau - \tau^\prime) \cos[k(\tau^\prime - \tau)]
\right] \,,
\label{eq:phiphivev}
\end{align}
which are used in appendices \ref{ap:cubic} and \ref{ap:nongaussianities} as well.  From \eqref{eq:quartic2} we evaluate the fermionic vacuum expectation value, either by substuting the mode expansion eq.~\eqref{eq:deco_psi} and using eq.~\eqref{eq:def_uv}, or by rotating $\psi$ to $Y$ before evaluting the expectation value.  The result in either case is eq.~\eqref{amp1}.  As noted in the text, the $\exd x_1$ integral can be performed but has an infrared divergence.  After this step, the power spectrum is given by
\begin{align}
\dfrac{\delta P_{\zeta}}{P^0}
&\approx   \dfrac{4 m H}{3 \pi^2 f^2} \ln \left( x \right)    \int  \exd y \, y \,  
 \sum_r \mathrm{Re}[s_r(y) d_r^*(y)].
 \label{eq:P_zeta1v1}
\end{align}
The remaining integral can be performed analytically following identical steps as those in appendix \ref{ap:backreaction} for the backreaction, in which we evaluated $\sum_r r \int \exd y \, y \Im[d_r^*(y) s_r(y)]$.

As before, we write the Whittaker functions in the integrand using the Mellin-Barnes representation (see eq.~\eqref{eq:MellinBarnes}) and identify the poles.  Everything proceeds as above until the sum over $r$; in place of \eqref{eq:intermediate1} one finds instead
\begin{align}
& \sum_r \int y s_r(y) d_r^*(y) \, \exd y
= \mu \left[ 
\frac{1}{2} \left(
\left(\mu ^2- 2 \xi (4 \xi - 3 i) +1\right) \left( H_{-i \left(\sqrt{\mu ^2+4 \xi ^2}-2 \xi \right)} + H_{i \left(2 \xi +\sqrt{\mu ^2+4 \xi ^2}\right)} \right)  \right. \right. \nonumber \\
& \qquad \left.  \left.
+\left(\mu ^2-2 \xi  (4 \xi +3 i)+1\right) \left(H_{i \left(\sqrt{\mu ^2+4 \xi ^2}-2 \xi \right)}+H_{-i \left(2 \xi +\sqrt{\mu ^2+4 \xi ^2}\right)}\right) \right) \right. \nonumber \\
& \qquad \left. 
+\Lambda ^2- (\ln (2 \Lambda )+ \gamma_E) \left(\mu ^2-8 \xi ^2+1\right)+i \Lambda +\frac{1}{8} \left(\mu ^4-4 i \pi  \left(\mu ^2-8 \xi ^2+1\right)-7 \mu ^2+12\right)
\right] \nonumber \\
& \qquad + \sum_r \mathcal{A}_r \,, 
\end{align}
with $\mathcal{A}_r$ given as in eq.~\eqref{eq:Ar}.  The remaining integral, which does not depend on the cutoff $\Lambda$, proceeds as in appendix \ref{ap:backreaction} with the same definitions for $f_1$, $g$, and $h$.  In place of eq.~\eqref{eq:quintic_last_integral_part1} we have
\begin{align}
&\sum_r \int_{C} ds \, \Re(f_1(s) g(s)) 
= \left[ 
\frac{1}{8} \left(\mathfrak{a}^2 \left(2 \mathfrak{b}^2-1\right)-\mathfrak{a}^4-\mathfrak{b}^4+\mathfrak{b}^2-4\right)+\frac{1}{4} e^{-\pi  (\mathfrak{a}+\mathfrak{b})} \text{csch}(\pi  \mathfrak{a}) \text{sech}(\pi  \mathfrak{a}) \right. \nonumber \\
& \left.
\left(\frac{\left((\mathfrak{a}-\mathfrak{b}) \left(3 \mathfrak{a}^4 \mathfrak{b}+\mathfrak{a}^3 \left(2-9 \mathfrak{b}^2\right)+9 \mathfrak{a}^2 \left(\mathfrak{b}^3+\mathfrak{b}\right)+\mathfrak{a} \left(4-3 \mathfrak{b}^2 \left(\mathfrak{b}^2+5\right)\right)+4 \mathfrak{b}^3+5 \mathfrak{b}\right)+2\right) }{(\mathfrak{a}-\mathfrak{b}-i) (\mathfrak{a}-\mathfrak{b}+i) (\mathfrak{a}-\mathfrak{b}-2 i) (\mathfrak{a}-\mathfrak{b}+2 i)} \right. \right. \nonumber \\
& \qquad \left. \left. \cdot \left(e^{2 \pi  (\mathfrak{a}+\mathfrak{b})}-1\right) \cosh (\pi  (\mathfrak{a}-\mathfrak{b})) \right. \right. \nonumber \\
&  \left. \left. -\frac{\left((\mathfrak{a}+\mathfrak{b}) \left(3 \mathfrak{a}^4 \mathfrak{b}+\mathfrak{a}^3 \left(9 \mathfrak{b}^2-2\right)+9 \mathfrak{a}^2 \left(\mathfrak{b}^3+\mathfrak{b}\right)+\mathfrak{a} \left(3 \mathfrak{b}^2 \left(\mathfrak{b}^2+5\right)-4\right)+4 \mathfrak{b}^3+5 \mathfrak{b}\right)-2\right) }{(\mathfrak{a}+\mathfrak{b}-i) (\mathfrak{a}+\mathfrak{b}+i) (\mathfrak{a}+\mathfrak{b}-2 i) (\mathfrak{a}+\mathfrak{b}+2 i)} \right. \right. \nonumber \\
& \qquad \left. \left. \cdot \left(e^{2 \pi  (\mathfrak{a}+\mathfrak{b})}+1\right) \sinh (\pi  (\mathfrak{a}-\mathfrak{b})) \right)
\right].
\end{align}
The $\int ds \, f_1(s) h(s)$ integral similarly proceeds along the lines of appendix \ref{ap:backreaction}; again, it is convenient to take the real part and sum over $r$ before performing the sum that results from the infinite number of poles.  The final integral contributes 
\begin{align}
&\sum_r \int_{-i \infty}^{i \infty} ds \, \Re(f_1(s) h(s))
=\frac{1}{4} \text{csch}(\pi  a) \text{sech}(\pi  a) \nonumber \\
& \qquad
\left[\left(3 \left(\mathfrak{a}^2-3 \mathfrak{b}^2+1\right)-\left(\mathfrak{a}^2-3 \mathfrak{b} (\mathfrak{b}-i)+1\right) H_{i (\mathfrak{a}+\mathfrak{b}-2 i)}  -\left(\mathfrak{a}^2-3 \mathfrak{b} (\mathfrak{b}+i)+1\right) H_{-i (\mathfrak{a}+\mathfrak{b}+2 i)}\right) \right. \nonumber \\
& \qquad  \qquad \left. \cdot  \sinh (\pi  (\mathfrak{a}-\mathfrak{b})) \cosh (\pi  (\mathfrak{a}+\mathfrak{b})) \right. \nonumber \\
& \qquad  
\left. +\left(3 \left(\mathfrak{a}^2-3 \mathfrak{b}^2+1\right)-\left(\mathfrak{a}^2-3 \mathfrak{b} (\mathfrak{b}-i)+1\right) H_{2-i \mathfrak{a}+i \mathfrak{b}}+\left(-\mathfrak{a}^2+3 \mathfrak{b} (\mathfrak{b}+i)-1\right) H_{2+i \mathfrak{a}-i \mathfrak{b}}\right) \right. \nonumber \\
& \qquad \qquad \left. \cdot  \sinh (\pi  (\mathfrak{a}+\mathfrak{b})) \cosh (\pi  (\mathfrak{a}-\mathfrak{b}))\right].
\end{align}
Combining all the pieces gives the full analytic result 
\begin{small}
\begin{align}\label{eq:full_quartic}
& \sum_r \int y \Re \left( s_r(y) d_r^*(y)\right) \, \exd y
= \mu \left[ \frac{1}{2} \left(2 \Lambda ^2+\frac{1}{4} \left(-8 (\log (2 \Lambda )+\gamma_E ) \left(\mu ^2-8 \xi ^2+1\right)+\mu ^4-7 \mu ^2+12\right)\right) \right. \nonumber \\
& \qquad
\left.
+ \frac{1}{4} \left(\mu ^2- 8 \xi^2 - 6 i \xi +1\right) 
\left[H_{-i \left(2 \xi +\sqrt{\mu ^2+4 \xi ^2}\right)} \left(\sinh (4 \pi  \xi ) \text{csch}\left(2 \pi  \sqrt{\mu ^2+4 \xi ^2}\right)+1\right)
\right. \right. \nonumber \\
& \qquad \qquad \left. \left.
+H_{i \left(\sqrt{\mu ^2+4 \xi ^2}-2 \xi \right)} \left(1-\sinh (4 \pi  \xi ) \text{csch}\left(2 \pi  \sqrt{\mu ^2+4 \xi ^2}\right)\right)\right] \right. \nonumber \\
& \qquad
\left.
+\frac{1}{4} \left(\mu ^2-8 \xi ^2+6 i \xi +1\right) 
\left[H_{i \left(2 \xi +\sqrt{\mu ^2+4 \xi ^2}\right)} \left(\sinh (4 \pi  \xi ) \text{csch}\left(2 \pi  \sqrt{\mu ^2+4 \xi ^2}\right)+1\right)
\right. \right. \nonumber \\
& \qquad \qquad \left. \left.
+H_{-i \left(\sqrt{\mu ^2+4 \xi ^2}-2 \xi \right)} \left(1-\sinh (4 \pi  \xi ) \text{csch}\left(2 \pi  \sqrt{\mu ^2+4 \xi ^2}\right)\right)\right] \right. \nonumber \\
& \qquad \left. +6 \xi  \sqrt{\mu ^2+4 \xi ^2} \sinh (4 \pi  \xi ) \text{csch}\left(2 \pi  \sqrt{\mu ^2+4 \xi ^2}\right)
-\frac{\mu ^4}{8} +\frac{11 \mu ^2}{8}-12 \xi ^2
\right],
\end{align}
\end{small}
which is quadratically divergent.  If $\mu \ll 1$, this becomes
\begin{align}\label{eq:integral_mull1}
\sum_r \int y \Re \left( s_r(y) d_r^*(y)\right) \, \exd y 
& \approx \mu \left[ 
\Lambda ^2+(8 \xi ^2-1) \log (2 \Lambda )+\left(-4 \xi ^2+3 i \xi +\frac{1}{2}\right) \psi ^{(0)}(4 i \xi +1) \right. \nonumber \\
& \qquad \left. +\left(-4 \xi ^2-3 i \xi +\frac{1}{2}\right) \psi ^{(0)}(1-4 i \xi )+\frac{3}{2}
\right].
\end{align}
In the $\xi \gg 1$ limit, the finite piece simplifies further to $\sum_r \int y \Re \left( s_r(y) d_r^*(y) \right) \approx -8 \mu \ln(\xi) \xi^2$.  Substituting this into eq.~\eqref{eq:P_zeta1v1} gives

\begin{align}
\dfrac{\delta P_{\zeta}}{P^0}
&\approx   \dfrac{32 m^2 \ln(\xi) \xi^2}{3 \pi^2 f^2} |\ln \left( x \right)|   \,, 
\end{align}
where we have noted that for $x \ll 1$, $\ln(x) < 0$.  This is in agreement with eq.~\eqref{eq:p_quartic}.

\section{Cubic loop integral}
\label{ap:cubic}

This appendix presents the calculations relevant to the evaluation of the right diagram in figure~\ref{fig:diagrams}, leading to eq.~\eqref{eq:p_cubic}.  In the in-in formalism, this diagram corresponds to 
\begin{align}
& \!\!\!\! \!\!\!\! 
\left< \delta \phi_{\bk_1}(\tau) \, \delta \phi_{\bk_2}(\tau) \right> 
% ~~~~~~~~~~~~~~~~~~~~~~~~~~~~~
= - \int^\tau \exd \tau_1 \left( - \dfrac{2 a(\tau_1)}{f} \right)
\int^{\tau_1} \exd \tau_2 \left( - \dfrac{2 a(\tau_2)}{f} \right)
\int \dfrac{\exd^3 p_1 \, \exd^3 q_1}{(2\pi)^{3 \slash 2}}
\int \dfrac{\exd^3 p_2 \, \exd^3 q_2}{(2\pi)^{3 \slash 2}} \nonumber \\
& \left< \left[ \left[ 
\delta \phi_{\bk_1}(\tau) \, \delta \phi_{\bk_2}(\tau),
\bar{\psi}_{\bp_1}(\tau_1) \left[ m_s + i \gamma^5 m_c \right] \psi_{\bq_1}(\tau_1)\, \delta \phi_{\bp_1 - \bq_1}(\tau_1) \right],  \right. \right. \nonumber \\
& \qquad \left. \left.
\bar{\psi}_{\bp_2}(\tau_2) \left[ m_s + i \gamma^5 m_c \right] \psi_{\bq_2}(\tau_2)\, \delta \phi_{\bp_2 - \bq_2}(\tau_2) \right]
\right> \,, 
\end{align}
where $m_s = m_I \slash a$ and $m_c = m_R \slash a$.  Because the $\delta \phi$ creation and annihilation operators commute with the fermionic creation and annihilation operators, we can use \eqref{phi-deco} to find

\begin{align}
&
\left< \delta \phi_{\bk_1}(\tau) \, \delta \phi_{\bk_2}(\tau) \right>
= - i \dfrac{H^4}{2 k_1^3 k_2^3} \int^\tau \exd \tau_1 \left( - \dfrac{2 a(\tau_1)}{f} \right)
\int^{\tau_1} \exd \tau_2 \left( - \dfrac{2 a(\tau_2)}{f} \right)
\int \dfrac{\exd^3 p_1 }{(2\pi)^{3 \slash 2}}
\int \dfrac{\exd^3 p_2 }{(2\pi)^{3 \slash 2}} \nonumber \\
& \qquad 
\left\lbrace
\left[  \sin[k_2 \tau_1] - k_2\tau_1 \cos[k_2\tau_1] \right]
(1 - i k_1\tau_2) \left( \cos[k_1 \tau_2] + i \sin[k_1 \tau_2 ] \right)
\right. \nonumber \\
& \qquad \qquad \cdot \left. 
\left< 
\bar{\psi}_{\bp_1}(\tau_1) \left[ m_s + i \gamma^5 m_c \right] \psi_{\bk_2 + \bp_1}(\tau_1)
\bar{\psi}_{\bp_2}(\tau_2) \left[ m_s + i \gamma^5 m_c \right] \psi_{\bk_1 + \bp_2}(\tau_2)
\right> 
\right.
\nonumber \\
% ~~~~~~~~~~~~~~~~~~~~~~~~~~~~~~~~~~~~~~~~~~~~~~~~~~~~~~~~~~~~~
& \qquad \left. - 
\left[ \sin[k_2\tau_1] - k_2 \tau_1 \cos[k_2\tau_1] \right] 
\left( 1 + i k_1\tau_2 \right)
\left( \cos[k_1\tau_2] - i \sin[k_1\tau_2] \right) \right. \nonumber \\
& \qquad \qquad \left. \cdot 
\left< 
\bar{\psi}_{\bp_2}(\tau_2) \left[ m_s + i \gamma^5 m_c \right] \psi_{\bk_1 + \bp_2}(\tau_2)
\bar{\psi}_{\bp_1}(\tau_1) \left[ m_s + i \gamma^5 m_c \right] \psi_{\bk_2 + \bp_1}(\tau_1)
\right> \right.
\nonumber \\
& \qquad  \left. + (k_1 \leftrightarrow k_2, p_1 \leftrightarrow p_2)
\right\rbrace.
\end{align}

Using the decomposition of eq.~\eqref{eq:deco_psi}, this expression simplifies  to 
\begin{align}
&\!\!\!\! \left< \delta \phi_{\bk_1}(\tau) \, \delta \phi_{\bk_2}(\tau) \right>^\prime 
= - i \dfrac{H^4}{2 k_1^6} \int^\tau \exd \tau_1 \left( - \dfrac{2 a(\tau_1)}{f} \right)
\left[  \sin[k_1 \tau_1] - k_1\tau_1 \cos[k_1\tau_1] \right]
\int^{\tau_1} \exd \tau_2 \left( - \dfrac{2 a(\tau_2)}{f} \right)
 \nonumber \\
& \int \dfrac{\exd^3 p_1 }{(2\pi)^{3}}
\left\lbrace
(1 - i k_1\tau_2) \left( \cos[k_1 \tau_2] + i \sin[k_1 \tau_2 ] \right) \right. \nonumber \\
& 
\left.
\cdot \left( 
[ V^\dagger_{r,-\bp_1} \mathbb{O}  U_{s,\bp_1-\bk_1}]_{\tau_1} \, 
[ U_{s,\bp_1-\bk_1}^\dagger \mathbb{O}   V_{r,-\bp_1}]_{\tau_2}
+ 
[ V^\dagger_{r,-\bp_1} \mathbb{O}  U_{s,\bp_1+\bk_1}]_{\tau_1} \, [ U_{s,\bp_1+\bk_1}^\dagger \mathbb{O}   V_{r,-\bp_1}]_{\tau_2} 
\right)  -  h.c. \right\rbrace \,, 
\end{align}
where
\begin{equation}
\mathbb{O} = \gamma^0 (m_s + i \gamma^5 m_c).
\end{equation}

Evaluating the remaining spinor product gives
\begin{align}
&
\left< \delta \phi_{\bk_1}(\tau) \, \delta \phi_{\bk_2}(\tau) \right>^\prime 
= - i \dfrac{H^2}{2 k_1^6 f^2} \int^\tau \dfrac{\exd \tau_1}{\tau_1} 
\left[  \sin[k_1 \tau_1] - k_1\tau_1 \cos[k_1\tau_1] \right]
\int^{\tau_1} \dfrac{\exd \tau_2}{\tau_2} 
\int \dfrac{\exd^3 p_1 }{(2\pi)^{3}} \sum_{rs}  \nonumber \\
&  \left\lbrace
\dfrac{1}{2}\left[  1 +rs \dfrac{\bp_1 \cdot (\bp_1 - \bk_1)}{p_1 |\bp_1 - \bk_1|}  \right] 
(1 - i k_1\tau_2) \left( \cos[k_1 \tau_2] + i \sin[k_1 \tau_2 ] \right)
\right. \nonumber \\
&\qquad \left. 
\cdot 
\left[
m_s (r  v_r(p_1) u_s(p_2) + s u_r(p_1) v_s(p_2))
+ i m_c ( u_r(p_1)  u_s(p_2) + rs  v_r(p_1)  v_s(p_2)
\right]_{\tau_1,p_2 = |\bp_1 - \bk_1|} \right.  \nonumber \\
& \qquad \left. 
\cdot \left[
m_s (r  v_r^*(p_1)  u_s^*(p_2) + s  u_r^*(p_1)  v_s^*(p_2))
- i m_c ( u_r^*(p_1)  u_s^*(p_2) + rs  v_r^*(p_1)  v_s^*(p_2)
\right]_{\tau_2,p_2 = |\bp_1 - \bk_1|} \right. \nonumber \\ 
& \qquad \qquad  \left.  + (\bk_1 \rightarrow - \bk_1)   \right\rbrace  + h.c. .
\end{align}

\begin{figure}
\centering
\includegraphics[scale=1]{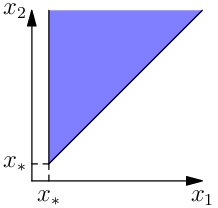}
\hspace*{1 cm}
\includegraphics[scale=1]{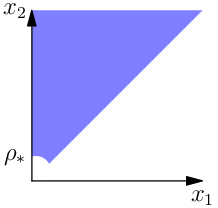}
\caption{
The true region of integration at finite $\tau$ (left) and the region we use in our approximation (right); as discussed, the integrand is exponentially suppressed as $x_1 \rightarrow 0$, justifying the approximation.}
\label{fig:integration_regions_not_same}
\end{figure}

We expect the remaining momentum integral to be dominated by values of $p_1$ signficantly larger than $k_1$; this approximation gives
\begin{align}
&\!\!\!\! \!\!\!\!
\left< \delta \phi_{\bk_1}(\tau) \, \delta \phi_{\bk_2}(\tau) \right>^\prime 
= - i \dfrac{H^2 m^2}{k_1^6 f^2} \int^\tau \dfrac{\exd \tau_1}{\tau_1} 
\left[  \sin[k_1 \tau_1] - k_1\tau_1 \cos[k_1\tau_1] \right]
\int^{\tau_1} \dfrac{\exd \tau_2}{\tau_2} 
\int \dfrac{\exd^3 p_1 }{(2\pi)^{3}} \sum_{r}  \nonumber \\
&  \left\lbrace (1 - i k_1\tau_2) \left( \cos[k_1 \tau_2] + i \sin[k_1 \tau_2 ] \right) 
[\left( u^2(p_1) + v^2(p_1) \right)]_{\tau_1}
[\left( u^{*2}(p_1) + v^{*2}(p_1) \right)]_{\tau_2}
\right\rbrace  + h.c. . 
\end{align}
The $u$ and $v$ are known in terms of the $s$ and $d$ functions via eq.~\eqref{eq:def_uv}, where $s$ and $d$ are given by eq.~\eqref{eq:def_sd}.  Due to the existence of two time variables ($\tau_1$ and $\tau_2$), the techniques used in appendix \ref{ap:backreaction} are not sufficient to allow us to evaluate this integral analytically.  As expected, we have a divergence as $p_1 \rightarrow \infty$; however, the integrand is also oscillating as $e^{-2 i p_1 (\tau_1 - \tau_2)}$ as $p_1 \rightarrow \infty$.  To make this a well-defined integral, we Wick rotate the time variables.  Because the upper limit of the $\tau_2$ integral is $\tau_1$, these must be rotated in the same direction in the complex plane, determined by $\tau_2$, which has the largest magnitude.  (Note that $-\infty < \tau_2 < \tau_1 < \tau$.).  Therefore, we take $\tau_i \rightarrow \tau_i (1 - i \epsilon)$ in the first term, while performing the opposite rotation in the hermitian conjugate term.  This gives
\begin{align}
& \dfrac{\delta P_{\zeta,2}}{P^0}
=  \dfrac{ m^2}{\pi^2 f^2} \int_x \dfrac{\exd x_1}{ x_1^4} 
\left[  -\sinh[ x_1] +  x_1 \cosh[ x_1] \right]
\int_{x_1} \dfrac{ \exd x_2}{ x_2} (1 +  x_2) \left( -\sinh[ x_2] + \cosh[ x_2] \right) 
  \nonumber \\
& \int  \exd y_1 \, y_1^2 \sum_{r}  
(u_r(-i y_1) u_r(-i y_1) +  v_r(-i y_1) v_r(-i y_1)) 
(u_r^*(-i y_2) u_r^*(-i y_2) + v_r^*(-i y_2) v_r^*(-i y_2)) \nonumber \\
&\qquad   +  h.c. ,
\end{align}
where we have introduced the variables $x_i = - k_1  \tau_i$ and $y_i = - p_1  \tau_i$ and normalized to the vacuum power spectrum.  Note that $u^*(- i y_i)$ and $v^*(-i y_i)$ are defined by first conjugating $u(y_i)$ and $v(y_i)$, and subsequently substituting $y_i \rightarrow -i y_i$.  In terms of the $s$ and $d$ functions, this is
\begin{align}
\dfrac{\delta P_{\zeta,2}}{P^0}
& =  \dfrac{m^2}{ \pi^2 f^2} \int_x \dfrac{ \exd x_1}{ x_1^3} 
\left[ \sinh[ x_1] - x_1 \cosh[x_1] \right]
\int_{x_1} \dfrac{\exd x_2}{x_2^2}  (1 +  x_2) \left( -\sinh[ x_2] + \cosh[ x_2] \right) 
\int  \exd y_1    
\nonumber \\
& \qquad 
\sum_{r}   (s_r^2(-i y_1) + d_r^2(-i y_1))  \cdot 
\left(s_r^{*2}\left(-i \dfrac{x_2}{x_1} y_1 \right)  + d_r^{*2}\left( -i \dfrac{x_2}{x_1} y_1 \right) \right) 
 + h.c.,
\end{align}
where $s^*$ and $d^*$ are defined in the same manner as $u^*$ and $d^*$ above.  Next we introduce in place of $x_1$ and $x_2$ the polar coordinates $\rho$ and $\alpha$, where $ x_1 = \rho \cos(\alpha)$ and  $x_2 = \rho \sin(\alpha)$.  In the $\tau \rightarrow 0$ limit, the integration region is $0 < \rho < \infty$, $\pi \slash 4 < \alpha < \pi \slash 2$.  If $\tau$ is finite, then the region of integration is as shown on the left of figure~\ref{fig:integration_regions_not_same}.  To achieve a partial decoupling of the integrals, we instead integrate over the region shown on the right of figure~\ref{fig:integration_regions_not_same}; below we show that the integrand is exponentially suppressed as $x_1 \rightarrow 0$ ($\alpha \rightarrow \pi \slash 2$), leading us to conclude that this does not affect our final result significantly.

After making the additional substitution $\beta = \tan(\alpha)$, this is
\begin{align}
& 
\dfrac{\delta P_{\zeta,2}}{P^0}
=  \dfrac{m^2}{ \pi^2 f^2} 
\int_{\rho_*}^\infty \dfrac{\exd \rho}{\rho^4} \int_{1}^{\infty} 
\exd\beta 
\frac{\sqrt{\beta ^2+1} e^{-\frac{\beta  \rho }{\sqrt{\beta ^2+1}}}  }{\beta ^2}
\left(\sqrt{\beta ^2+1}+\beta  \rho \right) \nonumber \\
& \qquad \qquad \times \left(\sqrt{\beta ^2+1} \sinh \left(\frac{\rho }{\sqrt{\beta ^2+1}}\right)-\rho  \cosh \left(\frac{\rho }{\sqrt{\beta ^2+1}}\right)\right) \nonumber \\
& \qquad 
\int  \exd y_1    \sum_{r}  
(s_r^2(i y_1) + d_r^2(i y_1))  \cdot 
\left(s_r^{*2}\left( i \beta y_1 \right)  + d_r^{*2}\left( i \beta y_1 \right) \right)
 + h.c. .
 \label{eq:full_rho_1}
\end{align}
Note that the $\rho$ dependence is entirely in the first two lines.  This integral can be done analytically for $\beta > 1$, with the result
\begin{align}
& \exd \rho \; \mathrm{integral} = 
\frac{e^{-\frac{(\beta +1) \rho_*}{\sqrt{\beta ^2+1}}} }{6 \beta ^2 \rho_*^3}
\left[
\rho_* \left( \beta^3 + \beta^2 + \beta + 1 \right)
- \rho_*^2 \sqrt{\beta ^2+1}
\left(\beta ^2 -\beta  +1 \right)
+\sqrt{\beta ^2+1} (\beta ^2+1)
\right. \nonumber \\
& \left. 
\rho^2 \sqrt{\beta ^2+1} e^{\frac{2 \rho_*}{\sqrt{\beta ^2+1}}}
\left(  \beta ^2 + \beta  +1 \right)
-\rho_* e^{\frac{2 \rho_*}{\sqrt{\beta ^2+1}}} 
\left( \beta^3 - \beta ^2  +\beta  -1 \right) 
-\sqrt{\beta ^2+1} e^{\frac{2 \rho_*}{\sqrt{\beta ^2+1}}} (\beta ^2  + 1) 
 \right. \nonumber \\
& \left.
+\left(\beta ^3-1\right) \rho_*^3 e^{\frac{(\beta +1) \rho_*}{\sqrt{\beta ^2+1}}} \text{Ei}\left(-\frac{(\beta -1) \rho_*}{\sqrt{\beta ^2+1}}\right)-\left(\beta ^3+1\right) \rho_*^3 e^{\frac{(\beta +1) \rho_*}{\sqrt{\beta ^2+1}}} \text{Ei}\left(-\frac{(\beta +1) \rho_*}{\sqrt{\beta ^2+1}}\right)\right],
\end{align}
where ${\rm Ei}$ denotes the exponential integral function. This is well-behaved as $\beta \rightarrow 1$ and has the expected logarithmic divergence as $\rho_* \rightarrow 0$ (equivalent to $\tau \rightarrow 0$).  Keeping only this piece gives
\begin{align} 
\dfrac{\delta P_{\zeta,2}}{P^0}
& = - \dfrac{m^2}{3 \pi^2 f^2} \log(\rho_*)
\int_{1}^{\infty} 
\dfrac{\exd \beta}{\beta^2} 
\int  \exd y_1    \sum_{r}  
(s_r^2(-i y_1) + d_r^2(-i y_1))  \cdot 
\left(s_r^{*2}\left( -i \beta y_1 \right)  + d_r^{*2}\left( -i \beta y_1 \right) \right)  \nonumber \\
& \qquad
 + h.c. .
\end{align}
After substituting our $s$ and $d$ functions in terms of Whittaker functions, this is
\begin{align}
& 
\dfrac{\delta P_{\zeta,2}}{P^0}
= - \dfrac{m^2}{3 \pi^2 f^2} \log(\rho_*)
\int_{1}^{\infty} 
\dfrac{\exd \beta}{\beta^2} 
\int  \exd y_1    \sum_{r} e^{-4 \pi r \xi}  \nonumber \\
& \qquad \cdot
\left(
 {W}^2_{\frac{1}{2} + 2 i r \xi, i \sqrt{ \mu^2 + 4 \xi^2}}(-2 y_1) 
- \mu^2   {W}^2_{ -\frac{1}{2} + 2 i r \xi, i \sqrt{ \mu^2 + 4 \xi^2}}( -2 y_1 )
\right)   \nonumber \\
&  \qquad
\cdot 
 \left(
{W}^2_{\frac{1}{2} - 2 i r \xi, i \sqrt{ \mu^2 + 4 \xi^2}}( 2 \beta y_1)
- \mu^2  {W}^2_{ -\frac{1}{2} - 2 i r \xi, i \sqrt{ \mu^2 + 4 \xi^2}}( 2 \beta y_1)
\right)
  + h.c.,
\label{eq:cubic_with_Whittaker}
\end{align}
noting that the Whittaker functions are even in their second argument.  Along the positive real axis (with positive second index), we approximate the Whittaker functions with
\begin{align}
 W_{\frac{1}{2} - 2 i r\xi, i \sqrt{\mu ^2+4 \xi ^2}}(2y)
& \approx \dfrac{(2y)^{-ir \sqrt{\mu ^2+4 \xi ^2}+1 \slash 2} e^{-y}}{\Gamma(i r \sqrt{\mu ^2+4 \xi ^2} + 2 i r \xi)} \Gamma(2 i r \sqrt{\mu ^2+4 \xi ^2}), \nonumber \\
 W_{-\frac{1}{2} - 2 i r \xi, i \sqrt{\mu ^2+4 \xi ^2}}( 2y )
& \approx \dfrac{(2y)^{-ir \sqrt{\mu ^2+4 \xi ^2}+1 \slash 2} e^{-y}}{\Gamma(i r \sqrt{\mu ^2+4 \xi ^2} + 2 i r \xi+1)} \Gamma(2 i r \sqrt{\mu ^2+4 \xi ^2}),
\label{eq:Whittaker_Approx_Positive}
\end{align}
which shows excellent agreement numerically along the positive real axis.  The situation along the negative real axis is complicated by a branch cut.  The above approximations work well for $x + i \epsilon$; however, due to the direction of the Wick rotation, the Whittaker functions on the second line of eq.~\eqref{eq:cubic_with_Whittaker} should be evaluated just below the real axis, at $x - i \epsilon$.  An approximation for these functions can be found using eq.~\eqref{eq:Whittaker_Approx_Positive} and the following recursion relation
\begin{align}
(-1)^m W_{\kappa, \mu}(z e^{2 m \pi i})
&= - \dfrac{e^{2 \kappa \pi i} \sin(2 m \mu \pi) + \sin((2m-2) \mu \pi)}{\sin(2 \mu \pi)} W_{\kappa, \mu}(z) \nonumber \\
& \qquad - \dfrac{\sin(2 m \mu \pi) 2 \pi i e^{\kappa \pi i}}{\sin(2 \mu \pi) \Gamma(1 \slash 2 + \mu - \kappa) \Gamma(1 \slash 2 - \mu - \kappa)} W_{-\kappa, \mu}(z e^{\pi i} ) \,. 
\end{align}
This leads to the approximations (valid along the negative axis)
\begin{align}
W_{\frac{1}{2}+2 i r\xi ,i \sqrt{\mu ^2+4 \xi ^2}}(2 y)
&\approx \mathcal{A}_{1r} e^{-2 \pi \xi} e^{-y} y^{\frac{1}{2}+ir \sqrt{\mu ^2+4 \xi ^2}}
+ \mathcal{B}_{1r} e^{-2 \pi  \xi } e^{y} (-y)^{\frac{1}{2}-ir \sqrt{\mu ^2+4 \xi ^2}}, \nonumber \\
W_{-\frac{1}{2}+ 2 i r\xi ,i \sqrt{\mu ^2+4 \xi ^2}}(2 y)
&\approx \mathcal{A}_{2r} e^{-2 \pi \xi} e^{-y} y^{\frac{1}{2}+ir \sqrt{\mu ^2+4 \xi ^2}}
+ \mathcal{B}_{2r} e^{-2 \pi \xi} e^y (-y)^{\frac{1}{2}-ir \sqrt{\mu ^2+4 \xi ^2}},
\end{align}
with
\begin{align}
\mathcal{A}_{1r} &= -\frac{2^{\frac{1}{2}+ir \sqrt{\mu ^2+4 \xi ^2}} e^{-2 \pi  \xi} \Gamma \left(-2 ir \sqrt{\mu ^2+4 \xi ^2}\right)}{\Gamma \left(-ir \left(2 \xi +\sqrt{\mu ^2+4 \xi ^2}\right)\right)}, \nonumber \\
\mathcal{A}_{2r} &= -\frac{2^{\frac{1}{2}+ir \sqrt{\mu ^2+4 \xi ^2}} e^{-2 \pi  \xi} \Gamma \left(-2 ir \sqrt{\mu ^2+4 \xi ^2}\right) }{\Gamma \left(-2 ir \xi -ir \sqrt{\mu ^2+4 \xi ^2}+1\right)}, \nonumber \\
\mathcal{B}_{1r} &= \frac{ir  2^{\frac{3}{2}-ir \sqrt{\mu ^2+4 \xi ^2}}  \Gamma \left(2 ir \sqrt{\mu ^2+4 \xi ^2}\right) \sinh \left(\pi  \left(\sqrt{\mu ^2+4 \xi ^2}+2 \xi \right)\right)}{\Gamma \left(ir \left( \sqrt{\mu ^2+4 \xi ^2}-2 \xi\right)\right)}, \nonumber \\
\mathcal{B}_{2r} &= \frac{ir  2^{\frac{3}{2}-ir \sqrt{\mu ^2+4 \xi ^2}} \Gamma \left(2 ir \sqrt{\mu ^2+4 \xi ^2}\right) \sinh \left(\pi  \left(\sqrt{\mu ^2+4 \xi ^2}+2 \xi \right)\right)}{\Gamma \left(-2 ir \xi +ir \sqrt{\mu ^2+4 \xi ^2}+1\right)}.
\end{align}
This is in agreement with eq.~\eqref{eq:approx_modes}; note that, in order to emphasize that these approximations are used along the negative axis, we have taken $y \rightarrow - y$ in eq. \eqref{eq:approx_modes}.  As noted below eq. \eqref{eq:approx_modes}, the positive frequency modes (the $\mathcal{A}$ part) corresponds to the vacuum part, while the negative frequency modes (the $\mathcal{B}$ part) correspond to the change resulting from the production of fermions.  Note that $\mathcal{A}_{1r} \mathcal{B}_{1r} - \mu^2 \mathcal{A}_{2r} \mathcal{B}_{2r} =0$.  Using these approximations gives
\begin{align}
& \dfrac{\delta P_{\zeta,2}}{P^0}
 = \dfrac{2 m^2}{3 \pi^2 f^2} \log(\rho_*)
\int_{1}^{\infty} 
\dfrac{d\beta}{\beta}  (2\beta)^{-2 ir \sqrt{ \mu^2 + 4 \xi^2}} 
\int  \exd y_1 \, e^{-4 \pi r\xi} e^{-4 \pi \xi}  y_1^2 
\mathcal{C}_r   \nonumber \\
&  \quad
\left[
e^{- 2 \pi r \sqrt{ \mu^2 + 4 \xi^2}}
(\mathcal{A}_{1r}^2 - \mu^2 \mathcal{A}_{2r}^2) e^{2 (1 - \beta) y_1} 
- 
(\mathcal{B}_{1r}^2 - \mu^2 \mathcal{B}_{2r}^2) e^{-2 (1 + \beta) y_1} 
(y_1)^{-4 i r \sqrt{ \mu^2 + 4 \xi^2}}
\right] 
 + h.c. \,, 
\end{align}
where
\begin{align}
\mathcal{C}_r &= \Gamma(2 ir \sqrt{ \mu^2 + 4 \xi^2})^2 \left(
\dfrac{1}{\Gamma(i r\sqrt{ \mu^2 + 4 \xi^2} + 2 ir \xi)^2} 
-  
 \dfrac{\mu^2}{\Gamma(i r\sqrt{ \mu^2 + 4 \xi^2} + 2 i r \xi+1)^2} \right).
\end{align}
Note that as $x_1 \rightarrow 0$, which corresponds to $\alpha \rightarrow \pi \slash 2$ and $\beta \rightarrow \infty$, the integrand is exponentially suppressed, as claimed above.  There is a divergence at $\beta = 1$, or $x_1 = x_2$; this is expected because along this line the Wick rotation does not suppress the exponential $e^{-2 i p_1 (\tau_1 - \tau_2)}$ and consequently the $y_1 \rightarrow \infty$ divergence remains.  We handle this divergence by setting the lower limit of the $\beta$ integral to $1 + \epsilon$ and consider the $\epsilon \rightarrow 0$ limit, finding
\begin{align}
& 
\dfrac{\delta P_{\zeta,2}}{P^0}
\approx  \dfrac{2 m^2}{3 \pi^2 f^2} \log(\rho_*) e^{-4 \pi \xi (1 + r)} \mathcal{C}_r  \nonumber \\
&
\left[e^{- 2 \pi r \sqrt{ \mu^2 + 4 \xi^2}}
(\mathcal{A}_{1r}^2 - \mu^2 \mathcal{A}_{2r}^2)  \left(
2^{-3-2 i r \sqrt{ \mu^2 + 4 \xi^2}} (-1 - 3 i r \sqrt{ \mu^2 + 4 \xi^2} + 2 (\sqrt{ \mu^2 + 4 \xi^2})^2) \right. \right. \nonumber \\
& \qquad \qquad \left. \left.
\cdot \left(2 H_{2 i r \sqrt{ \mu^2 + 4 \xi^2}+2}+2 \log (\epsilon )-3\right) \right. \right. \nonumber \\
& \qquad 
\left. \left.+\frac{2^{-3-2 ir \sqrt{ \mu^2 + 4 \xi^2}}}{\epsilon ^2}+\frac{4^{-1-i r\sqrt{ \mu^2 + 4 \xi^2}} (-1-2 i r\sqrt{ \mu^2 + 4 \xi^2})}{\epsilon } \right)
\right. \nonumber \\
& 
\left. 
- (\mathcal{B}_{1r}^2 - \mu^2 \mathcal{B}_{2r}^2)  \cdot 
(-2^{-3+2 i r\sqrt{ \mu^2 + 4 \xi^2}}) e^{-2 \pi  r\sqrt{ \mu^2 + 4 \xi^2}} \Gamma (3-4 i r\sqrt{ \mu^2 + 4 \xi^2}) \right. \nonumber \\
& \qquad \left. \cdot B_{-1}(3-2 i r\sqrt{ \mu^2 + 4 \xi^2},4 i r\sqrt{ \mu^2 + 4 \xi^2}-2)
\right]  \nonumber \\
&  + h.c..
\end{align}
All of the terms that are divergent as $\epsilon \rightarrow 0$ appear in the vacuum ($\mathcal{A}_r$) contribution, which should be removed by renormalization.  Furthermore, the $r = +1$ piece is exponentially suppressed by a factor of $e^{-8 \pi \xi}$, which accords with our earlier results that at large $\xi$, only the $r = -1$ modes are produced appreciably.  Dropping that piece, in the $\mu \rightarrow 0$ this goes as
\begin{align}
&\dfrac{P_{\zeta,2}}{P_0}
= \frac{2 m^4 e^{4 \pi  \xi } \sinh ^2(4 \pi  \xi ) \log (\rho_*)}{3 \pi ^2 f^2 H^2}  \left(\Gamma (4 i \xi )^2 \Gamma (3-8 i \xi ) B_{-1}^*(4 i \xi +3,-8 i \xi -2) \right. \nonumber \\
& \qquad \left. 
 +\Gamma (8 i \xi +3) \Gamma (-4 i \xi )^2 B_{-1}(4 i \xi +3,-8 i \xi -2)\right),
\end{align}
which at large $\xi$ has the scaling of eq.~\eqref{eq:p_cubic}.  For completeness, we note that the finite part of the vacuum contribution (the $\mathcal{A}$ piece) is
\begin{align}
\dfrac{\delta P_{\zeta,2}}{P^0} \bigg|_{\mathrm{vacuum},\mathrm{finite}}
&= \frac{m^2 \log (\rho_*) \left(12 i \xi  H_{2-4 i \xi }+2 \left(8 \xi ^2-6 i \xi -1\right) \Re\left(H_{2-4 i \xi }\right)-24 \xi ^2+3\right)}{3 \pi ^2 f^2} \nonumber \\
&\approx \frac{16 m^2 \xi ^2 \log (\xi ) \log (\rho_*)}{3 \pi ^2 f^2}
\end{align}
as $\xi \rightarrow \infty$.  This is not as strongly suppressed as $\mu \rightarrow 0$; however, as a vacuum contribution, it must be removed by renormalization.

%%%%%%%%%%%%%%%%%%%%%%%%%%%%%%%%%%%%%%
\section{Remarks on renormalization}%%
\label{ap:renormalization}%%%%%%%%%%%%
%%%%%%%%%%%%%%%%%%%%%%%%%%%%%%%%%%%%%%

In this appendix we motivate and describe our approach in dealing with the infinities that appear at almost every stage in our calculations. 

The use of Bogolyubov coefficients in the calculation of the occupation number provides a physically intuitive way of renormalizing such a quantity. By defining the number operators as $\hat{a}_r^\dagger(\bk)\,\hat{a}_r(\bk)$  and $\hat{b}_r^\dagger(\bk)\,\hat{b}_r(\bk)$ in eq.~(\ref{eq:nr}), we impose normal-ordering with respect to the time-dependent ladder operators $\hat{a}_r(\bk)$ and $\hat{b}_r(\bk)$; we remind the reader that these diagonalize the Hamiltonian at any given time and therefore annihilate the physical particle states. A direct calculation shows that such a time-dependent normal ordering is equivalent to adiabatic subtraction at the appropriate adiabatic order. In the case of the occupation number, this operation is sufficient to provide a result that is finite in the ultraviolet.

It is thus natural to use a similar approach when computing other quantities that would be otherwise divergent. In this appendix we will study in detail the case of the quartic diagram discussed in subsection~\ref{subsec:quartic}, which is  explicitly calculable in terms of analytic functions. The relevant integral is the one appearing in eq.~(\ref{amp1}), which reads
\begin{align}\label{eq:quartic_unreg}
\int \exd y \, y\, 
\sum_r \Re \left[ d_r^* \left( y \right) s_r \left( y \right) \right] \,, 
\end{align}
and as one can see from eq.~(\ref{eq:full_quartic}) is quadratically divergent. If one applies to this integral the prescription of time-dependent normal ordering (i.e., the use of Bogolyubov coefficients) which we have used in the calculation of the occupation numbers, one finds the integral~(\ref{eq:quartic_unreg}) to be replaced by
\begin{align}\label{eq:quartic_0thorder}
\sum_r \int \exd y \, y\, \left\{
\Re \left[ d_r^* \left( y \right) s_r \left( y \right) \right]-\frac{\mu\,y}{\sqrt{\mu^2+y^2}}\right\} \,, 
\end{align}
which turns out, indeed, to be equivalent to the result one would get by using zero-th order adiabatic regularization, as we will show below. The term subtracted in eq~(\ref{eq:quartic_0thorder}) does succeed in eliminating the quadratic divergence in eq.~(\ref{eq:quartic_unreg}), but this turns out not to be enough, as it leaves us with a logarithmic divergence. One then expects that, by going to higher order adiabatic regularization, the logarithmic divergence will also be eliminated.

Adiabatic regularization to arbitrary order for massive fermions in de Sitter space has been studied in detail in~\cite{Landete:2013lpa}. We extend their formalism to include the psudoscalar interaction. By rewriting the latter in terms of the parameter $\xi$ times $H$, we can continue organizing the adiabatic expansion as an expansion in the single parameter $H$. It is convenient to work with the mode functions $\tilde{u}$ and $\tilde{v}$  introduced in eq.~\eqref{eq:def_uvtilde}.  In physical time, these satisfy the equations
\begin{align}
i \dot{\tilde u}_r &= m \tilde u_r + \left( \dfrac{k}{a} + 2 H r \xi \right) \tilde v_r, \nonumber \\
i \dot{\tilde v}_r &= - m \tilde  v_r + \left( \dfrac{k}{a} + 2 H r \xi \right) \tilde  u_r,
\end{align}
which we solve as an expansion in the Hubble parameter $H$.  We introduce the Ansatz
\begin{align}
\tilde u_r &= \sqrt{ 1 + \dfrac{m}{\omega}} \exp \left(
-i \int^t d\tilde{t} \left[ \omega(\tilde{t}) + H \omega_1(\tilde{t}) + H^2 \omega_2 (\tilde{t}) \right] \right)
\left[ 1 + H F_{1,r} + H^2 F_{2,r} \right], \nonumber \\
\tilde v_r &= \sqrt{ 1 - \dfrac{m}{\omega}} \exp \left(
-i \int^t d\tilde{t} \left[ \omega(\tilde{t}) + H \omega_1(\tilde{t}) + H^2 \omega_2 (\tilde{t}) \right] \right)
\left[ 1 + H G_{1,r} + H^2 G_{2,r} \right],
\end{align}
where we have the freedom to choose $F_{i,r}$ to be real, and where $\omega = \sqrt{ k^2/a^2 + m^2 }$.  Solving the equations of motion iteratively, we find
\begin{align}
F_{1,r} &= - \dfrac{r \xi m}{\omega^2} \sqrt{ \dfrac{\omega - m}{\omega + m} }\,,\qquad G_{1,r} = \dfrac{r \xi m}{\omega^2} \sqrt{ \dfrac{\omega + m}{\omega - m} }  + i\dfrac{m}{2  \omega^2}, \nonumber \\
\omega_1 &= 2 r \xi \sqrt{ 1 - \dfrac{m^2}{\omega^2}}\,,\qquad \omega_2= \dfrac{m (4 \omega - 5 m) (\omega^2 - m^2)}{8 \omega^5} + \dfrac{2 m^2 \xi^2}{\omega^3}, \nonumber \\
F_2 &= \dfrac{m (\omega - m)(5 m^2 - 4 \omega^2)}{16 \omega^6} + \dfrac{m \xi^2 (4 \omega - 5 m)}{2 \omega^4}, \nonumber \\
G_2 &= \dfrac{m (-5 m^3 - 5 m^2 \omega + 2 m \omega^2 + 4 \omega^3)}{16 \omega^6}
-i \dfrac{m r \xi (5 m + 6 \omega) }{2 \omega^4} \sqrt{
\dfrac{ \omega - m}{\omega + m} } - \dfrac{m \xi^2 (5 m + 4 \omega)}{2 \omega^4} \,, 
\end{align}
so that
\begin{align}\label{eq:quartic_2ndorder}
\Re &\left[ d_r^* \left( y \right) s_r \left( y \right) \right]\Big|_{\rm adiab}  =\frac{m}{\omega} -\frac{2 m r \xi \sqrt{\omega^2-m^2}}{\omega^3} H\nonumber\\
&\qquad\qquad\qquad\qquad - \left[ \frac{4 \omega^4-9 \omega^2 m^2+5 m^4}{8 \omega^2} - 2 \left( 2 \omega^2 - 3 m^2 \right) \xi^2 \right] \frac{m}{\omega^5}\,H^2 + \dots  \nonumber\\ 
&= \frac{\mu}{\sqrt{y^2+\mu^2} }- \frac{2 r y \mu \xi}{\left( y^2 + \mu^2 \right)^{3/2}}   - \frac{\mu}{8} \frac{16 \mu^4 \xi^2 + 4 y^4 \left( 1 - 8 \xi^2 \right) - y^2 \mu^2 \left( 1 + 16 \xi^2 \right)}{\left( y^2 + \mu^2 \right)^{7/2}} + \dots  \,, 
\end{align} 
where we recognize the first term as zero-th order adiabatic expansion of the integrand that we found in eq.~(\ref{eq:quartic_0thorder}). One can then show that subtraction of eq.~(\ref{eq:quartic_2ndorder}) from eq.~(\ref{eq:quartic_unreg}) does yield, as desired, a result that is finite in the ultraviolet.

Subtraction at second adiabatic order, however, generates a different, undesired feature. A direct computation shows that for $y={\cal O}(1)$, the regularizing term~(\ref{eq:quartic_2ndorder}) is of order $\xi^2\gg1$, whereas the original integrand in eq.~(\ref{eq:quartic_unreg}) is of the order of unity. This implies that the regularizing term gives a large contribution to our integral in a regime where we would expect it to be irrelevant, since we would desire the adiabatic part to play a relevant role only in taming the ultraviolet divergences.  Moreover, the presence of this term depends on the order of the adiabatic regularization. This behavior was noted in the past, see e.g.~\cite{Durrer:2009ii}, and casts some doubts on the use of adiabatic regularization when dealing with modes that are not in the ultraviolet.

The discussion above emphasizes the difficulties that emerge when we try to extract a finite result from the quadratically divergent integral~(\ref{eq:quartic_unreg}). However, if we go ahead and ignore the presence of the unphysical behavior at $y={\cal O}(1)$ of the regularizing function, we obtain that numerical integration yields
\begin{align}\label{eq:integrated_quartic}
\sum_r \int \exd y \, y\, \left\{
\Re \left[ d_r^* \left( y \right) s_r \left( y \right) \right]-\Re\left[ d_r^* \left( y \right) s_r \left( y \right) \right]\Big|_{\rm adiab}\right\}\approx -8.5\,\mu\,\xi^2\,\log\xi\,.
\end{align}
We now crucially note that the spurious contribution from the adiabatic expansion~(\ref{eq:quartic_2ndorder}) at $y\approx 1$ can be evaluated to be $\approx 4\,\mu\,\xi^2$. Such a contribution is subdominant with respect to the full result~(\ref{eq:integrated_quartic}) in the limit $\xi\gg 1$ we are interested in, which allows us to neglect the issues  about the behavior of eq.~(\ref{eq:quartic_2ndorder}) at $y\approx 1$ raised above. 

We finally note that the brute force subtraction of the divergent part of the integral~(\ref{eq:integral_mull1}) leaves us with
\begin{align}
\sum_r \int \exd y \, y\,
\Re \left[ d_r^* \left( y \right) s_r \left( y \right) \right]\Big|_{\rm subtract}= -8\,\mu\,\xi^2\,\log\xi\, \,, 
\end{align}
which is in excellent agreement with the result~(\ref{eq:integrated_quartic}). For this reason, in the remaining calculations of divergent quantities in this work, we have simply dealt with infinities by subtracting the UV-divergent parts by hand.

%%%%%%%%%%%%%%%%%%%%%%%%%%%%%%%%%%%%%%%%%%%%%%%%%%
\section{Non-Gaussianity: the quintic integral}%%%
\label{ap:nongaussianities}%%%%%%%%%%%%%%%%%%%%%%%
%%%%%%%%%%%%%%%%%%%%%%%%%%%%%%%%%%%%%%%%%%%%%%%%%%

This appendix presents the calculation for the quintic vertex loop in figure~\ref{fig:3ptdiagrams}, leading to eq.~\eqref{eq:fnl_quintic_divergent}.  First recall that the interaction vertices arise from the following term in the Lagrangian,
\begin{align}
\mathcal{L}_\mathrm{int} = - i a m \pi_\psi \gamma^0 \left[ \cos \left( \dfrac{2 \phi}{f} \right) - i \gamma^5 \sin \left( \dfrac{2 \phi}{f} \right) \right] \psi.
\end{align}
Expanding $\phi = \phi_0 + \delta \phi$ and isolating the cubic term leads to a contribution to the interaction Hamiltonian of the form
\begin{align}
\mathcal{H}_{\delta \phi^3}= - a m  \dfrac{4 (\delta \phi)^3}{3 f^3} \bar{\psi}  \left[ \sin \left( \dfrac{2 \phi_0}{f} \right) + i \gamma^5 \cos \left( \dfrac{2 \phi_0}{f} \right) \right] \psi.
\end{align}
Introducing the mode functions gives
\begin{align}
H_{\delta \phi^3} 
&=- \dfrac{4 a m}{ 3 f^3} \int \dfrac{\exd^3p_1 \, \exd^3p_2 \, \exd^3p_3 \, \exd^3 p_4 }{(2\pi)^{9 \slash 2}} \delta \phi_{\bp_1} \delta \phi_{\bp_2} \delta \phi_{\bp_3}  \bar{\psi}_{\bp_4} \left[ \sin \left( \dfrac{2 \phi_0}{f} \right) + i \gamma^5 \cos \left( \dfrac{2 \phi_0}{f} \right) \right] \nonumber \\
& \qquad \qquad \cdot \psi_{\bp_4 - \bp_1 - \bp_2 - \bp_3}.
\end{align}

Using the in-in formalism, the diagram corresponds to 
\begin{align}
\left< \delta \phi_{\bk_1}(\tau)
\delta \phi_{\bk_2}(\tau)
\delta \phi_{\bk_3}(\tau) \right>
& = - i \int^\tau \exd \tau_1 \left< 
[ \delta \phi_{\bk_1}(\tau)
\delta \phi_{\bk_2}(\tau)
\delta \phi_{\bk_3}(\tau) , 
H_{\delta \phi^3}(\tau_1) ] \right> \nonumber \\
&=  i \dfrac{4  m}{3 f^3} \int^\tau \exd \tau_1 a(\tau_1)\int \dfrac{\exd^3p_1 \, \exd^3p_2 \, \exd^3p_3 \, \exd^3 p_4 }{(2\pi)^{9 \slash 2}} \nonumber \\
&
\qquad \cdot \left< 
[ \delta \phi_{\bk_1}(\tau)
\delta \phi_{\bk_2}(\tau)
\delta \phi_{\bk_3}(\tau) ,
\delta \phi_{\bp_1}(\tau_1) \delta \phi_{\bp_2}(\tau_1) \delta \phi_{\bp_3}(\tau_1) ] \right> \nonumber \\
& \qquad \cdot \left< \bar{\psi}_{\bp_4}(\tau_1) \left[ \sin \left( \dfrac{2 \phi_0}{f} \right) + i \gamma^5 \cos \left( \dfrac{2 \phi_0}{f} \right) \right] \psi_{\bp_4 - \bp_1 - \bp_2 - \bp_3}(\tau_1) \right> \,, 
\end{align}
where we have noted that the fermionic and bosonic creation and annihilation operators commute.  The bosonic contribution can be evaluated using using eq.~\eqref{eq:phi_commutator} and eq.~\eqref{eq:phiphivev}; this gives
\begin{align}
\lim_{\tau \rightarrow 0} \left< \delta \phi_{\bk_1}(\tau)
\delta \phi_{\bk_2}(\tau)
\delta \phi_{\bk_3}(\tau) \right>
&=  -\dfrac{2 H^6 m}{3 f^3} \int^\tau \exd \tau_1 a(\tau_1) f(k_1,k_2,k_3,\tau_1) \int \dfrac{\exd^3 p_4}{(2\pi)^{9 \slash 2}} \nonumber \\
& \left< \bar{\psi}_{\bp_4}(\tau_1) \left[ \sin \left( \dfrac{2 \phi_0}{f} \right) + i \gamma^5 \cos \left( \dfrac{2 \phi_0}{f} \right) \right] \psi_{\bp_4 - \bk_1 - \bk_2 - \bk_3}(\tau_1) \right> \,, 
\end{align}
where 
\begin{align}
 f(k_1,k_2,k_3,\tau_1) &= \dfrac{1}{k_1^3 k_2^3 k_3^3} \cdot \left[ \tau_1 
\left(k_1 k_2 k_3 \tau_1^2- k_1 -k_2-k_3\right) \cos (\tau_1 (k_1+k_2+k_3)) \right. \nonumber \\
& \qquad \left. -\left(\tau_1^2 (k_1 k_2+ k_1 k_3+k_2 k_3)-1\right) \sin (\tau_1 (k_1+k_2+k_3)) \right].
\end{align}

Proceeding to the fermionic contribution, we expand $\psi$ using \eqref{eq:deco_psi}, finding 
\begin{align}
& \left< \bar{\psi}_{\bp_4}(\tau_1) \left[ \sin \left( \dfrac{2 \phi_0}{f} \right) + i \gamma^5 \cos \left( \dfrac{2 \phi_0}{f} \right) \right] \psi_{\bp_4 - \bk_1 - \bk_2 - \bk_3}(\tau_1) \right> \nonumber \\
& \qquad = \sum_r \dfrac{m}{2} \left[ |v|^2 m_s + i m_c r (u v^* - v u^*) - m_s |u|^2 \right]_{r,\tau_1,p_4} \nonumber \\
& \qquad = -i\sum_r  m r (\tilde u^* \tilde v - \tilde v^* \tilde u) \big|_{r, p_4,\tau_1} \;, 
\end{align}
where the last line is expressed in terms of the $\tilde{u}$ and $\tilde{v}$ functions introduced in appendix \ref{ap:new_basis}; this result can equilvalently be derived by expressing $\psi$ in terms of $Y$ prior to evaluating the expectation values.  We introduce the new variables $x = - k_1 \tau_1$, $y = - p_4 \tau_1$, and $\zeta_2 = k_2 \slash k_1$ with $\zeta_3 = k_3 \slash k_1$ to find
\begin{align}
\left< \delta \phi_{\bk_1}(\tau)
\delta \phi_{\bk_2}(\tau)
\delta \phi_{\bk_3}(\tau) \right>^\prime
&=   i \dfrac{2 H^5 m}{3 f^3} \int_x \dfrac{\exd x}{x} f(x,\zeta_2,\zeta_3) 
\left( \dfrac{k_1}{x} \right)^3
\int \dfrac{\exd^3 y}{(2\pi)^{9 \slash 2}} \sum_r  \dfrac{r(sd^* - ds^*)}{y} \bigg|_{r,y} \,, 
\end{align}
where
\begin{align}
f(x,\zeta_2,\zeta_3) & \equiv f(k_1,k_2,k_3,\tau_1) = \dfrac{1}{k_1^3 k_2^3 k_3^3} \cdot \left[ 
x(-x^2\zeta_2 \zeta_3+1+ \zeta_2 + \zeta_3) \cos(x (1 + \zeta_2 + \zeta_3))
 \right. \nonumber \\
& \qquad \left. 
+ \left[ x^2\left( \zeta_2  + \zeta_3  + \zeta_2 \zeta_3 \right) -1 \right]
\sin(x (1 + \zeta_2 + \zeta_3)) \right] \,,
\end{align}
and a prime again denotes that the overall delta function has been dropped.  We have introduced the $s$ and $d$ functions which are known in terms of Whittaker functions.  

The $\exd x$ and $\exd y$ integrals have again separated, as in the calculation in appendix~\ref{ap:backreaction}.  The $\exd x$ integral again has a logarithmic divergence as $x_* \rightarrow 0$, or $\tau \rightarrow 0$, and in the equilateral configuration we obtain
\begin{align}
\left< \delta \phi_{\bk_1}(\tau)
\delta \phi_{\bk_2}(\tau)
\delta \phi_{\bk_3}(\tau) \right>^\prime
&= -  \dfrac{2 H^5  m}{f^3 k^6} 
\log (x_*) \cdot \dfrac{4 \pi}{(2\pi)^{9 \slash 2}} \sum_r  \int \exd y \, y  \, r  \cdot 2 \Im(sd^*) \bigg|_{r,y}.
\label{eq:quintic_before_long_integral}
\end{align}

The remaining integral was evaluated in appendix \ref{ap:backreaction}; the full analytic result is given in eq.~\eqref{eq:quintic_exact}.  Taking the small $\mu$ limit (but not necessarily the large $\xi$ limit) gives 
\begin{align}
f_{NL}^{eq} &= 
\frac{20 H^2 \mu^2  \xi }{9 \pi ^2 f^2} \frac{\log(x_*)}{\left(1+\delta P_\zeta(k)/P_\zeta^{(0)}\right)^2}\left(\left(-8 i \xi ^2+6 \xi +i\right) H_{-4 i \xi }+\left(8 i \xi ^2+6 \xi -i\right) H_{4 i \xi } \right. \nonumber \\
& \qquad \left. -4 \xi  (3 \log (2 \Lambda )+3 \gamma_E -2)\right),
\end{align}
where $\delta P_\zeta$ is defined in eq.~(\ref{eq:deltapzeta}). Further expanding in large $\xi$ gives
\begin{align}
f_{NL}^{eq} &= -\frac{160 H^2 \log(x_*) \mu ^2 \xi ^3}{9 \pi  f^2\left(1+\delta P_\zeta(k)/P_\zeta^{(0)}\right)^2},
\end{align}
in agreement with eq.~\eqref{eq:fnl_quintic_divergent}. 

%%%%%%%%%%%%%%%%%%%%%%%%%%%%%%%%%%
\section{Fermion energy density}%%
\label{ap:rho_psi}%%%%%%%%%%%%%%%%
%%%%%%%%%%%%%%%%%%%%%%%%%%%%%%%%%%

\begin{figure}
\begin{center}
\includegraphics[scale=.6]{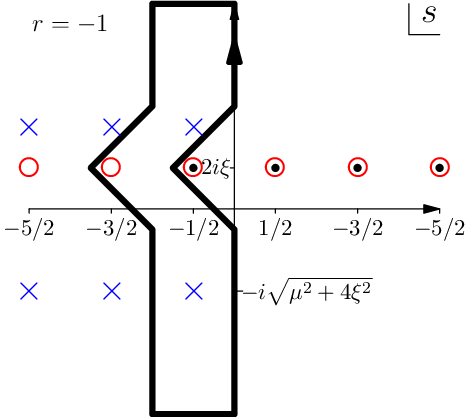}
\end{center}
\caption{The contour used to calculate the final contribution to $\int y^2 |s_r|^2$, in place of the contour in figure~\ref{fig:weird_contour}.  The small black dots are the poles of $g(s)$.  The poles of $f_1(s)$ result from the csc functions and are represented with red circles ($s = n - 1 \slash 2 + i \mathfrak{b} r$) and blue crosses ($s = n - 1 \slash 2 \pm i \mathfrak{a}$).  This is for helicity $r = -1$; for $r = +1$, the black dots and red circles reflect over the real axis.}
\label{fig:weirder_contour}
\end{figure}

In this appendix we make a few brief comments about the analytic calculation of the fermion energy density, relevant to the constraint \eqref{eq:rho_psi_constraint}.  Summing over both particles and antiparticles, the energy density in fermions is
\begin{align}
\rho_\psi &= 2 \sum_r \int \exd^3 k \,  \omega | \beta_r|^2,
\end{align}
where $|\beta_r|^2$ is given by eq.~\eqref{eq:nr}.  In terms of $s$ and $d$ functions, this is
\begin{align}
\rho_\psi
&=  \dfrac{\pi}{\tau^4}  \left(\Lambda  \left(2 \Lambda ^2+\mu ^2\right) \sqrt{\Lambda ^2+\mu ^2}+\mu ^4 \log \left(\frac{\mu }{\sqrt{\Lambda ^2+\mu ^2}+\Lambda }\right)\right) \nonumber \\
& \qquad -\dfrac{4 \pi }{2 \tau^4} \left[ \sum_r \int \exd y \, y 
\left(
2\mu \Re(s_r d_r^*)
+ 2 y |s_r|^2 \right) - \Lambda^4
\right],
\end{align}
where as is typical, $y = - k \tau$.  The first term arises from integrating $\int \omega \, \exd y$ with $\omega = \sqrt{ y^2 + \mu^2} \slash | \tau|$, and we have used the normalization condition $|s_r|^2 + |d_r|^2 = 2y$.  On the second line, note that $\sum_r \int \exd y \, y \Re(s_r d_r^*)$ was evaluated in appendix \ref{ap:quartic}, leaving only $\sum_r \int \exd y \, y^2 |s_r|^2$ remaining to evaluate.  We evaluate this analytically following the technique of appendix \ref{ap:backreaction}, observing that when we introduce the Mellin-Barnes representation for $s$ and $s^*$, both contours are deformed as shown on the left side of figure~\ref{fig:poles}.  Consequently, the $\exd s$ integral has poles at $n_2 - 1 \slash 2 + i r \mathfrak{b}$ (where $n_2 = 0, 1, \dots$).  This means not only that more $t$ and $s$ poles contribute as $\Lambda \rightarrow \infty$, but additional care must be used when evaluating the $O(\Lambda^0)$ contribution.  To evaluate this, we use the functions
\begin{align}
f_1(s) &= \frac{1}{16} \left(\mathfrak a^2-\mathfrak b^2\right) \csc \left(\pi  \left(-i \mathfrak a+s+\frac{1}{2}\right)\right) \csc \left(\pi  \left(i \mathfrak a+s+\frac{1}{2}\right)\right) \left(-\left((-1)^{-s} e^{\pi  (-\mathfrak b) r}\right)\right) \nonumber \\
& \qquad \times \sinh (\pi  (\mathfrak a-\mathfrak b r)) \sinh (\pi  (\mathfrak a+\mathfrak b r)) \csc \left(\pi  \left(-i \mathfrak b r+s+\frac{1}{2}\right)\right), \nonumber \\
f_2(s) &= \frac{\left(-i \mathfrak a-s+\frac{1}{2}\right) \left(-i \mathfrak a-s+\frac{3}{2}\right) \left(-i \mathfrak a-s+\frac{5}{2}\right) \left(i \mathfrak a-s+\frac{1}{2}\right) \left(i \mathfrak a-s+\frac{3}{2}\right) \left(i \mathfrak a-s+\frac{5}{2}\right)}{\left(i \mathfrak b r-s-\frac{1}{2}\right) \left(-i \mathfrak b r+s-\frac{7}{2}\right) \left(-i \mathfrak b r+s-\frac{5}{2}\right) \left(-i \mathfrak b r+s-\frac{3}{2}\right) \left(-i \mathfrak b r+s-\frac{1}{2}\right)}.
\end{align}
where the latter function is expressed in terms of
\begin{align}
h(s) &= \frac{C_{-7}}{s-i \mathfrak b r-\frac{7}{2}}, \nonumber \\
g(s) &= 
C_1 \left(-i b r+s-\frac{1}{2}\right)+C_2 \left(-i b r+s-\frac{1}{2}\right)^2+\frac{C_{-1}}{-i b r+s-\frac{1}{2}}+\frac{C_{-1}^\prime}{i b r-s-\frac{1}{2}} \nonumber \\
& \qquad +\frac{C_{-3}}{-i b r+s-\frac{3}{2}} +\frac{C_{-5}}{-i b r+s-\frac{5}{2}},
\end{align}
with the coefficients determined by $f_2(s) = g(s) - g(s-1) + h(s)$.  Perhaps most importantly, note that the contour equivalent to figure~\ref{fig:weird_contour} is shifted as shown in figure~\ref{fig:weirder_contour}.

Following the same procedure as in appendix \ref{ap:backreaction}, one arrives at the result
\begin{align}
& \sum_r \int dy \, y^2 |s_r|^2
= \Lambda ^4-\frac{\Lambda ^2 \mu ^2}{2}-\frac{7 \mu ^4}{16}+2 \mu ^2 \left(8 \xi ^2-\frac{19}{32}\right)-8 \xi ^4+\frac{11 \xi ^2}{2} \nonumber \\
&\qquad  + \frac{1x}{4} \xi  \sqrt{\mu ^2+4 \xi ^2} \left(-26 \mu ^2+16 \xi ^2-11\right) \sinh (4 \pi  \xi ) \text{csch}\left(2 \pi  \sqrt{\mu ^2+4 \xi ^2}\right) \nonumber \\
&\qquad  + \frac{3}{16} \mu ^2 \left(\mu ^2-16 \xi ^2+1\right) 
\left[ 4 (\log (2\Lambda )+\gamma_E ) +  \right. \nonumber \\
& \qquad \qquad \left.
\left( H_{-i \left(\sqrt{\mu ^2+4 \xi ^2}-2 \xi \right)} +
H_{i \left(\sqrt{\mu ^2+4 \xi ^2}-2 \xi \right)} \right)
\left(\sinh (4 \pi  \xi ) \text{csch}\left(2 \pi  \sqrt{\mu ^2+4 \xi ^2}\right)-1\right) \right. \nonumber \\
& \qquad \qquad \left. -
\left( H_{-i \left(2 \xi +\sqrt{\mu ^2+4 \xi ^2}\right)} 
+ H_{i \left(2 \xi +\sqrt{\mu ^2+4 \xi ^2}\right)} \right)
\left(\sinh (4 \pi  \xi ) \text{csch}\left(2 \pi  \sqrt{\mu ^2+4 \xi ^2}\right)+1\right)
\right].
\end{align}

\paragraph{}Neglecting divergent pieces, in the small $\mu$ and large $\xi$ limit the two relevant integrals go as
\begin{align}
\sum_r \int \exd y \, y \Re \left( s_r(y) d_r^*(y) \right) \approx -8 \mu \ln(\xi) \xi^2, \qquad
\sum_r \int \exd y \, y^2 |s_r|^2 \approx - 4 \pi \mu^2 \xi^3,
\end{align}
and so the second integral dominates the energy density at sufficiently large $\xi$, for $\mu \lesssim 1$.  This gives
\begin{align}
\rho_\psi \approx \dfrac{16 \pi^2 \mu^2 \xi^3}{\tau^4}.
\end{align}
To find the physical energy density we scale this by $a^{-4}$, giving
\begin{align}
\rho_\psi \big|_{\mathrm{phys}} \approx 16\, H^4 \pi^2 \mu^2 \xi^3,
\end{align}
as used in \eqref{eq:rho_psi_constraint}.

\bibliography{FermionLoops}

\providecommand{\href}[2]{#2}\begingroup\raggedright\begin{thebibliography}{10}

\bibitem{Guth:1980zm}
A.~H. Guth, \emph{{The Inflationary Universe: A Possible Solution to the
  Horizon and Flatness Problems}},
  \href{http://dx.doi.org/10.1103/PhysRevD.23.347}{\emph{Phys. Rev.} {\bf D23}
  (1981) 347--356}.

\bibitem{Linde:2005ht}
A.~D. Linde, \emph{{Particle physics and inflationary cosmology}},
  {\emph{Contemp. Concepts Phys.} {\bf 5} (1990) 1--362},
  [\href{http://arxiv.org/abs/hep-th/0503203}{{\tt hep-th/0503203}}].

\bibitem{Ade:2015xua}
{\scshape Planck} collaboration, P.~A.~R. Ade et~al., \emph{{Planck 2015
  results. XIII. Cosmological parameters}},
  \href{http://dx.doi.org/10.1051/0004-6361/201525830}{\emph{Astron.
  Astrophys.} {\bf 594} (2016) A13},
  [\href{http://arxiv.org/abs/1502.01589}{{\tt 1502.01589}}].

\bibitem{Ade:2013ydc}
{\scshape Planck} collaboration, P.~A.~R. Ade et~al., \emph{{Planck 2013
  Results. XXIV. Constraints on primordial non-Gaussianity}},
  \href{http://dx.doi.org/10.1051/0004-6361/201321554}{\emph{Astron.
  Astrophys.} {\bf 571} (2014) A24},
  [\href{http://arxiv.org/abs/1303.5084}{{\tt 1303.5084}}].

\bibitem{Ade:2015tva}
{\scshape BICEP2, Planck} collaboration, P.~A.~R. Ade et~al., \emph{{Joint
  Analysis of BICEP2/$Keck ?Array$ and $Planck$ Data}},
  \href{http://dx.doi.org/10.1103/PhysRevLett.114.101301}{\emph{Phys. Rev.
  Lett.} {\bf 114} (2015) 101301}, [\href{http://arxiv.org/abs/1502.00612}{{\tt
  1502.00612}}].

\bibitem{Array:2015xqh}
{\scshape BICEP2, Keck Array} collaboration, P.~A.~R. Ade et~al.,
  \emph{{Improved Constraints on Cosmology and Foregrounds from BICEP2 and Keck
  Array Cosmic Microwave Background Data with Inclusion of 95 GHz Band}},
  \href{http://dx.doi.org/10.1103/PhysRevLett.116.031302}{\emph{Phys. Rev.
  Lett.} {\bf 116} (2016) 031302}, [\href{http://arxiv.org/abs/1510.09217}{{\tt
  1510.09217}}].

\bibitem{Ade:2015lrj}
{\scshape Planck} collaboration, P.~A.~R. Ade et~al., \emph{{Planck 2015
  results. XX. Constraints on inflation}},
  \href{http://dx.doi.org/10.1051/0004-6361/201525898}{\emph{Astron.
  Astrophys.} {\bf 594} (2016) A20},
  [\href{http://arxiv.org/abs/1502.02114}{{\tt 1502.02114}}].

\bibitem{Abazajian:2016yjj}
{\scshape CMB-S4} collaboration, K.~N. Abazajian et~al., \emph{{CMB-S4 Science
  Book, First Edition}},  \href{http://arxiv.org/abs/1610.02743}{{\tt
  1610.02743}}.

\bibitem{Freese:1990rb}
K.~Freese, J.~A. Frieman and A.~V. Olinto, \emph{{Natural inflation with pseudo
  - Nambu-Goldstone bosons}},
  \href{http://dx.doi.org/10.1103/PhysRevLett.65.3233}{\emph{Phys. Rev. Lett.}
  {\bf 65} (1990) 3233--3236}.

\bibitem{Silverstein:2008sg}
E.~Silverstein and A.~Westphal, \emph{{Monodromy in the CMB: Gravity Waves and
  String Inflation}},
  \href{http://dx.doi.org/10.1103/PhysRevD.78.106003}{\emph{Phys. Rev.} {\bf
  D78} (2008) 106003}, [\href{http://arxiv.org/abs/0803.3085}{{\tt
  0803.3085}}].

\bibitem{McAllister:2008hb}
L.~McAllister, E.~Silverstein and A.~Westphal, \emph{{Gravity Waves and Linear
  Inflation from Axion Monodromy}},
  \href{http://dx.doi.org/10.1103/PhysRevD.82.046003}{\emph{Phys. Rev.} {\bf
  D82} (2010) 046003}, [\href{http://arxiv.org/abs/0808.0706}{{\tt
  0808.0706}}].

\bibitem{Peloso:2015dsa}
M.~Peloso and C.~Unal, \emph{{Trajectories with suppressed tensor-to-scalar
  ratio in Aligned Natural Inflation}},
  \href{http://dx.doi.org/10.1088/1475-7516/2015/06/040}{\emph{JCAP} {\bf 1506}
  (2015) 040}, [\href{http://arxiv.org/abs/1504.02784}{{\tt 1504.02784}}].

\bibitem{DAmico:2017cda}
G.~D'Amico, N.~Kaloper and A.~Lawrence, \emph{{Monodromy inflation at strong
  coupling: $4\pi$ in the sky}},  \href{http://arxiv.org/abs/1709.07014}{{\tt
  1709.07014}}.

\bibitem{Garretson:1992vt}
W.~D. Garretson, G.~B. Field and S.~M. Carroll, \emph{{Primordial magnetic
  fields from pseudoGoldstone bosons}},
  \href{http://dx.doi.org/10.1103/PhysRevD.46.5346}{\emph{Phys. Rev.} {\bf D46}
  (1992) 5346--5351}, [\href{http://arxiv.org/abs/hep-ph/9209238}{{\tt
  hep-ph/9209238}}].

\bibitem{Pajer:2013fsa}
E.~Pajer and M.~Peloso, \emph{{A review of Axion Inflation in the era of
  Planck}},
  \href{http://dx.doi.org/10.1088/0264-9381/30/21/214002}{\emph{Class. Quant.
  Grav.} {\bf 30} (2013) 214002}, [\href{http://arxiv.org/abs/1305.3557}{{\tt
  1305.3557}}].

\bibitem{Anber:2009ua}
M.~M. Anber and L.~Sorbo, \emph{{Naturally inflating on steep potentials
  through electromagnetic dissipation}},
  \href{http://dx.doi.org/10.1103/PhysRevD.81.043534}{\emph{Phys. Rev.} {\bf
  D81} (2010) 043534}, [\href{http://arxiv.org/abs/0908.4089}{{\tt
  0908.4089}}].

\bibitem{Ferreira:2017lnd}
R.~Z. Ferreira and A.~Notari, \emph{{Thermalized Axion Inflation}},
  \href{http://dx.doi.org/10.1088/1475-7516/2017/09/007}{\emph{JCAP} {\bf 1709}
  (2017) 007}, [\href{http://arxiv.org/abs/1706.00373}{{\tt 1706.00373}}].

\bibitem{Ferreira:2017wlx}
R.~Z. Ferreira and A.~Notari, \emph{{Thermalized axion inflation: natural and
  monomial inflation with small $r$}},
  \href{http://arxiv.org/abs/1711.07483}{{\tt 1711.07483}}.

\bibitem{Prokopec:2001nc}
T.~Prokopec, \emph{{Cosmological magnetic fields from photon coupling to
  fermions and bosons in inflation}},
  \href{http://arxiv.org/abs/astro-ph/0106247}{{\tt astro-ph/0106247}}.

\bibitem{Anber:2006xt}
M.~M. Anber and L.~Sorbo, \emph{{N-flationary magnetic fields}},
  \href{http://dx.doi.org/10.1088/1475-7516/2006/10/018}{\emph{JCAP} {\bf 0610}
  (2006) 018}, [\href{http://arxiv.org/abs/astro-ph/0606534}{{\tt
  astro-ph/0606534}}].

\bibitem{Caprini:2014mja}
C.~Caprini and L.~Sorbo, \emph{{Adding helicity to inflationary
  magnetogenesis}},
  \href{http://dx.doi.org/10.1088/1475-7516/2014/10/056}{\emph{JCAP} {\bf 1410}
  (2014) 056}, [\href{http://arxiv.org/abs/1407.2809}{{\tt 1407.2809}}].

\bibitem{Fujita:2015iga}
T.~Fujita, R.~Namba, Y.~Tada, N.~Takeda and H.~Tashiro, \emph{{Consistent
  generation of magnetic fields in axion inflation models}},
  \href{http://dx.doi.org/10.1088/1475-7516/2015/05/054}{\emph{JCAP} {\bf 1505}
  (2015) 054}, [\href{http://arxiv.org/abs/1503.05802}{{\tt 1503.05802}}].

\bibitem{Adshead:2016iae}
P.~Adshead, J.~T. Giblin, T.~R. Scully and E.~I. Sfakianakis,
  \emph{{Magnetogenesis from axion inflation}},
  \href{http://dx.doi.org/10.1088/1475-7516/2016/10/039}{\emph{JCAP} {\bf 1610}
  (2016) 039}, [\href{http://arxiv.org/abs/1606.08474}{{\tt 1606.08474}}].

\bibitem{Caprini:2017vnn}
C.~Caprini, M.~C. Guzzetti and L.~Sorbo, \emph{{Inflationary magnetogenesis
  with added helicity: constraints from non-gaussianities}},
  \href{http://arxiv.org/abs/1707.09750}{{\tt 1707.09750}}.

\bibitem{Barnaby:2010vf}
N.~Barnaby and M.~Peloso, \emph{{Large Nongaussianity in Axion Inflation}},
  \href{http://dx.doi.org/10.1103/PhysRevLett.106.181301}{\emph{Phys. Rev.
  Lett.} {\bf 106} (2011) 181301}, [\href{http://arxiv.org/abs/1011.1500}{{\tt
  1011.1500}}].

\bibitem{Barnaby:2011vw}
N.~Barnaby, R.~Namba and M.~Peloso, \emph{{Phenomenology of a Pseudo-Scalar
  Inflaton: Naturally Large Nongaussianity}},
  \href{http://dx.doi.org/10.1088/1475-7516/2011/04/009}{\emph{JCAP} {\bf 1104}
  (2011) 009}, [\href{http://arxiv.org/abs/1102.4333}{{\tt 1102.4333}}].

\bibitem{Barnaby:2011qe}
N.~Barnaby, E.~Pajer and M.~Peloso, \emph{{Gauge Field Production in Axion
  Inflation: Consequences for Monodromy, non-Gaussianity in the CMB, and
  Gravitational Waves at Interferometers}},
  \href{http://dx.doi.org/10.1103/PhysRevD.85.023525}{\emph{Phys. Rev.} {\bf
  D85} (2012) 023525}, [\href{http://arxiv.org/abs/1110.3327}{{\tt
  1110.3327}}].

\bibitem{Sorbo:2011rz}
L.~Sorbo, \emph{{Parity violation in the Cosmic Microwave Background from a
  pseudoscalar inflaton}},
  \href{http://dx.doi.org/10.1088/1475-7516/2011/06/003}{\emph{JCAP} {\bf 1106}
  (2011) 003}, [\href{http://arxiv.org/abs/1101.1525}{{\tt 1101.1525}}].

\bibitem{Cook:2013xea}
J.~L. Cook and L.~Sorbo, \emph{{An inflationary model with small scalar and
  large tensor nongaussianities}},
  \href{http://dx.doi.org/10.1088/1475-7516/2013/11/047}{\emph{JCAP} {\bf 1311}
  (2013) 047}, [\href{http://arxiv.org/abs/1307.7077}{{\tt 1307.7077}}].

\bibitem{Shiraishi:2013kxa}
M.~Shiraishi, A.~Ricciardone and S.~Saga, \emph{{Parity violation in the CMB
  bispectrum by a rolling pseudoscalar}},
  \href{http://dx.doi.org/10.1088/1475-7516/2013/11/051}{\emph{JCAP} {\bf 1311}
  (2013) 051}, [\href{http://arxiv.org/abs/1308.6769}{{\tt 1308.6769}}].

\bibitem{Linde:2012bt}
A.~Linde, S.~Mooij and E.~Pajer, \emph{{Gauge field production in supergravity
  inflation: Local non-Gaussianity and primordial black holes}},
  \href{http://dx.doi.org/10.1103/PhysRevD.87.103506}{\emph{Phys. Rev.} {\bf
  D87} (2013) 103506}, [\href{http://arxiv.org/abs/1212.1693}{{\tt
  1212.1693}}].

\bibitem{Bugaev:2013fya}
E.~Bugaev and P.~Klimai, \emph{{Axion inflation with gauge field production and
  primordial black holes}},
  \href{http://dx.doi.org/10.1103/PhysRevD.90.103501}{\emph{Phys. Rev.} {\bf
  D90} (2014) 103501}, [\href{http://arxiv.org/abs/1312.7435}{{\tt
  1312.7435}}].

\bibitem{Erfani:2015rqv}
E.~Erfani, \emph{{Primordial Black Holes Formation from Particle Production
  during Inflation}},
  \href{http://dx.doi.org/10.1088/1475-7516/2016/04/020}{\emph{JCAP} {\bf 1604}
  (2016) 020}, [\href{http://arxiv.org/abs/1511.08470}{{\tt 1511.08470}}].

\bibitem{Garcia-Bellido:2016dkw}
J.~Garcia-Bellido, M.~Peloso and C.~Unal, \emph{{Gravitational waves at
  interferometer scales and primordial black holes in axion inflation}},
  \href{http://dx.doi.org/10.1088/1475-7516/2016/12/031}{\emph{JCAP} {\bf 1612}
  (2016) 031}, [\href{http://arxiv.org/abs/1610.03763}{{\tt 1610.03763}}].

\bibitem{Parker:1968mv}
L.~Parker, \emph{{Particle creation in expanding universes}},
  \href{http://dx.doi.org/10.1103/PhysRevLett.21.562}{\emph{Phys. Rev. Lett.}
  {\bf 21} (1968) 562--564}.

\bibitem{Kuzmin:1998kk}
V.~Kuzmin and I.~Tkachev, \emph{{Matter creation via vacuum fluctuations in the
  early universe and observed ultrahigh-energy cosmic ray events}},
  \href{http://dx.doi.org/10.1103/PhysRevD.59.123006}{\emph{Phys. Rev.} {\bf
  D59} (1999) 123006}, [\href{http://arxiv.org/abs/hep-ph/9809547}{{\tt
  hep-ph/9809547}}].

\bibitem{Chung:2011ck}
D.~J.~H. Chung, L.~L. Everett, H.~Yoo and P.~Zhou, \emph{{Gravitational Fermion
  Production in Inflationary Cosmology}},
  \href{http://dx.doi.org/10.1016/j.physletb.2012.04.066}{\emph{Phys. Lett.}
  {\bf B712} (2012) 147--154}, [\href{http://arxiv.org/abs/1109.2524}{{\tt
  1109.2524}}].

\bibitem{Adshead:2015kza}
P.~Adshead and E.~I. Sfakianakis, \emph{{Fermion production during and after
  axion inflation}},
  \href{http://dx.doi.org/10.1088/1475-7516/2015/11/021}{\emph{JCAP} {\bf 1511}
  (2015) 021}, [\href{http://arxiv.org/abs/1508.00891}{{\tt 1508.00891}}].

\bibitem{Adshead:2015jza}
P.~Adshead and E.~I. Sfakianakis, \emph{{Leptogenesis from left-handed neutrino
  production during axion inflation}},
  \href{http://dx.doi.org/10.1103/PhysRevLett.116.091301}{\emph{Phys. Rev.
  Lett.} {\bf 116} (2016) 091301}, [\href{http://arxiv.org/abs/1508.00881}{{\tt
  1508.00881}}].

\bibitem{Anber:2016yqr}
M.~M. Anber and E.~Sabancilar, \emph{{Chiral Gravitational Waves from Chiral
  Fermions}}, \href{http://dx.doi.org/10.1103/PhysRevD.96.023501}{\emph{Phys.
  Rev.} {\bf D96} (2017) 023501}, [\href{http://arxiv.org/abs/1607.03916}{{\tt
  1607.03916}}].

\bibitem{Alexander:2017bxe}
S.~Alexander, E.~McDonough and R.~Sims, \emph{{V-mode Polarization in Axion
  Inflation and Preheating}},
  \href{http://dx.doi.org/10.1103/PhysRevD.96.063506}{\emph{Phys. Rev.} {\bf
  D96} (2017) 063506}, [\href{http://arxiv.org/abs/1704.00838}{{\tt
  1704.00838}}].

\bibitem{Weinberg:2005vy}
S.~Weinberg, \emph{{Quantum contributions to cosmological correlations}},
  \href{http://dx.doi.org/10.1103/PhysRevD.72.043514}{\emph{Phys. Rev.} {\bf
  D72} (2005) 043514}, [\href{http://arxiv.org/abs/hep-th/0506236}{{\tt
  hep-th/0506236}}].

\bibitem{Durrer:2009ii}
R.~Durrer, G.~Marozzi and M.~Rinaldi, \emph{{On Adiabatic Renormalization of
  Inflationary Perturbations}},
  \href{http://dx.doi.org/10.1103/PhysRevD.80.065024}{\emph{Phys. Rev.} {\bf
  D80} (2009) 065024}, [\href{http://arxiv.org/abs/0906.4772}{{\tt
  0906.4772}}].

\bibitem{Ade:2015ava}
{\scshape Planck} collaboration, P.~A.~R. Ade et~al., \emph{{Planck 2015
  results. XVII. Constraints on primordial non-Gaussianity}},
  \href{http://dx.doi.org/10.1051/0004-6361/201525836}{\emph{Astron.
  Astrophys.} {\bf 594} (2016) A17},
  [\href{http://arxiv.org/abs/1502.01592}{{\tt 1502.01592}}].

\bibitem{Kobayashi:2014zza}
T.~Kobayashi and N.~Afshordi, \emph{{Schwinger Effect in 4D de Sitter Space and
  Constraints on Magnetogenesis in the Early Universe}},
  \href{http://dx.doi.org/10.1007/JHEP10(2014)166}{\emph{JHEP} {\bf 10} (2014)
  166}, [\href{http://arxiv.org/abs/1408.4141}{{\tt 1408.4141}}].

\bibitem{Frob:2014zka}
M.~B. Fr{\"o}b, J.~Garriga, S.~Kanno, M.~Sasaki, J.~Soda, T.~Tanaka et~al.,
  \emph{{Schwinger effect in de Sitter space}},
  \href{http://dx.doi.org/10.1088/1475-7516/2014/04/009}{\emph{JCAP} {\bf 1404}
  (2014) 009}, [\href{http://arxiv.org/abs/1401.4137}{{\tt 1401.4137}}].

\bibitem{Landete:2013lpa}
A.~Landete, J.~Navarro-Salas and F.~Torrenti, \emph{{Adiabatic regularization
  and particle creation for spin one-half fields}},
  \href{http://dx.doi.org/10.1103/PhysRevD.89.044030}{\emph{Phys. Rev.} {\bf
  D89} (2014) 044030}, [\href{http://arxiv.org/abs/1311.4958}{{\tt
  1311.4958}}].

\end{thebibliography}\endgroup
\bibliographystyle{JHEP}

\end{document}